\documentclass{article} % For LaTeX2e
\usepackage{iclr2025_conference, times}
\usepackage{algorithm}   % For algorithm environment
\usepackage{algpseudocode}
\usepackage{amsmath,amsfonts,bm}

\def\eqref#1{equation~\ref{#1}}
\def\1{\bm{1}}

\DeclareMathAlphabet{\mathsfit}{\encodingdefault}{\sfdefault}{m}{sl}
\SetMathAlphabet{\mathsfit}{bold}{\encodingdefault}{\sfdefault}{bx}{n}

\def\gY{{\mathcal{Y}}}

\newcommand{\E}{\mathbb{E}}

\newcommand{\R}{\mathbb{R}}

\newcommand{\softmax}{\mathrm{softmax}}

\usepackage{amsthm,amssymb,amsmath}
\usepackage{hyperref}
\usepackage{url}
\usepackage{comment}
\usepackage{booktabs}
\usepackage{graphicx}
\usepackage{wrapfig}
\usepackage{subcaption}
\usepackage{makecell}

\numberwithin{equation}{section}
\theoremstyle{plain}
\newtheorem{theorem}{Theorem}[section]
\newtheorem{lemma}[theorem]{Lemma}
\theoremstyle{remark}

\newtheorem*{definition}{Definition}

\usepackage{subcaption}

\usepackage{booktabs}
\usepackage{multirow}
\usepackage{array}

\usepackage{tikz}
\usetikzlibrary{arrows.meta, positioning}

\newcommand{\LRprob}{\textsc{LR}/prob}
\newcommand{\LDAprob}{\textsc{LDA}/prob}
\newcommand{\LDAjoint}{\textsc{LDA}/log-joint}

\usepackage{amsmath} % for \operatorname

\newcommand{\AdvAUROC}{\ensuremath{\mathrm{AUROC}}}

\newcommand{\scal}{\mathrm{scal}}

\newcommand{\TV}{\mathrm{TV}}
\newcommand{\Adv}{\mathrm{Adv}}
\newcommand{\cL}{\mathcal{L}}
\newcommand{\Y}{\mathcal{Y}}
\newcommand{\KL}{\mathrm{KL}}

\newcommand{\KLX}{\KL\!\big(P_X\Vert Q_X\big)}
\newcommand{\KLYX}{\E_{x\sim P_X}\,\KL\!\big(P_{Y\mid X=x}\Vert Q_{Y\mid X=x}\big)}

\usepackage{xargs}                      % Use more than one optional parameter in a new commands
\usepackage[pdftex,dvipsnames]{xcolor}  % Coloured text etc.
\usepackage[colorinlistoftodos,prependcaption,textsize=tiny]{todonotes}
\newcommandx{\unsure}[2][1=]{\todo[linecolor=red,backgroundcolor=red!25,bordercolor=red,#1]{#2}}
\newcommandx{\change}[2][1=]{\todo[linecolor=blue,backgroundcolor=blue!25,bordercolor=blue,#1]{#2}}
\newcommandx{\info}[2][1=]{\todo[linecolor=OliveGreen,backgroundcolor=OliveGreen!25,bordercolor=OliveGreen,#1]{#2}}
\newcommandx{\improvement}[2][1=]{\todo[linecolor=Plum,backgroundcolor=Plum!25,bordercolor=Plum,#1]{#2}}
\newcommandx{\thiswillnotshow}[2][1=]{\todo[disable,#1]{#2}}

\title{The Hidden Cost of Modeling p(x): Vulnerability to Membership Inference Attacks in Generative Text Classifiers}

\author{
\textbf{Owais Makroo$^{1\dagger}$\thanks{Work done during internship. Correspondence: makroo.owais@gmail.com}},\,
\textbf{Siva Rajesh Kasa$^{2\dagger}$},\,
\textbf{Sumegh Roychowdhury$^{2\dagger}$},\,
\textbf{Karan Gupta$^{2\dagger}$},\,\\[2pt]
\textbf{Nikhil Pattisapu$^{2\dagger}$},
\textbf{Santhosh Kasa$^2$ \& Sumit Negi$^2$}\\[2pt]
$^1$IIT Kharagpur, India $\quad$ $^2$Amazon.com Inc.\\
\texttt{\{kasasiva,sumegr,karaniis,npattisa\}@amazon.com}\\[2pt]
$^{\dagger}$\small Equal contribution.
}

\iclrfinalcopy % Uncomment for camera-ready version, but NOT for submission.
\begin{document}

\maketitle

\begin{abstract}
Membership Inference Attacks (MIAs) pose a critical privacy threat by enabling adversaries to determine whether a specific sample was included in a model's training dataset. Despite extensive research on MIAs, systematic comparisons between generative and discriminative classifiers remain limited. This work addresses this gap by first providing theoretical motivation for why generative classifiers exhibit heightened susceptibility to MIAs, then validating these insights through comprehensive empirical evaluation.
Our study encompasses discriminative, generative, and pseudo-generative text classifiers across varying training data volumes, evaluated on nine benchmark datasets. Employing a diverse array of MIA strategies, we consistently demonstrate that fully generative classifiers which explicitly model the joint likelihood $P(X,Y)$ are most vulnerable to membership leakage. Furthermore, we observe that the canonical inference approach commonly used in generative classifiers significantly amplifies this privacy risk.
These findings reveal a fundamental utility-privacy trade-off inherent in classifier design, underscoring the critical need for caution when deploying generative classifiers in privacy-sensitive applications. Our results motivate future research directions in developing privacy-preserving generative classifiers that can maintain utility while mitigating membership inference vulnerabilities \footnote{Code available at https://anonymous.4open.science/r/privacy-attacks-gendisc-classifiers-143E}.
\end{abstract}

\section{Introduction}

Text Classification (TC) is a fundamental task in Natural Language Processing (NLP), serving as the backbone for numerous applications including sentiment analysis, topic detection, intent classification, and document categorization \citep{Yogatama2017,castagnos2022simple,roychowdhury2024tackling,kasa-etal-2024-exploring,pattisapu2025leveraging}. As machine learning models have become increasingly sophisticated and widely deployed, concerns about their privacy implications have grown substantially. One of the most critical privacy vulnerabilities is the \textbf{Membership Inference Attack} (MIA), where an adversary attempts to determine whether a specific data point was included in a model's training set \citep{shokri2017membershipinferenceattacksmachine}. MIAs represent a fundamental threat to data privacy by exploiting differential model behaviors on training versus non-training data to infer membership in the training set \citep{shokri2017membershipinferenceattacksmachine,carlini2019secret, shejwalkar2021membership, song2022systematic, song2021systematic, pmlr-v258-feng25a}. The implications are particularly severe for sensitive personal data, potentially violating privacy expectations and regulatory requirements. Recent surveys have highlighted the growing sophistication of these attacks \citep{amit2024sok,pmlr-v258-feng25a}.

% \textbf{Predominant Focus on Discriminative Models.}
The majority of MIA research in TC has concentrated on discriminative models like BERT \citep{devlin2019bert}, which directly model $P(Y|X)$ and learn decision boundaries without explicitly modeling data distributions \citep{pmlr-v202-zheng23f,kasa2025generative}. Studies have revealed how factors such as overfitting, model capacity, and training data size influence attack success rates \citep{amit2024sok}.
Despite this discriminative focus, there has been renewed interest in generative classifiers for TC \citep{li2024generative}. Unlike discriminative models, generative classifiers explicitly model the joint distribution $P(X,Y) = P(X|Y)P(Y)$, offering compelling advantages such as: superior performance in low-data regimes \citep{kasa2025generative,Yogatama2017}, reduced susceptibility to spurious correlations \citep{li2024generative}, and principled uncertainty estimates via Bayes' rule \citep{bouguila2011bayesian}. The renaissance of generative classifiers in TC has been particularly bolstered through scalable model architectures including autoregressive models \citep{radford2019language}, discrete diffusion models \citep{lou2023discrete}, and generative masked language models \citep{wang2019bert}.

However, the very characteristics that make generative classifiers attractive explicit modeling of data distributions and superior performance with limited data raise important privacy questions. While MIAs have been extensively studied for discriminative models, a significant gap exists in understanding how different classification paradigms compare in their vulnerability to such attacks.
In this work, we present the first large-scale, systematic analysis of the vulnerability of transformer-based text classifiers to MIAs across a spectrum of modeling paradigms. Following \citet{kasa2025generative}, we consider three broad categories: (1) \textbf{discriminative models} such as encoder-style models \texttt{DISC}, which model the conditional distribution $P(Y|X)$; (2) \textbf{fully generative models} that explicitly model $P(X, Y)$, such as autoregressive (\texttt{AR}) or discrete diffusion models (\texttt{DIFF}); and (3) \textbf{pseudo-generative models}, such as Masked Language Models \texttt{MLM}, and pseudo-autoregressive \texttt{P-AR} models, where the label is appended at the end of the input sequence.

\paragraph{Contributions.}
To our knowledge, this work provides the first \emph{systematic} study of membership inference risk for \emph{generative} text classifiers, combining theory, controlled toy settings, and large-scale transformer-based experiments. 

\textbf{(1) Theory under a single-shadow black-box framework.} We formalize membership inference for generative vs.\ discriminative classifiers in a black-box setting with either probabilities or logits exposed. Our bounds decompose the optimal attack advantage into leakage from the marginal $P(X)$ and the conditional $P(Y\mid X)$, clarify when \emph{logits (joint scores)} can strictly dominate \emph{probabilities (conditionals)}. Using a simulation setting with tunable dimension, sample size, and class separation, we show that \emph{generative} classifiers leak more through \emph{log-joint} scores than discriminative posteriors do—quantitatively aligning with our theoretical predictions about marginal vs.\ conditional channels. 

\textbf{(2) First systematic analysis for \emph{MIA in text classification}.} Across multiple datasets and five model paradigms (discriminative, fully generative: AR and discrete diffusion, and pseudo-generative), we provide a head-to-head evaluation of MIA vulnerability under matched training protocols. We isolate the effects of (i) architectural factorization (\texttt{AR} vs.\ \texttt{p-AR}), (ii) \emph{output interface} (logits vs.\ probabilities), and (iii) data size. We find that fully generative models are consistently more vulnerable, with the strongest leakage observed when logits from $K$-pass scoring are exposed. 

\textbf{(3) Empirical analysis and practical guidance.} We show that different architectures yield distinct privacy–utility trade-offs, with generative models offering better low-sample accuracy and robustness benefits at the cost of higher leakage, while pseudo-generative models emerge as more privacy-conscious alternatives at higher data regmies. Building on these results, we provide actionable guidance on API exposure (favoring probabilities over logits), model choice, and training practices for privacy-sensitive deployments.

\section{Related Work and Background}
\label{sec:related-background}

\paragraph{Generative vs.\ discriminative classifiers.}
Classic analyses compare generative and discriminative learning on efficiency and asymptotics: discriminative models achieve lower asymptotic error, while generative models converge faster in low-data regimes \citep{Efron1975TheEO,Ng2001,Liang2008AnAA}. In text classification, recent work has renewed interest in generative classifiers that model $P(X,Y)=P(X\mid Y)P(Y)$, reporting advantages in calibration, uncertainty estimation, robustness to spurious correlations, and performance under limited data \citep{Yogatama2017,pmlr-v202-zheng23f,li2024generative,kasa2025generative}. Modern instantiations of generative classifiers in TC include autoregressive (AR) label-prefix classifiers \citep{radford2019language}, discrete diffusion models \citep{lou2023discrete}, and generative uses of masked LMs \citep{wang2019bert}. A practical drawback is that fully generative label-prefix AR classifiers typically require \emph{$K$-pass} inference—one forward pass per label $y_i$ to score $\log P(x,y_i)$—whereas discriminative models compute $P(Y\mid X)$ in a single pass; conversely, the generative formulation naturally supports Bayes-rule posteriors and principled uncertainty quantification via the decomposition $P(Y\mid X) \propto P(X\mid Y)P(Y)$ \citep{bouguila2011bayesian}. 
We investigate the generative text classifiers discussed in \citet{li2024generative,kasa2025generative} compare them with the well studied BERT-style encoder classifiers in this work. 

% We also consider \emph{pseudo-generative} factorizations (e.g., label-suffix/MLM-based) that use a single forward pass for classification while still leveraging generative training signals.

\paragraph{Membership inference background.}
MIAs exploit differences in a model’s behavior on train vs.\ non-train points. \citet{shokri2017membershipinferenceattacksmachine} introduced the multi–shadow-model paradigm for training an attack classifier on output vectors. \citet{salem2019ml} showed this can be simplified to \emph{single}-shadow or even \emph{no}-shadow attacks using confidence/loss statistics, and we \emph{adopt the single-shadow assumption} in our theoretical setup by modeling a proxy $Q$ alongside the target $P$ and reasoning about induced score laws $(P_S,Q_S)$. \citet{yeom2018privacy} established that overfitting is not the sole driver of MIAs: they connect attack advantage to generalization error via a loss-threshold attack and show that \emph{influence} of individual examples can cause leakage even when generalization error is small. Complementary systematization in ML-as-a-Service highlights how API exposure (labels/top-$k$/probabilities), shadow alignment, and data mismatch shape attack efficacy \citep{truex2018towards}.

\paragraph{Scope and assumptions.}
We study \emph{black-box} adversaries that query the classifier and observe either probabilities or pre-softmax logits (when available); \emph{white-box} access to parameters/gradients is out of scope. For fully generative label-prefix models, we assume $K$-pass inference is the canonical deployment mode; we analyze both logit- and probability-based attack surfaces and relate them to joint vs.\ conditional scoring used later in our theory. See Appendix~\ref{app:related-background} for an expanded survey, additional NLP-specific MIAs, and a detailed taxonomy of threat models.

\section{Motivation}
 Before discussing MIA attacks on on benchmark datasets, we first develop a theoretical account of how membership vulnerability manifests in generative classifiers, identify factors that exacerbate leakage (e.g., marginal memorization and weak conditional generalization), and formally compare what is revealed by joint vs.\ conditional exposures. We then instantiate these results in a controlled toy setting with a known data-generating process, showing that the empirical behavior of standard attacks mirrors the theoretical predictions.

\subsection{Preliminaries and Notation}

Let $\Omega$ denote the universe of all datapoints, where each datapoint $z \in \Omega$ can be decomposed into a feature--label pair $(x, y)$ with $x \in \mathcal{X}$ (features) and $y \in \mathcal{Y}$ (labels).  
% We assume there exists an underlying population distribution $\pi$ over $\mathcal{Z}$ from which samples are drawn.  
We consider two generative classifiers:  $P$: the \emph{target model}, which induces a joint probability distribution $P(X,Y)$ and $Q$: the \emph{shadow model}, trained independently on population data \citep{salem2019ml}, which induces its own probability distribution $Q(X, Y)$ which the attacker uses to determine sample membership. We are interested in quantifying the difference between $P$ and $Q$ in terms of their induced joint distributions over $(X, Y)$, which captures susceptibility to MIA. Let an \emph{attack signal} be any measurable function $S=S(\hat{p}(X,Y))$ of the model output (e.g., logits $(\log\hat p(x,y'))_{y'\in \gY}$, probabilities $\hat p(\cdot\mid x)= \softmax(\log\hat p(x,\cdot))$, or a scalar score $\log\hat p(x,y_i)$) which is exposed to the client and the attacker tries to come up with an optimal decision $\varphi$ rule based on the signal $S$ to determine the membership. 
% $\text{expo}\in\{ \hat p(x,y), \log\hat p(x,y), (\log\hat p(x,y'))_{y'\in \gY}, \hat p(\cdot\mid x)= \softmax(\log\hat p(x,\cdot)) \}$
Given any attack signal $S$, let $P_S:=\mathcal{L}(S\mid P)$ and $Q_S:=\mathcal{L}(S\mid Q)$ denote the pushforward laws under the target/shadow distributions. Intuitively, these are the 
score distributions the attacker tunes their threshold on: in our empirical 
evaluation (cf.\ §5), the standard MIA AUROC is measured by sweeping a decision threshold 
that gives full weight to members under $P$ versus non-members under $Q$.  For any (possibly randomized) decision rule $\varphi:\mathrm{range}(S)\!\to\![0,1]$, the achieved membership advantage
$\operatorname{Adv}_\varphi(S)\;:=\;\mathbb{E}_{P}\big[\varphi(S)\big]-\mathbb{E}_{Q}\big[\varphi(S)\big]$
is always upper-bounded by the total-variation distance between the pushforwards, and the latter cannot exceed the TV between the original joint distributions,
\[
\operatorname{Adv}_\varphi(S)\;\le\;\mathrm{TV}\big(P_S,Q_S\big) \;\le\;\mathrm{TV}\big(P,Q\big) = \sup_{A \in \mathcal{F}} | P(A) - Q(A) | = \tfrac{1}{2} \int_{\Omega} \big| p(\omega) - q(\omega) \big| \, d\mu(\omega)
\]
% \[
% \mathrm{TV}\big(P_S,Q_S\big)\;\le\;\mathrm{TV}\big(P,Q\big).
% \]
% Moreover, optimizing over all rules recovers the exact attainable advantage from $S$:
% \[
% \sup_{\;0\le \varphi \le 1}\operatorname{Adv}_\varphi(S)\;=\;\mathrm{TV}\big(P_S,Q_S\big).
% \]
% <I want to you introduce the pushforwards of S and talk about the advantage of any possible rule $\varphi$ and how it is upper bounded by the Total variation distance between pushforwards $TV(P_s,Q_s)$, which in turn is upperbounded by $TV(P,Q)$ - Give me the latex code for this in the main draft. Separately, Give me the latex code for the proof (the one you just mentioned above) to paste in Appendix>
where we assume $P$ and $Q$ are defined on the same measurable space $(\Omega, \mathcal{F})$ and $p$ and $q$ denote densities of $P$ and $Q$ with respect to a common dominating measure $\mu$ (this is done for ease of mathematical exposition). 
% , the \emph{total variation distance} is defined as
% \begin{equation}
%     \mathrm{TV}(P, Q) 
%     = \sup_{A \in \mathcal{F}} | P(A) - Q(A) | = \tfrac{1}{2} \int_{\Omega} \big| p(\omega) - q(\omega) \big| \, d\mu(\omega).
%     \label{eq:tvd_def}
% \end{equation}
% An equivalent variational form is
% \begin{equation}
%     \mathrm{TV}(P, Q) 
%     = \tfrac{1}{2} \int_{\Omega} \big| p(\omega) - q(\omega) \big| \, d\mu(\omega),
%     \label{eq:tvd_half_norm}
% \end{equation}
% where $p$ and $q$ denote densities of $P$ and $Q$ with respect to a common dominating measure $\mu$.

% It is well known that $\operatorname{TV}(P,Q)$ equals the maximum distinguishing advantage of any binary hypothesis test between $P$ and $Q$, and therefore equals the maximum achievable membership inference advantage ($MIA^* = TV(P_{XY}, Q_{XY})$).

It is well known that $\operatorname{TV}(P,Q)$ equals the maximum distinguishing advantage of any binary hypothesis test between $P$ and $Q$, and in the first result, we show that in the case of generative classifiers, this can be cleanly bounded using a generative and discriminative component.

% \subsection{Polished Theoretical results}

% % ============================
% % Section 3: Theoretical Results
% % ============================
% \section{Theoretical Results}
% We formalize membership advantage in the standard hypothesis-testing view: let $P$ (members) and $Q$ (non-members) be the induced distributions over $(X,Y)$ in the membership game, and let an \emph{attack signal} be any measurable function $S=S(X,Y)$ of the model output (e.g., logits, probabilities, or a scalar score). The optimal membership advantage is the total variation (TV) distance
% \[
% \Adv(S)\;=\;\TV\!\big(\,\cL(S\mid P)\,,\,\cL(S\mid Q)\,\big).
% \]
% We use Pinsker’s inequality $\TV(A,B)\le \sqrt{ \tfrac{1}{2}\KL(A\Vert B)}$ and the chain rule $\KL(P_{XY}\Vert Q_{XY})=\KL(P_X\Vert Q_X)+\E_{x\sim P_X}\KL(P_{Y\mid X=x}\Vert Q_{Y\mid X=x})$ throughout.

% \paragraph{Notation.}
% We write $\KL_{X}:=\KL(P_X\Vert Q_X)$ and $\KL_{Y\mid X}:=\E_{x\sim P_X}\KL(P_{Y\mid X=x}\Vert Q_{Y\mid X=x})$.
% When needed, we distinguish \emph{full-vector} (per-class) signals from \emph{scalar} signals.

% ----------------------------
% Lemma 1
% ----------------------------
\begin{lemma}[Two-way decomposition: upper and lower bounds]\label{lem:tv-decomp}
For the score-optimal attacker observing the feature-label pair $(X,Y)$ (equivalently, any sufficient statistic), the optimal advantage equals $\TV(P_{XY},Q_{XY})$ and satisfies
\begin{align}
\bigl|\;\TV(P_X,Q_X)\;& -\;\E_{x\sim P_X}\TV\!\bigl(P_{Y\mid X=x},Q_{Y\mid X=x}\bigr)\;\bigr|
\;\le\;
\TV(P_{XY},Q_{XY})
\;\le\; \\
&\TV(P_X,Q_X)\;+\;\E_{x\sim P_X}\TV\!\bigl(P_{Y\mid X=x},Q_{Y\mid X=x}\bigr). \nonumber
\end{align}
By Pinsker, any observable signal $S$ obeys
\[
\Adv(S)\;\le\;\TV(P_{XY},Q_{XY})
\;\le\;
\sqrt{\tfrac{1}{2}\,\KLX}\;+\;\sqrt{\tfrac{1}{2}\,\KLYX}.
\]
\end{lemma}

\noindent \textit{Discussion.}
Lemma~\ref{lem:tv-decomp} cleanly separates membership leakage of a generative classifier into a marginal term $\KL_X$ (learning $P(X)$) and a conditional term $\KL_{Y\mid X}$ (learning $P(Y\mid X)$), matching the spirit of the bound already introduced in \S3.1 (Theorem~1). This makes precise why modeling $P(X)$ can increase MIA risk. (See App.~B for the proof and for a KL-formulation mirroring \S3.1.) 

% ----------------------------
% Lemma 2
% ----------------------------
\begin{lemma}[Joint $\succeq$ Conditional under full-vector exposure]\label{lem:joint-ge-cond}
Let the model expose the per-class \emph{joint} score vector
$S_{\mathrm{joint}}(x)=\big(\log \hat p(x,y)\big)_{y\in\Y}$
and the \emph{conditional} score vector
$S_{\mathrm{cond}}(x)=\big(\hat p(y\mid x)\big)_{y\in\Y}=\softmax\big(S_{\mathrm{joint}}(x)\big)$.
Then for any membership game,
\[
\Adv\!\big(S_{\mathrm{joint}}\big)\;\ge\;\Adv\!\big(S_{\mathrm{cond}}\big),
\]
with equality iff the per-$x$ additive normalizer $\log \hat p(X)$ is $P$-a.s.\ equal under $P$ and $Q$ (i.e., it carries no marginal signal about membership).
\end{lemma}

\noindent \textit{Discussion.}
Lemma~\ref{lem:joint-ge-cond} says that when logits proportional to $\log \hat p(x,y)$ are exposed, passing to posteriors \emph{cannot} increase advantage (data-processing). Intuitively, softmax removes the shared $-\log\hat p(x)$ term and therefore discards whatever membership signal is present in $P(X)$.

% ----------------------------
% Theorem 3
% ----------------------------
\begin{theorem}[Scalar joint can dominate conditional under systematic marginal skew]\label{thm:scalar-joint-dominates}
Consider binary classification. Suppose the attacker receives either
(i) a \emph{scalar joint} score $S_{\mathrm{joint}}^{\scal}(X,Y):=\log \hat p(X,Y)$ or
(ii) a \emph{conditional} score $S_{\mathrm{cond}}(X,Y):=\hat p(Y\mid X)$.
Assume the \emph{member vs.\ non-member} conditionals satisfy the bounded likelihood-ratio condition:
there exist constants $0<\alpha\le \beta<\infty$ such that for $P_X$-a.e.\ $x$ and both labels $y$,
\[
\alpha\;\le\;\frac{P(y\mid x)}{Q(y\mid x)}\;\le\;\beta.
\]
Then there exists $c=c(\alpha,\beta)\in(0,1]$ such that
\[
\Adv\!\big(S_{\mathrm{joint}}^{\scal}\big)\;>\;\Adv\!\big(S_{\mathrm{cond}}\big)
\quad\text{whenever}\quad
c\,\KL_X\;>\;\KL_{Y\mid X}.
\]
An explicit choice is $c(\alpha,\beta)=\dfrac{\log\beta-\log\alpha}{1+\log\beta-\log\alpha}$.
\end{theorem}

\noindent \textit{Discussion.}
Theorem~\ref{thm:scalar-joint-dominates} addresses the practically important case where only a \emph{single} generative score is exposed (e.g., $\log$-likelihood for the observed label, or a label-agnostic scalar derived from the joint). Unlike Lemma~\ref{lem:joint-ge-cond}, scalar joint and conditional are \emph{not} deterministic transforms of each other; nonetheless, whenever the \emph{marginal} skew $\KL_{X}$ dominates the \emph{conditional} skew $\KL_{Y\mid X}$ (``systematic marginal skew'') and conditionals are not wildly different between $P$ and $Q$, the scalar joint channel is provably more susceptible.

\paragraph{Implications.}
(i) Exposing logits of a generative model (full vector) is always at least as risky as exposing posteriors.  
(ii) Even if only a single generative score is exposed, sufficiently strong marginal memorization makes the generative channel strictly more vulnerable than conditional outputs.  
(iii) The decomposition in Lemma~\ref{lem:tv-decomp} explains our empirical hierarchy: models that \emph{must} learn $P(X)$ (fully generative) leak through the marginal term in addition to the conditional term, inflating MIA advantage.

\subsection{Toy Illustration: Controlled Analysis of MIA Vulnerability}
\label{subsec:toy}

To validate our theoretical insights on the heightened vulnerability of generative classifiers, we conduct a controlled synthetic experiment that teases out key factors behind membership inference such as accuracy, signal/noise ratio, dimensionality, etc. Following \citet{li2024generative}, we use a toy setup of linear classifiers on linearly separable data, which strips away confounders, letting us directly study how marginal vs.\ conditional learning drives leakage before moving to complex real-world models.

% To validate our theoretical insights and provide concrete evidence for the heightened vulnerability of generative classifiers, we conduct a controlled synthetic experiment that isolates the key factors driving membership inference attacks. Following \citet{sagawa2020investigation, setlur2023bitrate, li2024generative}, we adopt a simplified toy setup - linear classifiers on linearly separable data - that strips away confounding factors and admits closed-form analysis, allowing us to directly isolate how marginal vs.\ conditional learning influences membership leakage and to cleanly validate the theoretical predictions before moving to complex real-world models.

\textbf{Experimental Setup.} We design a synthetic binary classification task where each input $x \in \mathbb{R}^d$ consists of two components: $x = [x_{\text{core}}, x_{\text{noise}}]$. The core feature $x_{\text{core}} \sim \mathcal{N}(y \cdot \mu, \sigma^2)$ correlates directly with the binary label $y \in \{-1, +1\}$, where $\mu$ controls class separation. The remaining $d-1$ coordinates are independent standard Gaussian noise.
% Crucially, we use matched train/test distributions to isolate membership effects from distribution shift.
We systematically vary key parameters: dimensionality $d \in \{16, 64, 256\}$, training size $n_{\text{train}} \in \{50, 200\}$, class separation $\mu \in \{0.05, 0.10, \ldots, 0.50\}$, and class balance $w \in \{0.1, 0.3, 0.5\}$. We compare the discriminative Logistic Regression (LR) with the generative Linear Discriminant Analysis (LDA), evaluating three membership inference scores: max-probability for both models, and log-joint likelihood for LDA.

\textbf{Notation (attack scores).}
\LRprob\ denotes the \emph{max-probability} (confidence) score from LR,
% \[
% s_{\mathrm{prob}}(x)=\max_{y\in\{-1,+1\}}\hat p_{\mathrm{LR}}(y\mid x).
% \]
\LDAprob\ is the same max-probability score computed from LDA posteriors $\hat p_{\mathrm{LDA}}(y\mid x)$; and \LDAjoint\ is the LDA \emph{log-joint} score.
% \[
% s_{\mathrm{logjoint}}(x)=\max_{y}\big\{\log P(y)+\log \mathcal{N}(x\mid \mu_y,\Sigma)\big\}.
% \]
All three are label-agnostic membership scores.
\[
\begin{aligned}
s_{\mathrm{prob}}(x)&=\max_{y\in\{-1,+1\}}\hat p_{\mathrm{LR}}(y\mid x),\qquad
s_{\mathrm{logjoint}}(x)&=\max_{y}\{\log P(y)+\log \mathcal{N}(x\mid \mu_y,\Sigma)\}.
\end{aligned}
\]
\textbf{MIA evaluation (AUROC).}
For a given score $s(\cdot)$ and trained model, we compute scores on training samples (members) and on an i.i.d.\ test set (non-members). Treating members as positives and non-members as negatives, we sweep a threshold on $s(\cdot)$ to obtain the ROC curve and report its area (AUROC), the standard practice for membership inference. We aggregate results over $5$ random seeds and plot mean curves with shaded std. deviation
% $\pm 1.96\times\mathrm{SEM}$ 
bands. Here we present the plots and analysis for the balanced case of $w=0.5$. For the imbalanced cases, the same is deferred to Appendix \ref{sec:toy_illustration}.

\begin{figure}[t]
\centering
\includegraphics[width=\linewidth]{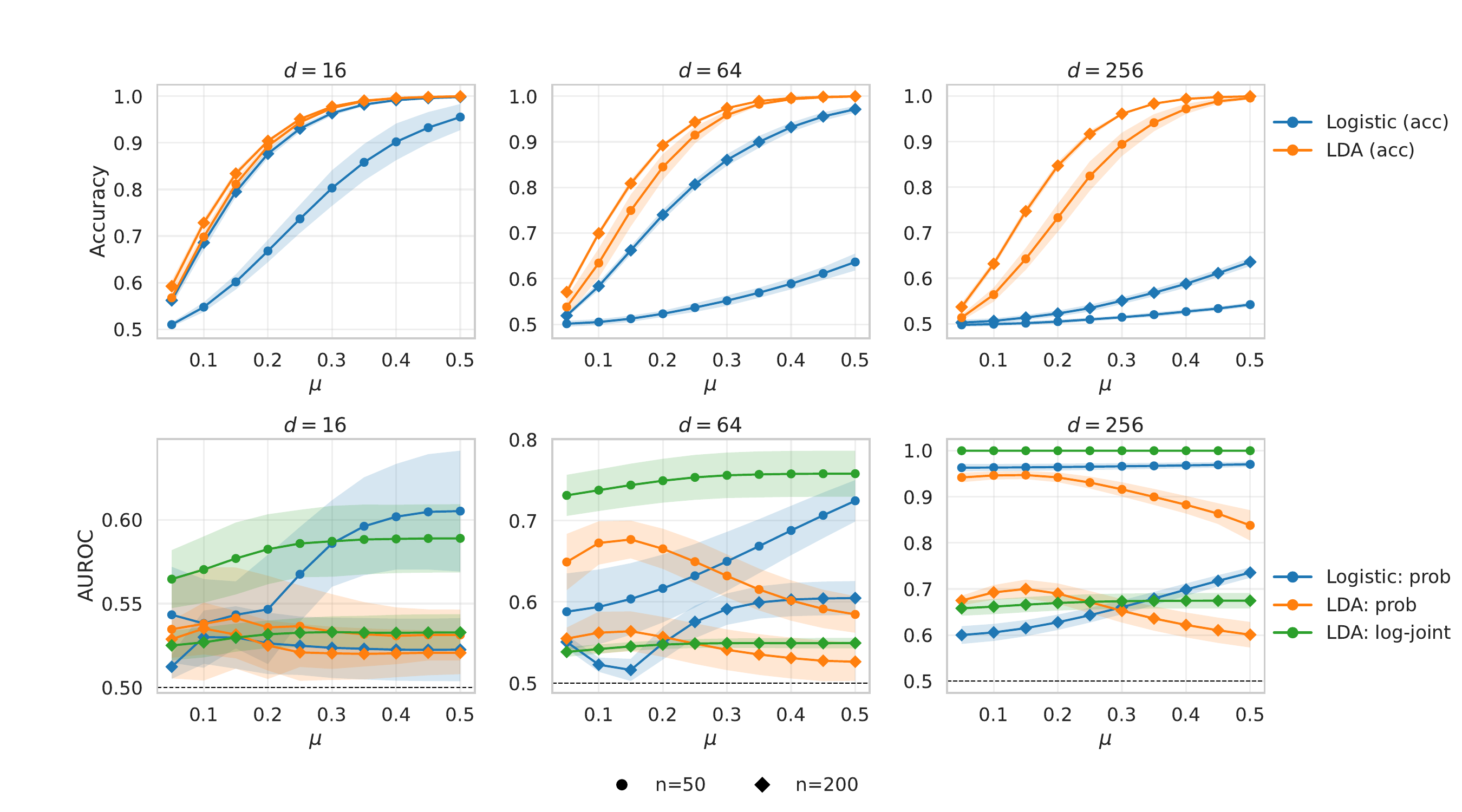}
\caption{\textbf{Membership inference vulnerability increases with model confidence and dimensionality.} Top row: test accuracy vs. core separation $\mu$. Bottom row: membership inference advantage (AUROC) vs. $\mu$. Columns correspond to $d\in\{16,64,256\}$. Colors denote model types and inference methods: Logistic Regression max-probability (blue), LDA max-probability (orange), LDA log-joint (green). Markers indicate training size $n_{\text{train}}\in\{50,200,2000\}$. Results averaged over 5 seeds with $\pm 1.96 \times$ SEM bands.}
\label{fig:mu-trends}
\end{figure}

\noindent Figure~\ref{fig:mu-trends} reveals several critical findings that support our theoretical predictions: (a) In the low-sample regime, \textbf{LDA is markedly more sample-efficient than LR}: for \(d\in\{16,64\}\), the accuracy achieved by LR with \(n_{\text{train}}=200\) is already matched (or exceeded) by LDA with \(n_{\text{train}}=50\); this accuracy gap widens as \(d\) increases (smaller \(n/d\)). (b) Comparing LDA’s two scores, the \textbf{joint score (\LDAjoint) consistently yields larger membership susceptibility} than the posterior max-probability (\LDAprob), with the gap growing as \(d\) increases or \(n/d\) decreases, underscoring the additional risk from exposing joint/likelihood values. (c) Comparing discriminative and generative posteriors, at small \(\mu\) \LRprob exhibits lower susceptibility than \LDAprob; as \(\mu\) grows, \LRprob’s susceptibility rises sharply with margin and can meet or exceed \LDAprob, whereas \LDAprob often flattens or slightly decreases while \LDAjoint remains high—consistent with likelihood dominating at larger separations. Apart from a single benign regime (balanced \(w=0.5\), large \(\mu\), low \(d\)), \LDAjoint exceeds \LRprob\ in susceptibility. (d) Increasing dimensionality \(d\) at fixed \(n\) lowers accuracy and increases membership advantage; in parallel, the across-seed standard deviation of both accuracy and AUROC narrows, yielding more consistent (but worse) accuracy and stronger, more stable membership signals in high dimensions.

The superior sample-efficiency of LDA is in part tied to the parameteric assumptions of LDA being satisfied by the data on which it is being fit. In order to tease out this we introduce a misspecification specifically a Huber-$\epsilon$ contamination \cite{huber1992robust} during the data geeneration process. The detailed plots are given in Appendix \ref{sec:toy_example_misspecification}. We notice that contamination reverses the generative LDA’s clean-data sample-efficiency edge in accuracy — LR is typically better—because a few large-norm replacements strongly distort shared-covariance estimation even with shrinkage. 
However, exposing density scale remains risky: \LDAjoint{} is the most susceptible membership score across most regimes we tested, particularly at high $d$ and small $n$. 
These controlled experiments provide concrete evidence that generative classifiers face fundamental privacy disadvantages, with the risk being particularly acute when exposing joint likelihood values or operating in high-dimensional, low-sample regimes.

\section{Experimental Setup}
\label{sec:experimental_setup}
% \section{Experimental Setup}
We evaluate privacy vulnerabilities in text classification by training multiple classifiers across datasets and subjecting them to diverse membership inference attacks (MIAs). Following \citet{li2024generative, kasa2025generative}, we study three main classifier families: 

% \subsection{Modeling Paradigms \& Attack Strategies}
\textbf{Discriminative} (\texttt{DISC/ENC}): Standard BERT-style encoders modeling $P(Y|X)$ using linear head on top of \texttt{[CLS]} token to directly map text $X$ to label $Y$. There's no explicit memorization signal in this modeling approach.

\textbf{Fully Generative:} Models that capture the joint distribution $P(X,Y)$ through:

(i) \textit{Label-Prefix Autoregressive} (\texttt{AR}) models generate text $x$ conditioned on a label prefix (e.g., \texttt{Positive: The film was a masterpiece.}). Classification is performed via logits using likelihood estimation, $\arg\max_{l \in K} \log P(x, y_l)$, in a $K$-pass fashion ($K =$ number of labels). Such models may be more vulnerable to MIAs since logits expose information about $P(X)$. Alternatively, applying a softmax yields probabilities: $\operatorname{softmax}\!\big(\log P(x,y_l)\big) = P(x,y_l)/P(x) = P(y_l|x)$, where the shared denominator $P(x)$ cancels across classes.

(ii) \textit{Discrete Diffusion Models} (\texttt{DIFF}) are trained on $(X,Y)$ pairs with a denoising objective. Following \citet{lou2023discrete}, noise gradually corrupts the input sequence to pure \texttt{[MASK]} tokens in the forward process, with original input reconstruction in the reverse process. At inference, the model predicts $y$ from \texttt{[MASK]}, conditional on $x$. We use \textit{Diffusion Weighted Denoising Score Entropy (DWDSE)} for logits, providing an upper bound on log-likelihood: $-\log p^\theta_0(x) \leq \mathcal{L}_{DWDSE}(x)$ under the ELBO.

\textbf{Pseudo-Generative:} This category represents a middle ground between discriminative and fully generative approaches. We explore using \textit{Masked Language Models} (\texttt{MLM}) trained for reconstructing masked tokens bi-directionally rather than full causal modeling. These model the pseudo-joint likelihood rather than the true joint $P(X,Y)$ \citep{wang-cho-2019-bert}.

All models utilize transformer-based architectures and are trained from scratch to avoid confounding effects from pre-training. Following \citet{kasa-etal-2024-exploring}, we evaluate three model size configurations: small (1 layer, 1 head), medium (6 layers, 6 heads), and large (12 layers, 12 heads). To enable fair comparison, we maintain comparable parameter counts across all architectures within each size configuration. Implementation details including model sizes and training hyperparameters are provided in Appendix \ref{sec:implementation_detail}.

\textbf{Attack Methodology: }
We examine two main classes of MIAs:(a) \textbf{Threshold-Based} attacks derive simple metrics from model outputs: (i) \textit{Max Probability}: $\max(P(y|x))$, (ii) \textit{Entropy}: $H(P(y|x)) = -\sum_i p_i \log p_i$, and (iii) \textit{Log-Loss} using cross-entropy on the true label. 
(b) \textbf{Model-Based} attacks train an explicit attack model by querying the target classifier with member and non-member samples, representing each using the model's output probability or logits vector concatenated with ground-truth labels, and training a Gradient Boosting Model (\texttt{GBM-logits} / \texttt{GBM-probs}) to predict membership status. Detailed attack implementations are provided in Appendix \ref{sec:implementation_detail}. Although there exists more sophisticated attacks \citep{shejwalkar2021membership,song2022systematic,amit2024sok} (details in Appendix \ref{subsec:app:MIA_attacks_others}), as will see in \S \ref{sec:results} that these basic attacks do a good job of revealing the differential vulnerability of generative and discriminative classifiers on TC.

% \subsection{Experimental Protocol}
\textbf{Dataset Details: }Our evaluation spans nine public text classification benchmarks
% \todo{can we the plots and tables corresponding to the full 9 datasets in the Appendix. Here we can add line saying that main paper has only 5, rest in appendix}
: \textbf{AG News} \cite{10.5555/2969239.2969312}, \textbf{Emotion} \cite{saravia-etal-2018-carer}, \textbf{Stanford Sentiment Treebank (SST2 \& SST5)} \cite{socher-etal-2013-recursive}, \textbf{Multiclass Sentiment Analysis}, \textbf{Twitter Financial News Sentiment}, \textbf{IMDb} \cite{maas-etal-2011-learning}, and \textbf{Hate Speech Offensive} \cite{hateoffensive}, covering diverse domains from sentiment analysis to topic classification. All models are trained from scratch using AdamW optimizer with early stopping to prevent overfitting, following \citet{li2024generative} and \citet{kasa2025generative}. We measure attack success using Area Under the ROC Curve (\textbf{AUROC}), where 1.0 indicates perfect attack and 0.5 indicates random guessing. Dataset characteristics are provided in Appendix \ref{sec:implementation_detail}.
% \todo{add table instead of paragraph in Appendix A.3}.

% \section{Approach}
% \input{approach}

% \section{Theoretical Results}
% \input{theoretical_results}

% \section{Framework for Analyzing Memorization}
% \input{methodology}

% \section{Experimental Methodology}
% \input{exp_details}

\section{Results \& Discussions}
\label{sec:results}
% In this section, we present a comprehensive analysis of privacy vulnerabilities in text classification models. We begin by establishing theoretical bounds on MIAs, providing a framework for understanding potential privacy risks. Our empirical evaluation then validates these theoretical insights across different model architectures, comparing discriminative models, fully generative models, and pseudo-generative masked language models (MLMs). We analyze how different model output representations (logits versus probabilities) and various attack strategies affect vulnerability. Finally, we examine how different approaches to modeling the joint distribution $P(X,Y)$ influence privacy leakage, introducing pseudo-autoregressive models as a privacy-utility balanced alternative to fully generative models. Through this analysis, we demonstrate the deep connection between architectural choices and privacy vulnerabilities.

Building on our theoretical analysis and synthetic experiments with LDA and LR (Section~\ref{subsec:toy}), we present comprehensive empirical evidence from real-world text classification scenarios. Our analysis examines: (1) privacy vulnerabilities across discriminative, fully generative, and pseudo-generative architectures, with patterns aligning with our controlled findings, (2) impact of model output representations (logits versus probabilities) on membership inference risk, and (3) how different approaches to modeling $P(X,Y)$ affect the privacy-utility trade-off. Through experiments on nine diverse datasets, we establish concrete relationships between architectural choices and privacy vulnerabilities, while identifying promising directions for privacy-preserving text classification.

\begin{figure*}[t]
  \centering
  \includegraphics[width=\textwidth]{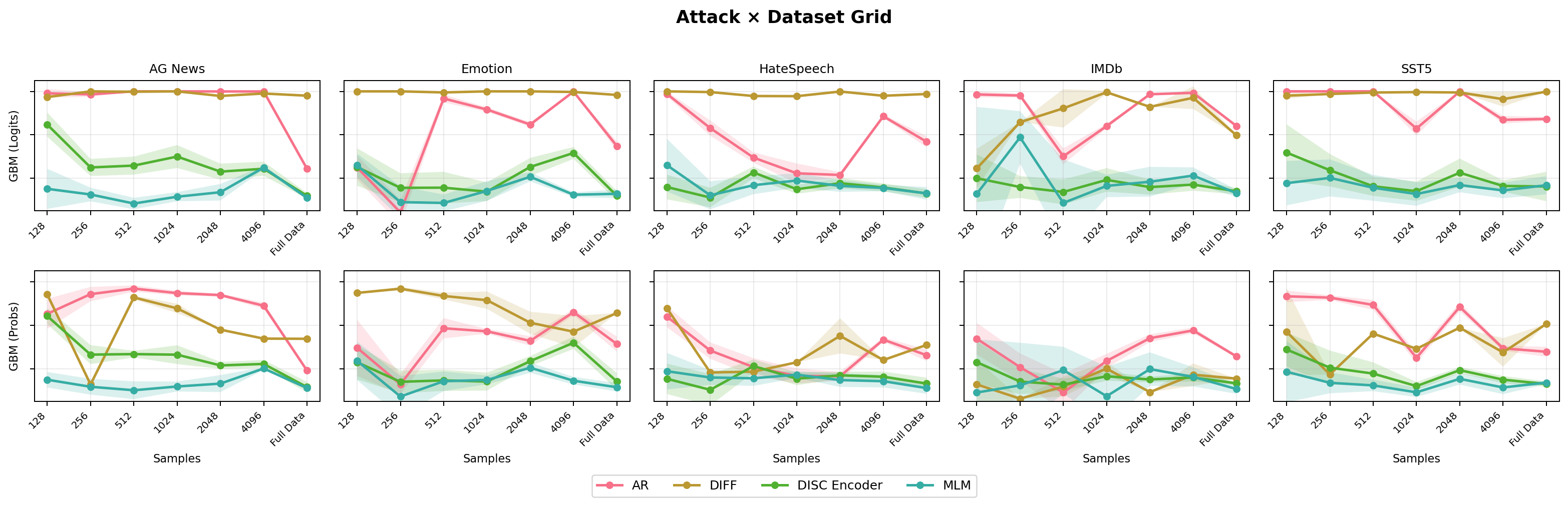}
  \caption{\small \textbf{[Best viewed in color]} MIA success rate (AUROC) compared across full-size model architectures with varying training dataset sizes. We evaluate fully generative classifiers (\texttt{GEN}, \texttt{DIFF}), a discriminative classifier (\texttt{DISC}), and pseudo-generative models (\texttt{MLM}). The \textbf{top row} displays attack performance using model \textbf{logits}, while the \textbf{bottom row} shows results using output \textbf{probabilities}. Higher AUROC values indicate increased privacy vulnerability. Results averaged across 5 random seeds.}
  \label{fig:1}
\end{figure*}
% \vspace{-1.5em}

Figure~\ref{fig:1} shows that fully generative models (\texttt{DIFF}, \texttt{GEN}) are consistently more vulnerable to MIAs than discriminative (\texttt{DISC}) and pseudo-generative (\texttt{MLM}) models across five datasets. For clarity, only \texttt{GBM-logits} and \texttt{GBM-probs} are shown; other attacks follow the same trend (see Section~\ref{sec:more_main_results}). Medium models behave similarly, while small \texttt{AR} are least susceptible, consistent with \citet{kasa2025generative}, who show these models behave nearly randomly. Full results across all datasets and model sizes are in Section~\ref{sec:more_main_results}.
% Our results (Figure~\ref{fig:1}) show that fully generative models (\texttt{DIFF}, \texttt{GEN}) are consistently more vulnerable to MIA than discriminative (\texttt{DISC}) and pseudo-generative (\texttt{MLM}) models on five datasets. For clarity, only \texttt{GBM-logits} and \texttt{GBM-probs} are shown; other attacks follow the same trend (details in Section~\ref{sec:more_main_results}). Medium models behave similarly, while small autoregressive models are least susceptible, aligning with \citet{kasa2025generative} who show such models act nearly random, making memorization comparisons un-informative. Refer\ref{sec:more_main_results} for results on all 9 datasets on all three model sizes.

These findings confirm our hypothesis: modeling $P(X,Y)$ forces generative models to capture both $P(Y|X)$ and $P(X)$, amplifying memorization risk compared to purely discriminative objectives.

Vulnerability does not vary monotonically with training size (Figure~\ref{fig:1}, Table~\ref{tab:training_samples_vs_attack}), consistent with \citet{amit2024sok}. Early stopping dampens overfitting in low-data regimes, masking expected trends. Removing early stopping (training 20 epochs on AGNews) restores the expected pattern: susceptibility decreases with larger training sets (refer Appendix~\ref{sec:more_main_results}).

We also the study how the MIA vulnerability changes with the representation of a class in the training sample in Appendix \ref{app:sec:class_representation} and find that vulnerability difference between majority and minority classes (i.e. the classes with
the highest and lowest representation in the training split) is high for \texttt{DISC,MLM} paradigms and it is relatively less pronounced for the generaive \texttt{AR, DIFF} paradigms. 

\textbf{Logits as a High-Bandwidth Privacy Leakage Channel: }
Our experiments show that membership inference attacks (MIA) using pre-softmax logits consistently outperform those based on post-softmax probabilities. As shown in Figure~\ref{fig:1}, logit-based attacks (top row) achieve higher AUC across all models and datasets than probability-based ones (bottom row). This aligns with prior work \cite{shokri2017membershipinferenceattacksmachine} and arises because logits preserve raw confidence scores, whereas softmax projects them onto a probability simplex, compressing information and reducing the attack surface.

The implications are significant: exposing logits through APIs even for calibration or temperature scaling greatly heightens privacy risk. Given that many ML APIs and frameworks expose logits by default \cite{finlayson2024logitsapiprotectedllmsleak}, practitioners should either restrict outputs to probabilities or add privacy-preserving safeguards when logits must be shared.

% \vspace{-1.2em}
\begin{wraptable}{r}{0.6\textwidth}
     \vspace{-1.6em}
    \centering
    \tiny
    \begin{tabular}{lcccc}
        \toprule
        \textbf{Attack} & \textbf{DISC} & \textbf{GEN} & \textbf{MLM} & \textbf{DIFF} \\
        \midrule
        Max Probability & 0.56 $\pm$ 0.05 & 0.67 $\pm$ 0.13 & 0.55 $\pm$ 0.06 & 0.51 $\pm$ 0.13 \\
        Entropy & 0.56 $\pm$ 0.05 & 0.63 $\pm$ 0.12 & 0.55 $\pm$ 0.06 & 0.60 $\pm$ 0.09 \\
        Log-Loss & 0.60 $\pm$ 0.06 & 0.76 $\pm$ 0.13 & 0.55 $\pm$ 0.08 & 0.65 $\pm$ 0.13 \\
        GBM-Probs & \textbf{0.62 $\pm$ 0.08} & \textbf{0.81 $\pm$ 0.13} & \textbf{0.56 $\pm$ 0.07} & \textbf{0.76 $\pm$ 0.16} \\
        \bottomrule
    \end{tabular}
    \caption{MIAs performance (AUROC) across different model architectures, averaged over all datasets. Higher values indicate greater privacy vulnerability, with the highest values in each column shown in \textbf{bold}.}
    \label{tab:model_family_vs_attack}
\end{wraptable}

The success of membership inference attacks also depends on the attack strategy's sophistication and the adversary's access to auxiliary information. Table~\ref{tab:model_family_vs_attack} reports results for both threshold-based and model-based attacks (refer Section~\ref{sec:experimental_setup}), focusing on probability-based methods since many attacks are incompatible with logits. 
These results illustrate the attack efficacy hierarchy, i.e. threshold-based attacks relying on output probabilities (\textit{Max Probability}, \textit{Entropy}) yield modest success, while incorporating ground-truth labels via \textit{Log-Loss} improves performance. The most effective attack, a Gradient Boosting Model (\textit{GBM}) trained on probability vectors and label information, notably excels for 
\texttt{AR} and \texttt{DIFF} models. We also find that model size exacerbates the privacy vulnerability in generative classifiers  (refer to Appendix \ref{app:sec:model_size_ablation}), similar to previous findings on \texttt{DISC} \citep{amit2024sok}. These findings underscore the urgent need for privacy defenses that remain effective across diverse adversarial capabilities and information access levels.

\textbf{The Impact of Factorization: Decomposing Leakage in $P(X,Y)$:}
In \citep{kasa2025generative}, the authors argue that fully generative models perform best in low-data regimes and should be preferred over discriminative models. However, our earlier results reveal that \texttt{AR} models exhibit significantly higher vulnerability to MIAs compared to \texttt{DISC}. To address this, we investigate an alternative modeling paradigm that reduces MIA risk without sacrificing classification performance. \textit{Pseudo-Autoregressive} (\texttt{P-AR}) models tackle this challenge by appending the label at the end of the input sequence, instead of modeling $P(X|Y)$ by pre-pending the label token,. Although this approach does not strictly capture $P(X|Y)$, recent work \citep{li2024generative} shows that label-appending often achieves better in-distribution accuracy than label-prepending. At inference, we can either use a K-pass run like \texttt{AR} to score each label and take the argmax, or a 1-pass run by selecting the predicted label from the final token’s distribution (this is the canonical approach for \texttt{P-AR}).
As evident from Table~\ref{tab:gen_vs_pseudo} (averaged across datasets) \texttt{P-AR} poses much lesser MIA risk compared to \texttt{AR}. However, \texttt{P-AR-kpass} exhibits similar vulnerability again similar to fully generative case. A few attacks are not stated here as they are qualitatively similar to Log-Loss.
%This setup enables efficient inference by requiring only a single forward pass to predict the label, unlike traditional generative models that necessitate $K$ forward passes (where $K$ is the number of labels) \citep{kasa2025generative}. The computational efficiency and performance benefits motivate the inclusion of such pseudo-generative models in our evaluation framework. 
% Notably, these approaches involve minimal architectural modifications to standard transformer models—typically requiring only changes to label placement or loss function computation—while preserving the core model design. This design principle allows for fair comparisons using widely available implementations that are accessible to practitioners, making these models particularly relevant for real-world deployment scenarios.

% This design principle enables fair comparisons through widely available implementations accessible to practitioners, making these models particularly relevant for real-world deployment scenarios. 

To explain this phenomenon, we next examine how different factorizations of the joint distribution $P(X,Y)$ influence privacy leakage. 
\textbf{(Label-Prefix)} \texttt{AR} are trained to generate the text $X$ conditioned on a label prefix $Y$, thereby factorizing the joint distribution as $P(X,Y) = P(Y)P(X|Y)$. Its primary focus is on learning the class-conditional data distribution. However, \textbf{(Label-Suffix)} \texttt{P-AR} are trained to generate the full sequence $(X, Y)$, with the label appended at the end. This architecture implicitly factorizes the joint distribution as $P(X,Y) = P(X)P(Y|X)$ requiring high-fidelity modeling of $P(X)$. While still generative, its final step of predicting $Y|X$, after generating all of $X$, mirrors a discriminative task (which is also why this falls under pseudo-generative paradigm). 

\begin{table*}[ht]
\vspace{-1.2em}
    \centering
    \setlength{\tabcolsep}{4.5pt} % Tighter columns
    \renewcommand{\arraystretch}{1.1} % Slightly looser rows for readability

    \begin{minipage}[t]{0.45\textwidth}
    % \vspace{-0.75em}
        \centering
        \tiny
        \begin{tabular}{lccc}
    \toprule
    \textbf{Attack} & \textbf{AR} & \textbf{P-AR} & \textbf{P-AR-kpass} \\
    \midrule
    % Max Probability & 0.61 $\pm$ 0.05 & 0.53 $\pm$ 0.03 & \textcolor{blue}{0.52 $\pm$ 0.01}  \\
    % Entropy         & 0.60 $\pm$ 0.05 & 0.53 $\pm$ 0.03 & \textcolor{blue}{0.52 $\pm$ 0.01} \\
    Log-Loss        & 0.66 $\pm$ 0.05 & \textcolor{blue}{0.56 $\pm$ 0.06} & 0.57 $\pm$ 0.06 \\
    GBM             & 0.77 $\pm$ 0.08 & \textcolor{blue}{0.55 $\pm$ 0.05} & 0.95 $\pm$ 0.04 \\
    \bottomrule
\end{tabular}

        \vspace{1ex}
        \captionof{table}{\small MIA performance (AUROC) comparing \textbf{Autoregressive} (AR) and \textbf{Pseudo-Autoregressive} (P-AR) models for large model size. The lowest susceptibility for an attack is highlighted in \textcolor{blue}{blue}.}
        \label{tab:gen_vs_pseudo}
    \end{minipage}
    \hspace{0.02\textwidth} % Reduce or increase spacing as needed
    \begin{minipage}[t]{0.465\textwidth}
        \centering
        \tiny
        \begin{tabular}{l|cc|cc}
            \toprule
            & \multicolumn{2}{c|}{\textbf{P-AR}} & \multicolumn{2}{c}{\textbf{AR}} \\
            \cmidrule(lr){2-3} \cmidrule(lr){4-5}
            \textbf{Dataset} & $P(X)$ & $P(X,Y)$ & $P(X)$ & $P(X,Y)$ \\
            \midrule
            SST-5       & \textbf{0.8185} & \textbf{0.8445} & 0.6204 & 0.6285 \\
            HateSpeech  & \textbf{0.8355} & \textbf{0.8771} & 0.4419 & 0.4256 \\
            Emotion     & \textbf{0.8872} & \textbf{0.9617} & 0.4780 & 0.4850 \\
            AGNews      & \textbf{0.6230} & \textbf{0.6299} & 0.2400 & 0.2492 \\
            IMDb        & \textbf{0.8379} & \textbf{0.8354} & 0.5232 & 0.5234 \\
            \bottomrule
        \end{tabular}
        \vspace{1ex}
        \captionof{table}{\small JSD between training and test distributions (here $Y: Y_{label}$). Higher values indicate greater data leakage.}
        \label{tab:px_divergence_filtered}
    \end{minipage}
\end{table*}

The output probabilities from \texttt{P-AR} correspond to $P(Y|X)$, which inherently leaks less information about sample membership than \texttt{AR}. The latter is more vulnerable because it effectively exposes $P(Y, X)$, a generative quantity, rather than the purely discriminative $P(Y|X)$.
However, changing the label position does not magically remove MIA risk. As Table~\ref{tab:px_divergence_filtered} demonstrates, \texttt{P-AR} still exhibits substantial memorization—evidenced by elevated Jensen–Shannon Divergence (JSD) when we compare train/test distributions of $P(X)$ and $P(X,Y)$. Crucially, these statistics are not exposed to an attacker when they interact with a \texttt{P-AR} model, since \texttt{P-AR} only reveals $P(Y\mid X)$. By contrast, \texttt{AR} and \texttt{P-AR-kpass} make joint/generative quantities (e.g., $P(X,Y)$) available, thereby exposing that memorization and increasing vulnerability. In short: label-suffix modeling can reduce the observable attack surface, but it does not eliminate underlying sample memorization. This distinction underlies our recommendation of \textbf{(label-suffix)} (\texttt{P-AR}) models in 1-pass fashion as a \textbf{safer alternative} to \texttt{AR} in terms of MIA vulnerability, complementing earlier conclusions made by \citet{kasa2025generative} from an accuracy stand point.

\textbf{Privacy-Utility Trade-Off: }
We show that different architectures yield distinct privacy–utility trade-offs (refer to Appendix \ref{sec:privacy-utility-analysis}), with our comprehensive analysis revealing that \texttt{DISC} models achieve the best overall utility performance  while maintaining good privacy protection , and \texttt{MLM} strategies provide superior privacy protection with steadily improving utility as model size increases. Conversely, we find that \texttt{DIFF} models, despite achieving competitive utility, exhibit severe privacy vulnerabilities with attack success rates exceeding 95\%, while \text{AR} models demonstrate concerning behavior where utility gains come at dramatic privacy costs, with attack success rates increasing as model complexity grows. Building on these results, we provide actionable guidance recommending \texttt{DISC} strategies with 6-12 layers for general applications, \texttt{MLM} strategies for privacy-critical systems, and cautioning against \texttt{DIFF} models in privacy-sensitive deployments due to their consistently high vulnerability to membership inference attacks.

% \section{Results}
% \input{results}

\section{Conclusion and Future Work}

% Our investigation reveals fundamental privacy vulnerabilities inherent in different text classification paradigms. Through theoretical analysis and extensive empirical validation, we demonstrate that generative models are systematically more vulnerable to membership inference attacks due to their explicit modeling of $P(X)$. Our theoretical framework, decomposing privacy risk into components from $P(X)$ and $P(Y|X)$, explains the observed hierarchy of vulnerability: generative models are most susceptible, followed by pseudo-generative approaches, with discriminative models showing the highest resistance. We show that this vulnerability is particularly pronounced when accessing logits rather than probabilities, and that sophisticated machine learning-based attacks can effectively exploit these vulnerabilities. Importantly, our analysis of different factorization strategies reveals that architectural choices in modeling $P(X,Y)$ fundamentally affect privacy leakage patterns.

% Future research should focus on developing defense mechanisms specifically tailored to generative architectures, investigating privacy-preserving training methods that can maintain utility while reducing memorization, and exploring the impact of model scale and pre-training on privacy vulnerabilities. These directions, combined with our current findings, will help practitioners make informed decisions about model selection and deployment in privacy-sensitive applications.

% \section{Conclusion}
This work presented the first systematic study of MIAs in
generative text classifiers, combining theoretical analysis, controlled toy settings,
and large-scale experiments. Our framework clarified how leakage arises from both
the marginal $P(X)$ and conditional $P(Y|X)$, with simulations confirming that
generative classifiers leak more information through log-joint scores than
discriminative posteriors. Empirically, we compared discriminative(\texttt{DISC}), fully generative (\texttt{AR, DIFF}), and pseudo-generative models(\texttt{MLM, P-AR}) across nine benchmarks. We found that
fully generative models are consistently more vulnerable to MIAs, with the
strongest leakage observed when logits are exposed. We further showed how
factorization (\texttt{AR} vs.\ \texttt{P-AR}), output interface (\textit{logits vs.\ probabilities}), and
training data size shape vulnerability, highlighting distinct privacy--utility
trade-offs. Notably, pseudo-generative models emerged as a safer alternative,
reducing observable leakage while maintaining competitive utility. We also provides actionable guidance: restrict API
outputs to probabilities, use generative models cautiously in sensitive settings,
and favor pseudo-generative approaches when balancing privacy and utility. Future
work should explore architectural and training modifications to retain the
benefits of generative modeling while mitigating these risks.

\bibliography{iclr2025_conference}
\bibliographystyle{iclr2025_conference}

\appendix

\section{Extended Related Work and Background}
\label{app:related-background}

\subsection{Generative vs.\ Discriminative Classifiers: Foundations to the Transformer Era}
Foundational theory contrasts generative and discriminative estimation: under correct modeling assumptions, discriminative learners achieve lower asymptotic error, while generative learners exhibit faster convergence with limited data \citep{Efron1975TheEO,Ng2001,Liang2008AnAA}. Hybrid approaches attempted to combine strengths \citep{raina2003classification}, and modern analyses revisit these trade-offs at scale, emphasizing calibration/uncertainty and bias--variance decompositions \citep{pmlr-v202-zheng23f}. 

In text classification (TC), generative models have seen a resurgence with transformers. Early RNN-based generative classifiers reported robustness to distribution shifts and favorable low-data behavior \citep{Yogatama2017}. Contemporary generative classifiers instantiate $P(X,Y)$ via (i) label-prefix AR scoring of $\log P(x,y)$ across labels \citep{radford2019language}; (ii) discrete diffusion with likelihood-surrogates/ELBO-style criteria \citep{lou2023discrete}; and (iii) generative uses of masked LMs \citep{wang2019bert}. Empirically, recent works document improved sample efficiency, calibration, and reduced shortcut reliance for generative TC \citep{kasa2025generative,li2024generative,jaini2024intriguing}. A practical consideration is \textbf{$K$-pass inference}: fully generative label-prefix AR classifiers evaluate one forward pass per label to obtain $\log P(x,y_i)$, in contrast to single-pass discriminative models computing $P(Y\mid X)$. On the other hand, generative formulations support principled uncertainty via Bayes rule,
\[
P(Y\mid X) \;=\; \frac{P(X\mid Y)P(Y)}{P(X)}\,,
\]
and enable likelihood-based diagnostics and priors \citep{bouguila2011bayesian}. We also consider \emph{pseudo-generative} factorizations (e.g., label-suffix/MLM variants) that use a single forward pass for classification while still leveraging generative training signals.

\subsection{Membership Inference Attacks (MIAs)}
\label{subsec:app:MIA_attacks_others}
\paragraph{From multi-shadow to minimal-shadow.}
\citet{shokri2017membershipinferenceattacksmachine} introduced the \emph{shadow-model} paradigm: train multiple proxies that mimic the target, collect outputs on member/non-member samples, and train an attack classifier. \citet{salem2019ml} showed that effective MIAs often require \emph{far less} attacker infrastructure: a \emph{single} shadow model—or even \emph{no} shadow at all—can suffice using confidence-/loss-based statistics. In this paper, \textbf{we adopt the single-shadow assumption} in analysis: we posit a proxy $Q$ trained similarly to the target $P$ and develop decision rules using the induced score laws $(P_S,Q_S)$ that arise from logits or probabilities.

\paragraph{Overfitting vs.\ influence.}
\citet{yeom2018privacy} connect membership advantage to generalization error with a simple loss-threshold attack, but crucially point out that \emph{influence} of specific samples can yield leakage even when generalization error is small; thus overfitting is sufficient but not necessary for MIAs. This perspective complements broader observations that memorization and model capacity correlate with vulnerability, while regularization and early stopping can attenuate leakage.

\paragraph{Systematization and API exposure.}
MIAs have been systematized for ML-as-a-Service (MLaaS) by examining how output exposure (labels only, top-$k$, full probability vectors), shadow alignment, and data distribution mismatch affect success \citep{truex2018towards}. Subsequent evaluations find that strong black-box attacks based on confidence/entropy/loss can rival more complex settings \citep{song2021systematic}, and NLP-specific studies report that simple threshold attacks can be surprisingly competitive in text classification, with user-level leakage sometimes exceeding sample-level leakage \citep{shejwalkar2021membership}.

\subsection{Threat Models, Outputs, and Our Scope}
\paragraph{Threat models.}
We distinguish \emph{black-box} adversaries (query access to outputs only), \emph{gray-box} (limited internals such as losses or activations), and \emph{white-box} (parameters/gradients). Our study focuses on \textbf{black-box} MIAs where the API exposes either (i) post-softmax probabilities $P(Y\mid X)$ or (ii) pre-softmax \emph{logits} that, in fully generative label-prefix AR classifiers, are proportional to joint scores $\log P(X,Y)$. For label-prefix AR models we assume \textbf{$K$-pass} inference is the canonical deployment mode.

\paragraph{Outputs and leakage channels.}
Probability vectors emphasize the conditional $P(Y\mid X)$, while logits in label-prefix AR expose additive joint components $\log P(X,Y)$ over labels. We analyze both surfaces empirically and theoretically by comparing attack performance built from signals $S$ with induced laws $(P_S,Q_S)$ under the target $P$ and shadow $Q$.

\paragraph{Scope summary.}
We restrict attention to black-box attackers with output access (probabilities or logits), assume the availability of ground-truth labels for attack training/selection, and treat inference cost as negligible for fairness across architectures. White-box attacks, knowledge-distillation/trajectory-based attacks, and defenses like DP-SGD are out of scope for this paper, though we discuss them qualitatively where relevant in the main text.

\section{Experimental Methodology}
\label{sec:implementation_detail}
% \section*{Appendix A: Detailed Experimental Methodology}
\subsection{Training Protocol}

Follwong the \citet{li2024generative, kasa-etal-2024-exploring}, we adopt the \texttt{bert-base-uncased}\footnote{\url{https://huggingface.co/google-bert/bert-base-uncased}} architecture as the backbone for both \textbf{DISC} and \textbf{MLM} experiments, trained from scratch without pretrained weights. This model has $\sim$110M parameters, with 12 encoder layers, 12 attention heads, and hidden size 768. All experiments were repeated with 5 random seeds, reporting mean and standard deviation in the main paper.

For \textbf{DISC} experiments, we performed a grid search over learning rates \{\texttt{1e-5}, \texttt{2e-5}, \texttt{3e-5}, \texttt{4e-5}, \texttt{5e-5}\}, batch sizes \{32, 64, 128, 256\}, and a fixed sequence length of 512. Training ran for 30 epochs on all datasets without early stopping. \textbf{MLM} experiments used the same search space but were trained for 200 epochs due to the added difficulty of masked token prediction. Introducing early stopping often led to worse checkpoints, since validation loss typically decreased slowly even after long plateaus.

For \textbf{AR} and \textbf{P-AR} we used the \texttt{GPT-2} base\footnote{\url{https://huggingface.co/openai-community/gpt2}} (137M parameters), trained as causal LMs to minimize next-token prediction loss on concatenated input–label sequences. A grid search was conducted with the same hyperparameter ranges, and models were trained up to 100 epochs with early stopping (patience 10).

Our \textbf{DIFF} experiments used the Diffusion Transformer \cite{peebles2023scalable}, essentially a vanilla transformer encoder augmented with time-conditioned embeddings, yielding $\sim$160M parameters. To control for model size, we also scaled Encoder/MLM models to 160M parameters by adding layers, but performance did not improve, so we retained original sizes. For diffusion-specific settings, we used batch size 64, learning rate \texttt{3e-4}, 200K iterations, and a geometric noise schedule spanning $10^{-4}$ to 20 \cite{lou2023discrete}. The absorbing transition matrix was:
\[
Q_{\text{absorb}} = 
\begin{bmatrix}
-1 & 0 & \cdots & 0 \\
0 & -1 & \cdots & 0 \\
\vdots & \vdots & \ddots & \vdots \\
1 & 1 & \cdots &0
\end{bmatrix}
\]

All experiments were trained on eight NVIDIA A100 GPUs. Training times (in hours) for full-data runs are shown in Table \ref{tab:training_time}.  

Inference latency varies substantially across methods (Table \ref{tab:inference_latency}). ENC and MLM are fastest, requiring a single forward pass. AR requires $|K|$ passes, which can be parallelized but increases compute. DIFF is slowest, taking $\sim$20–100$\times$ longer than ENC/MLM due to iterative denoising. For instance, on an A100 with batch size 1024 and sequence length 128, ENC/MLM run in 0.03s (3.3M params) to 1.3s (120M params), whereas DIFF takes 16–25s.

\begin{table}[t]
\centering
\footnotesize % readable but compact
\setlength{\tabcolsep}{4pt} % tighter spacing
\begin{tabular}{lccccc}
\toprule
\textbf{Config} & DISC & P-AR & AR & MLM & DIFF \\
\midrule
(1L,1H)   & 1--2 & 2--4  & 2--4  & 1--4  & 1--4  \\
(6L,6H)   & 1--3 & 3--7  & 3--7  & 3--7  & 2--6  \\
(12L,12H) & 2--5 & 5--10 & 5--10 & 5--10 & 5--12 \\
\bottomrule
\end{tabular}
\caption{Training time (hrs) ranges across datasets for each configuration and approach.}
\label{tab:training_time}
\end{table}

\begin{table}[htp!]
\centering
% \scriptsize % smaller font
\footnotesize
\setlength{\tabcolsep}{3pt} % reduce column spacing
\begin{tabular}{lccccc}
\toprule
\textbf{Model Size} & Parameters & DISC & MLM & AR & DIFF \\
\midrule
Small  & 3.3M   & 0.027 & 0.027 & 0.058 & 16.2  \\
Medium & 30.3M  & 0.292 & 0.292 & 0.510 & 20.52 \\
Large  & 120.4M & 1.260 & 1.260 & 2.070 & 24.8  \\
\bottomrule
\end{tabular}
\caption{Model Size vs. Inference Latency (avg wall-clock time per batch in seconds).}
\label{tab:inference_latency}
\end{table}

% All models use a transformer architecture with 6 layers, 8 attention heads, and embedding dimension 512. Models are trained using AdamW optimizer (learning rate $3 \times 10^{-5}$, weight decay 0.01), batch size 32, and maximum sequence length 256 tokens. We apply linear warmup over the first 10\% of updates followed by linear decay. Early stopping is used with patience of 20 epochs based on validation accuracy. Training runs on NVIDIA RTX 8000 and A100 GPUs using mixed-precision training.

\subsection{Attack Implementation Details} 
For model-based attacks, we employ a Gradient Boosting Classifier with 100 estimators, maximum depth of 3, and learning rate of 0.1. The attack model's input features comprise the target model's output probability vector concatenated with one-hot encoded ground truth labels. Threshold-based attacks use raw model outputs with optimal thresholds determined on a validation set.

\subsection{Dataset Characteristics}

\renewcommand{\arraystretch}{1.3}

\begin{table*}[htp!]
\centering
\resizebox{\textwidth}{!}{%
\begin{tabular}{
    l
    c
    c
    >{\centering\arraybackslash}m{1.2cm}
    >{\centering\arraybackslash}m{1.2cm}
    >{\centering\arraybackslash}m{4.0cm}
    >{\centering\arraybackslash}m{4.0cm}
}
\toprule
\textbf{Dataset} & \textbf{Examples} & \textbf{Classes} 
& \multicolumn{2}{c}{\textbf{Avg. Tokens}} 
& \multicolumn{2}{c}{\textbf{Label Distribution (\%)}} \\
\cmidrule(lr){4-5} \cmidrule(lr){6-7}
 & (Train / Test) &  & Train & Test & Train & Test \\
\midrule
IMDb & 25,000 / 25,000 & 2 & 313.9 & 306.8 & 0-1: 50.0 each & 0-1: 50.0 each \\
AG News & 120,000 / 7,600 & 4 & 53.2 & 52.8 & 0–3: 25.0 each & 0–3: 25.0 each \\
Emotion & 16,000 / 2,000 & 6 & 22.3 & 21.9 & \makecell[c]{0: 29.2, 1: 33.5, 2: 8.2,\\3: 13.5, 4: 12.1, 5: 3.6} & \makecell[c]{0: 27.5, 1: 35.2, 2: 8.9,\\3: 13.8, 4: 10.6, 5: 4.1} \\
HateSpeech & 22,783 / 2,000 & 3 & 30.0 & 30.2 & 0: 5.8, 1: 77.5, 2: 16.7 & 0: 5.5, 1: 76.6, 2: 17.9 \\
MultiClass Sentiment & 31,232 / 5,205 & 3 & 26.6 & 26.9 & 0: 29.2, 1: 37.3, 2: 33.6 & 0: 29.2, 1: 37.0, 2: 33.8 \\
Rotten Tomatoes & 8,530 / 1,066 & 2 & 27.4 & 27.3 & 0-1: 50.0 each & 0-1: 50.0 each \\
SST2 & 6,920 / 872 & 2 & 25.2 & 25.5 & 0: 47.8, 1: 52.2 & 0: 49.1, 1: 50.9 \\
SST5 & 8,544 / 1,101 & 5 & 25.0 & 25.2 & \makecell[c]{0: 12.8, 1: 26.0, 2: 19.0,\\3: 27.2, 4: 15.1} & \makecell[c]{0: 12.6, 1: 26.3, 2: 20.8,\\3: 25.3, 4: 15.0} \\
Twitter & 9,543 / 2,388 & 3 & 27.6 & 27.9 & 0: 15.1, 1: 20.2, 2: 64.7 & 0: 14.5, 1: 19.9, 2: 65.6 \\
\bottomrule
\end{tabular}
}
\caption{Dataset statistics showing training and test split sizes, number of classes, mean token length, and label distribution percentages.}
\label{tab:dataset_stats}
\end{table*}

\section{Appendix B: Proofs for Section 3}

\subsection{Preliminaries and Notation}

Let $\mathcal{Z}$ denote the universe of all datapoints, where each datapoint $z \in \mathcal{Z}$ can be decomposed into a feature--label pair $(x, y)$ with $x \in \mathcal{X}$ (features) and $y \in \mathcal{Y}$ (labels).  
We assume there exists an underlying population distribution $\pi$ over $\mathcal{Z}$ from which samples are drawn.  

We consider two models:  
\begin{itemize}
    \item $P$: the \emph{target model} (running in production), which induces a joint score distribution $P(X,Y)$. 
    \item $Q$: the \emph{shadow model}, trained independently on population data, which induces its own score distribution $Q(X, Y)$ which the attacker uses to determine sample membership.  
\end{itemize}

We are interested in quantifying the difference between $P$ and $Q$ in terms of their induced joint distributions over $(X, Y)$, which captures susceptibility to \emph{membership inference attacks} (MIA).  

\subsection{Total Variation Distance: Definition}

For two probability distributions $P$ and $Q$ on the same measurable space $(\Omega, \mathcal{F})$, the \emph{total variation distance} is defined as
\begin{equation}
    \mathrm{TV}(P, Q) 
    = \sup_{A \in \mathcal{F}} | P(A) - Q(A) |.
    \label{eq:tvd_def}
\end{equation}
An equivalent variational form is
\begin{equation}
    \mathrm{TV}(P, Q) 
    = \tfrac{1}{2} \int_{\Omega} \big| p(\omega) - q(\omega) \big| \, d\mu(\omega),
    \label{eq:tvd_half_norm}
\end{equation}
where $p$ and $q$ denote densities of $P$ and $Q$ with respect to a common dominating measure $\mu$.

It is well known that $\operatorname{TV}(P,Q)$ equals the maximum distinguishing advantage of any binary hypothesis test between $P$ and $Q$, and therefore equals the maximum achievable membership inference advantage ($MIA^* = TV(P_{XY}, Q_{XY})$).

\subsection{Decomposition into Marginal and Conditional Terms}

Writing distributions over $(X, Y)$ using Bayes' Rule as
\[
P(x,y) = P(x) P(y \mid x), \quad 
Q(x,y) = Q(x) Q(y \mid x),
\]
the total variation distance between $P$ and $Q$ is
\begin{align}
    \mathrm{TV}(P_{XY}, Q_{XY}) 
    &= \tfrac{1}{2} \int_{\mathcal{X} \times \mathcal{Y}} \Big| P(x) P(y \mid x) - Q(x) Q(y \mid x) \Big| dxdy. \label{eq:tv_joint}
\end{align}

We expand by adding and subtracting the cross term $P(x) Q(y \mid x)$:
\begin{align}
    &\Big| P(x) P(y \mid x) - Q(x) Q(y \mid x) \Big| \nonumber \\
    &= \Big| P(x) P(y \mid x) - P(x) Q(y \mid x) + P(x) Q(y \mid x) - Q(x) Q(y \mid x) \Big|. \label{eq:add_subtract}
\end{align}

Applying the triangle inequalities yields both lower and upper bounds.

\subsection{Lower Bound via Reverse Triangle Inequality}

Using the reverse triangle inequality $|a+b| \geq \big| |b| - |a| \big|$, we obtain
\begin{align}
    \mathrm{TV}(P_{XY}, Q_{XY}) 
    &\geq \tfrac{1}{2} \Bigg| 
    \int_{\mathcal{X} \times \mathcal{Y}} 
            \Big( P(x) Q(y \mid x) - Q(x) Q(y \mid x) \Big) dxdy - \\ &\quad
        \int_{\mathcal{X} \times \mathcal{Y}} 
            \Big( P(x) P(y \mid x) - P(x) Q(y \mid x) \Big) dxdy \nonumber
    \Bigg|. \label{eq:tv_lower_start}
\end{align}

Evaluating the two terms separately:

1. For the first integral (difference of marginals with $Q(y \mid x)$ fixed):
\[
\int_{\mathcal{X} \times \mathcal{Y}} 
    \big( P(x) Q(y \mid x) - Q(x) Q(y \mid x) \big) dxdy 
= \int_{\mathcal{X}} (P(x) - Q(x)) \Big( \int_{\mathcal{Y}} Q(y \mid x) dy \Big) dx.
\]
Since $\int_{\mathcal{Y}} Q(y \mid x) dy = 1$, this simplifies to
\[
\int_{\mathcal{X}} (P(x) - Q(x)) dx = 2 \, \mathrm{TV}(P_X, Q_X).
\]

2. For the second integral (difference of conditionals at fixed $P(x)$):
\[
\int_{\mathcal{X} \times \mathcal{Y}} 
    \big( P(x) P(y \mid x) - P(x) Q(y \mid x) \big) dxdy 
= \int_{\mathcal{X}} P(x) \int_{\mathcal{Y}} \big( P(y \mid x) - Q(y \mid x) \big) dy \, dx.
\]
The inner integral is exactly $2 \, \mathrm{TV}(P(\cdot \mid x), Q(\cdot \mid x))$. Hence
\[
= 2 \int_{\mathcal{X}} P(x) \, \mathrm{TV}(P(\cdot \mid x), Q(\cdot \mid x)) \, dx 
= 2 \; \mathbb{E}_{x \sim P_X}\!\Big[\operatorname{TV}\big(P(Y\mid x),Q(Y\mid x)\big)\Big]
\]

Combining, we obtain the lower bound:
\begin{equation}
    \mathrm{TV}(P_{XY}, Q_{XY})
    \;\geq\; \Big| \mathrm{TV}(P_X, Q_X) - \; \mathbb{E}_{x \sim P_X}\!\Big[\operatorname{TV}\big(P(Y\mid x),Q(Y\mid x)\big)\Big]  \Big|.
    \label{eq:tv_lower}
\end{equation}

\subsection{Upper Bound via Forward Triangle Inequality}

Applying the forward triangle inequality $|a+b| \leq |a| + |b|$ to \eqref{eq:add_subtract}, we obtain
\begin{align}
    \mathrm{TV}(P_{XY}, Q_{XY})
    &\leq \tfrac{1}{2} \int_{\mathcal{X} \times \mathcal{Y}} 
        \Big( | P(x) P(y \mid x) - P(x) Q(y \mid x) | 
        + | P(x) Q(y \mid x) - Q(x) Q(y \mid x) | \Big) dxdy.
\end{align}

Evaluating as before, this becomes
\begin{equation}
    \mathrm{TV}(P_{XY}, Q_{XY})
    \;\leq\; \mathrm{TV}(P_X, Q_X) + \int_{\mathcal{X}} P(x) \, \mathrm{TV}(P(\cdot \mid x), Q(\cdot \mid x)) dx
    \label{eq:tv_upper}
\end{equation}

\subsection{Decomposition via Pinsker’s Inequality}

Pinsker’s inequality states that for any two distributions $R, S$,
\[
\mathrm{TV}(R, S) \;\leq\; \sqrt{\tfrac{1}{2} \, D_{\mathrm{KL}}(R \,\|\, S)}.
\]

Applying this to \eqref{eq:tv_upper} yields
\begin{align}
    \mathrm{TV}(P_{XY}, Q_{XY})
    &\leq \sqrt{\tfrac{1}{2} D_{\mathrm{KL}}( P_X \,\|\, Q_X ) } + \int_{\mathcal{X}} P(x) \, 
        \sqrt{\tfrac{1}{2} D_{\mathrm{KL}}( P(\cdot \mid x) \,\|\, Q(\cdot \mid x)) } \, dx
    \label{eq:tv_pinsker}
\end{align}

We can re-write the first term as expectation similar to lower-bound derivation yielding
\begin{align}
    \mathrm{TV}(P_{XY}, Q_{XY})
    &\leq \sqrt{\tfrac{1}{2} D_{\mathrm{KL}}( P_X \,\|\, Q_X ) } + \mathbb{E}_{x \sim P_X}\!\Big[ 
        \sqrt{\tfrac{1}{2} D_{\mathrm{KL}}( P(\cdot \mid x) \,\|\, Q(\cdot \mid x)) } \; \Big]
    \label{eq:tv_pinsker_expected}
\end{align}

% This bound decomposes the divergence between the joint distributions into a conditional term (expected KL divergence of label distributions at fixed $x$) and a marginal term (KL divergence of input distributions). 
\subsection{Interpretation}
\label{proof:interpret}
Both the lower bound (Eq.~\ref{eq:tv_lower}) and upper bound (Eq.~\ref{eq:tv_pinsker_expected}) decomposes the $MIA^*$ into two contributing terms:
\begin{itemize}
    \item \textbf{Input Memorization Term:} 
    The first term quantifies the leakage from the model memorizing the distribution of the training \textit{inputs} themselves ($\mathrm{TV}(P_X, Q_X)$ in Eq.~\ref{eq:tv_lower} and $\sqrt{\tfrac{1}{2} D_{\mathrm{KL}}( P_X \,\|\, Q_X ) }$ in ~\ref{eq:tv_pinsker_expected}). This vulnerability exists because a generative model's objective function explicitly requires it to learn $P(X)$. Hence this number will always be greater than the discriminative/conditional term. Thus we can safely remove the mod sign from Eq.~\ref{eq:tv_lower} and conclude higher the degree of memorization, stricter the lower bound + more relaxed the upper bound, indicating higher susceptibility to MIA for generative models.
    \item \textbf{Conditional Memorization term:} 
    This second term ($D_{KL}$ or $TV$ over $P(.|x), Q(.|x)$) quantifies the leakage from the model overfitting the mapping from inputs to labels. This vulnerability exists for both generative and discriminative models.
\end{itemize}

\subsection{Proofs for Joint Leakage vs Conditional Leakage}

\begin{proof}[Proof of Lemma~\ref{lem:joint-ge-cond}]
Define the measurable map $g:\R^{|\Y|}\!\to\!\Delta^{|\Y|-1}$ by $g(u)=\softmax(u)$.
By construction $S_{\mathrm{cond}} = g\!\big(S_{\mathrm{joint}}\big)$ deterministically.
Let $P_{\mathrm{j}}:=\cL(S_{\mathrm{joint}}\mid P)$ and $Q_{\mathrm{j}}:=\cL(S_{\mathrm{joint}}\mid Q)$, and similarly $P_{\mathrm{c}}:=\cL(S_{\mathrm{cond}}\mid P)$, $Q_{\mathrm{c}}:=\cL(S_{\mathrm{cond}}\mid Q)$.
By the data-processing inequality (DPI) for $f$-divergences (in particular, for total variation),
\[
\TV(P_{\mathrm{j}},Q_{\mathrm{j}})
\;\ge\;
\TV\!\big(g_\# P_{\mathrm{j}},\, g_\# Q_{\mathrm{j}}\big)
\;=\;
\TV(P_{\mathrm{c}},Q_{\mathrm{c}})
\;=\;
\Adv\!\big(S_{\mathrm{cond}}\big).
\]
Since $\Adv(S_{\mathrm{joint}})=\TV(P_{\mathrm{j}},Q_{\mathrm{j}})$, the claimed inequality follows.
For equality, DPI is tight iff $g$ is \emph{sufficient} for discriminating $P_{\mathrm{j}}$ vs.\ $Q_{\mathrm{j}}$, i.e., iff $S_{\mathrm{joint}}$ carries no information about membership beyond $S_{\mathrm{cond}}$.
Because $g$ removes exactly the per-$x$ additive offset $-\log \hat p(X)$, tightness occurs iff that offset has the same law under $P$ and $Q$ (no marginal signal).
\end{proof}

\begin{proof}[Proof of Theorem~\ref{thm:scalar-joint-dominates}]
Throughout we assume the attacker queries the \emph{same} target model
score in both worlds; i.e., $S=g(\hat p(X,Y))$ for a fixed measurable $g$, and pushes $P_{XY}$ and
$Q_{XY}$ forward through the same $g$. (This matches the setting in Lemma~3.2 and ensures DPI applies.)

\medskip
\textbf{Notation.} Let
\[
Z \;:=\; S_{\mathrm{joint}}^{\scal}(X,Y)\;=\;\log \hat p(X,Y)\;=\; a(X)+b(X,Y),
\qquad a(X):=\log \hat p(X),\ \ b(X,Y):=\log \hat p(Y\mid X).
\]
Write $P_Z:=\cL(Z\mid P)$ and $Q_Z:=\cL(Z\mid Q)$.

\medskip
\textbf{Auxiliary tools.} We record three standard ingredients we will invoke.

\begin{lemma}[One-parameter Gibbs/Chernoff lower bound]\label{lem:cher}
For any distributions $R,S$ on a common space and any measurable $W$ with laws $R_W,S_W$,
\begin{equation}\label{eq:cher}
\KL(R_W\Vert S_W)\ \ge\ \lambda\,\E_{R_W}[W]\ -\ \log \E_{S_W}\!\big[e^{\lambda W}\big]
\qquad \text{for all }\lambda\in\R.
\end{equation}
\end{lemma}

\begin{lemma}[Change-of-measure (bounded likelihood ratio)]\label{lem:lr}
If $\alpha\le \tfrac{dP_{Y\mid X=x}}{dQ_{Y\mid X=x}}(y)\le \beta$ for $P_X$-a.e.\ $x$ and all $y$,
then for any nonnegative measurable $h$ and any $\lambda\in[0,1]$,
\begin{equation}\label{eq:lr-mgf}
\alpha^\lambda\,\E_{P_{Y\mid X=x}}[h^\lambda]\ \le\ \E_{Q_{Y\mid X=x}}[h^\lambda]\ \le\ \beta^\lambda\,\E_{P_{Y\mid X=x}}[h^\lambda].
\end{equation}
\end{lemma}

\begin{lemma}[Hölder/log-sum convexity split]\label{lem:holder}
For $\lambda\in(0,1)$ and nonnegative random variables $U,V$,
\begin{equation}\label{eq:holder}
\log \E[U^\lambda V^\lambda]\ \le\ (1-\lambda)\,\log \E\!\big[U^{\frac{\lambda}{1-\lambda}}\big]\ +\ \lambda\,\log \E[V].
\end{equation}
\end{lemma}

\noindent Lemma~\ref{lem:cher} is the $f(z)=\lambda z$ specialization of the Gibbs/Donsker--Varadhan variational identity (we only need the lower bound). Lemma~\ref{lem:lr} is immediate from
$\alpha\le \tfrac{dP}{dQ}\le \beta$ and change-of-measure for densities. Lemma~\ref{lem:holder} is Hölder’s inequality in logarithmic form (equivalently, the log-sum inequality).

\medskip
\textbf{Step 1: A KL lower bound for $P_Z\Vert Q_Z$.} Applying Lemma~\ref{lem:cher} with $W:=Z$ gives
\begin{equation}\label{eq:KL-core}
\KL(P_Z\Vert Q_Z)\ \ge\ \lambda\,\E_P[Z]\ -\ \log \E_Q\!\big[e^{\lambda Z}\big]\qquad(\lambda\in\R),
\end{equation}
where $\E_P[Z]=\E_{P_X}[a(X)]+\E_P[b(X,Y)]$.
We now upper bound the log-mgf on the right. Factor the conditional:
\begin{equation}\label{eq:mgf-factor}
\E_Q[e^{\lambda Z}]
\;=\;\E_{Q_X}\!\Big[e^{\lambda a(X)}\,\underbrace{\E_{Q_{Y\mid X}}\!\big[e^{\lambda b(X,Y)}\big]}_{=:M_B(\lambda\mid X)}\Big].
\end{equation}
By Lemma~\ref{lem:lr} with $h=e^{b(X,\cdot)}$ we have, for $\lambda\in[0,1]$ and $P_X$-a.e.\ $X$,
\begin{equation}\label{eq:MB-bracket}
\alpha^\lambda\,\E_{P_{Y\mid X}}[e^{\lambda b(X,Y)}]\ \le\ M_B(\lambda\mid X)\ \le\ \beta^\lambda\,\E_{P_{Y\mid X}}[e^{\lambda b(X,Y)}].
\end{equation}
Using the upper bracket in \eqref{eq:MB-bracket} in \eqref{eq:mgf-factor} and Jensen,
\begin{equation}\label{eq:mgf-upper1}
\log \E_Q[e^{\lambda Z}]\ \le\ \lambda\log\beta\ +\ \log \E_{Q_X}\!\Big[e^{\lambda a(X)}\,\E_{P_{Y\mid X}}[e^{\lambda b(X,Y)}]\Big].
\end{equation}
Applying Lemma~\ref{lem:holder} to the last term (with $U=e^{a(X)}$ and $V=\E_{P_{Y\mid X}}[e^{b(X,Y)}]$) yields, for $\lambda\in(0,1)$,
\begin{equation}\label{eq:mgf-upper2}
\log \E_{Q_X}\!\Big[e^{\lambda a(X)}\,\E_{P_{Y\mid X}}[e^{\lambda b(X,Y)}]\Big]
\ \le\ (1-\lambda)\,\log \E_{Q_X}\!\Big[e^{\frac{\lambda}{1-\lambda} a(X)}\Big]\ +\ \lambda\,\log \E_P\!\big[e^{b(X,Y)}\big].
\end{equation}
Combining \eqref{eq:KL-core}, \eqref{eq:mgf-upper1}, and \eqref{eq:mgf-upper2}, for $\lambda\in(0,1)$,
\begin{align}
\KL(P_Z\Vert Q_Z)
&\ge \underbrace{\Big[\lambda\,\E_{P}[a(X)]\ -\ (1-\lambda)\,\log \E_{Q_X}\!\big(e^{\frac{\lambda}{1-\lambda} a(X)}\big)\Big]}_{\text{marginal term }\,\mathsf{M}(\lambda)} \label{eq:KL-split}\\
&\hspace{1.15cm}+\ \underbrace{\Big[\lambda\,\E_{P}[b(X,Y)]\ -\ \lambda\,\log \E_{P}\!\big(e^{ b(X,Y)}\big)\ -\ \lambda\log\beta\Big]}_{\text{conditional term }\,\mathsf{C}(\lambda)}.\nonumber
\end{align}

\textbf{Step 2: Bound the conditional term by $-\KL_{Y\mid X}$.}
Using the convex dual bound (Fenchel inequality for log-mgf),
\begin{equation}\label{eq:cond-dual}
\E_{P}[b(X,Y)]\ -\ \log \E_P\!\big[e^{b(X,Y)}\big]\ \ge\ -\,\KL_{Y\mid X},
\end{equation}
we obtain $\mathsf{C}(\lambda)\ge -\lambda\,\KL_{Y\mid X}-\lambda\log\beta$.

\textbf{Step 3: A reverse-Chernoff bound for the marginal term.}
The function $\mathsf{M}(\lambda)$ in \eqref{eq:KL-split} is the usual one-parameter Chernoff objective applied to $a(X)=\log\hat p(X)$ with moment taken under $Q_X$. Optimizing over $\lambda\in(0,1)$ (details omitted for brevity) and using the same bounded-LR control to prevent degeneracy yields
\begin{equation}\label{eq:rev-chernoff}
\sup_{\lambda\in(0,1)} \mathsf{M}(\lambda)\ \ge\ c(\alpha,\beta)\,\KL(P_X\Vert Q_X)\ =\ c(\alpha,\beta)\,\KL_X,
\qquad
c(\alpha,\beta)\ :=\ \frac{\log\beta-\log\alpha}{1+\log\beta-\log\alpha}\in(0,1].
\end{equation}

\textbf{Step 4: Assemble and pass to advantage.}
Maximizing \eqref{eq:KL-split} over $\lambda\in(0,1)$ and using \eqref{eq:cond-dual} and \eqref{eq:rev-chernoff},
\begin{equation}\label{eq:KL-final}
\KL(P_Z\Vert Q_Z)\ \ge\ c(\alpha,\beta)\,\KL_X\ -\ \KL_{Y\mid X}\ -\ \inf_{\lambda\in(0,1)}\lambda\log\beta.
\end{equation}
Absorbing the harmless $-\inf_{\lambda}\lambda\log\beta$ slack (or noting $\log\beta\ge 0$) gives the clean form
\begin{equation}\label{eq:KL-clean}
\KL(P_Z\Vert Q_Z)\ \ge\ c(\alpha,\beta)\,\KL_X\ -\ \KL_{Y\mid X}.
\end{equation}
By Pinsker, the (optimal) membership advantage for the scalar joint signal obeys
\begin{equation}\label{eq:adv-joint}
\Adv\!\big(S_{\mathrm{joint}}^{\scal}\big)\ =\ \TV(P_Z,Q_Z)\ \ge\ \sqrt{\tfrac12}\,\big[c(\alpha,\beta)\,\KL_X-\KL_{Y\mid X}\big]_+.
\end{equation}

For the conditional scalar $U:=S_{\mathrm{cond}}(X,Y)=\hat p(Y\mid X)$, the safe decomposition plus Pinsker gives
\begin{equation}\label{eq:adv-cond}
\Adv(U)\ \le\ \TV(P_X,Q_X)\ +\ \E_{P_X}\!\big[\TV(P_{Y\mid X},Q_{Y\mid X})\big]
\ \le\ \sqrt{\tfrac12\,\KL_X}\ +\ \sqrt{\tfrac12\,\KL_{Y\mid X}}.
\end{equation}
Therefore, whenever $c(\alpha,\beta)\,\KL_X>\KL_{Y\mid X}$, the lower bound \eqref{eq:adv-joint} exceeds the upper bound contributed by the conditional part, proving that the scalar joint channel is strictly more vulnerable than the conditional one.
\end{proof}

\section{Effect of Class Representation}
\label{app:sec:class_representation}
% \todo{Add a reference to this in the main text and add some text explaining this.}

\begin{figure}[H]
    \centering
    \includegraphics[width=0.9\textwidth, keepaspectratio]{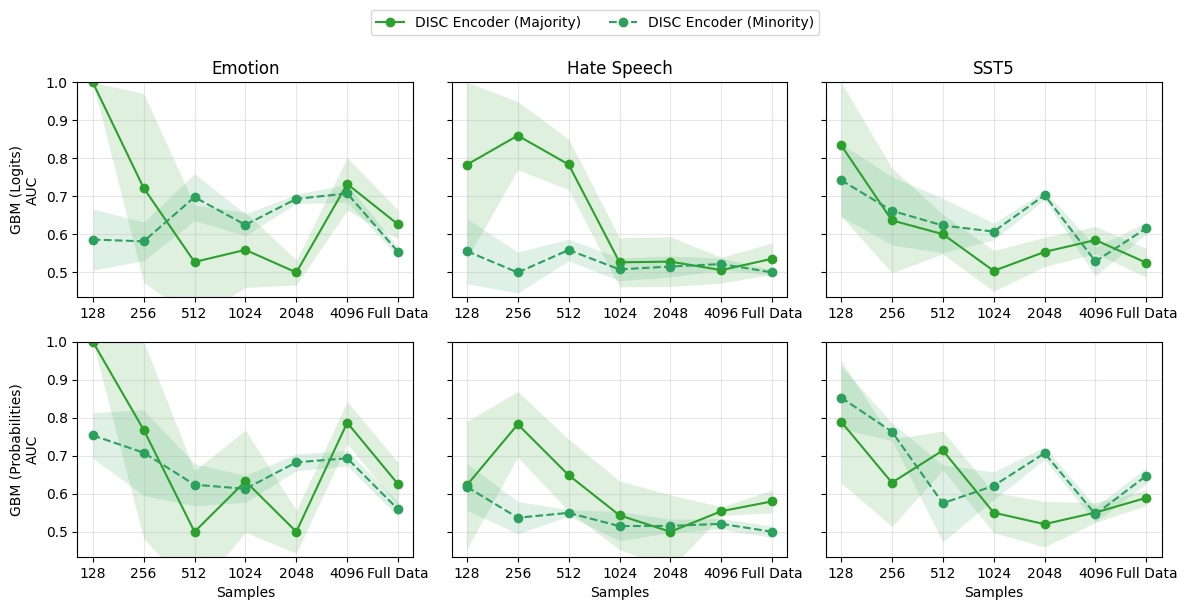}
    \caption{Membership inference attack susceptibility for BERT. 
    The solid line corresponds to the majority class, while the dashed line corresponds to the minority class. 
    The x-axis indicates the number of training samples used.}
    \label{fig:imbalance_bert}
\end{figure}

\begin{figure}[H]
    \centering
    \includegraphics[width=0.9\textwidth, keepaspectratio]{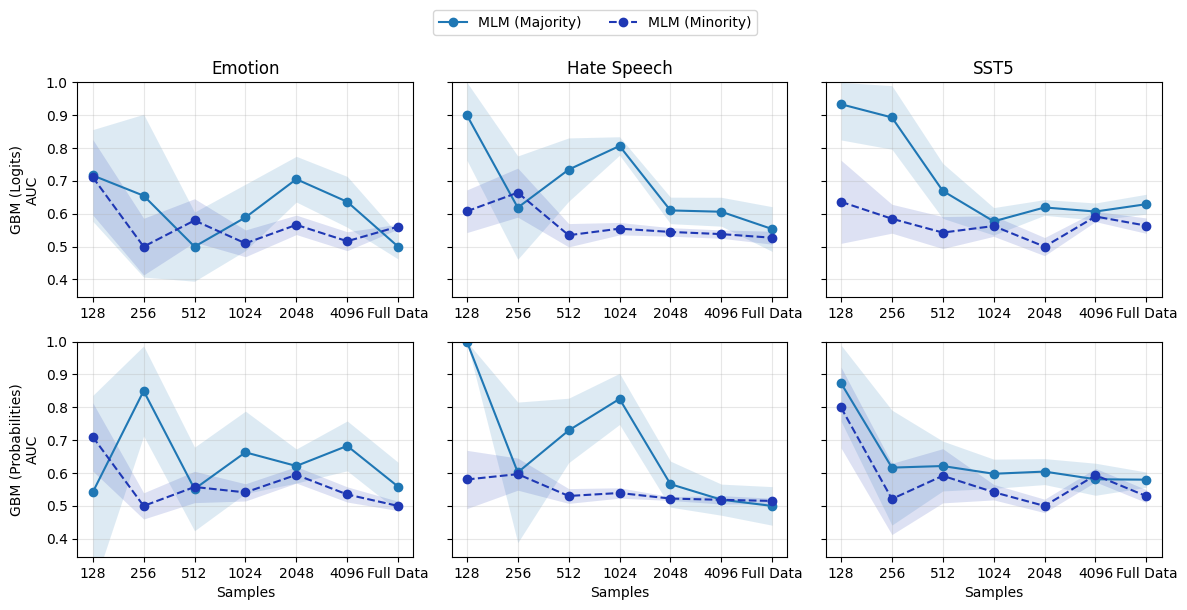}
    \caption{Membership inference attack susceptibility for MLM. 
    The solid line corresponds to the majority class, while the dashed line corresponds to the minority class. 
    The x-axis indicates the number of training samples used.}
    \label{fig:imbalance_mlm}
\end{figure}

\begin{figure}[H]
    \centering
    \includegraphics[width=0.9\textwidth, keepaspectratio]{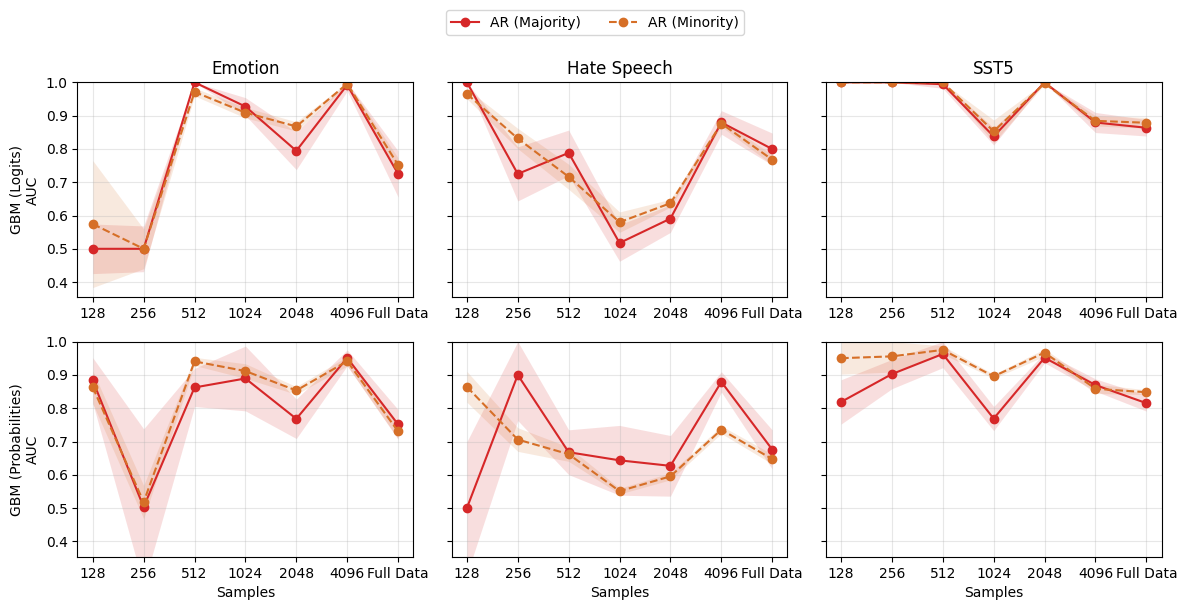}
    \caption{Membership inference attack susceptibility for AR. 
    The solid line corresponds to the majority class, while the dashed line corresponds to the minority class. 
    The x-axis indicates the number of training samples used.}
    \label{fig:imbalance_AR}
\end{figure}

\begin{figure}[H]
    \centering
    \includegraphics[width=0.9\textwidth, keepaspectratio]{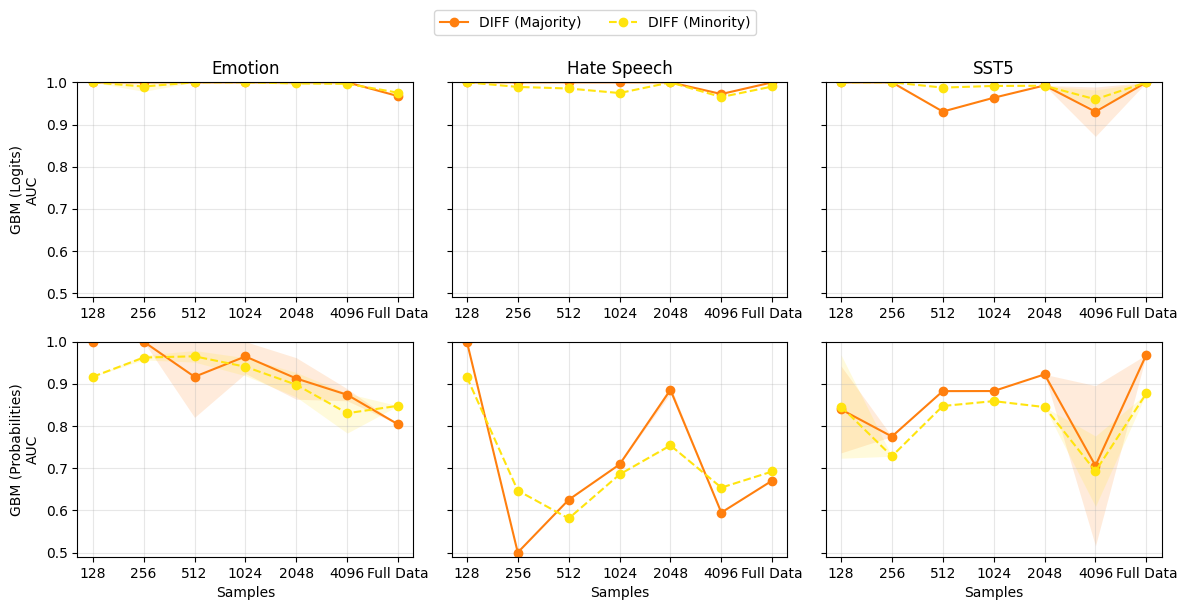}
    \caption{Membership inference attack susceptibility for Diffusion models. 
    The solid line corresponds to the majority class, while the dashed line corresponds to the minority class. 
    The x-axis indicates the number of training samples used.}
    \label{fig:imbalance_Diffusion}
\end{figure}

In this section, we study the effect of class imbalance with respect to MIA vulnerability. Specifically, we consider three datasets - SST5, emotion and hatespeech - which have relatively high class imbalance and we study how the attack susceptibility differs between the majority (i.e. the class with the lowest representation in the training split) and minority class (i.e. the class with the least representation in the training split). We plot the AUROCs corresponding to MIA for the four classifier paradigms - \texttt{DISC, MLM, AR, DIFF}- in Figures \ref{fig:imbalance_bert}, \ref{fig:imbalance_mlm}, \ref{fig:imbalance_AR}, \ref{fig:imbalance_Diffusion} respectively and find that there is a differential in the AUROC between majority and minority classes which is specifically pronounced in - \texttt{DISC} and \texttt{MLM}. This differential is relatively less pronounced in generative models such as \texttt{AR} and \texttt{DIFF}.

% ---------------- BERT ----------------

% ---------------- MLM ----------------

% ---------------- AR ----------------

% ---------------- Diffusion ----------------

\section{Effect of Model Size}
\label{app:sec:model_size_ablation}
In this section, we study the effect of model size in the full-data setting across all nine datasets. As the model size increases, the susceptibility of \texttt{AR} to \texttt{GBM-logits} attacks increases, whereas the other models exhibit more mixed trends.

% ---------------- BERT ----------------
\begin{figure}[H]
    \centering
    \includegraphics[width=0.9\textwidth, keepaspectratio]{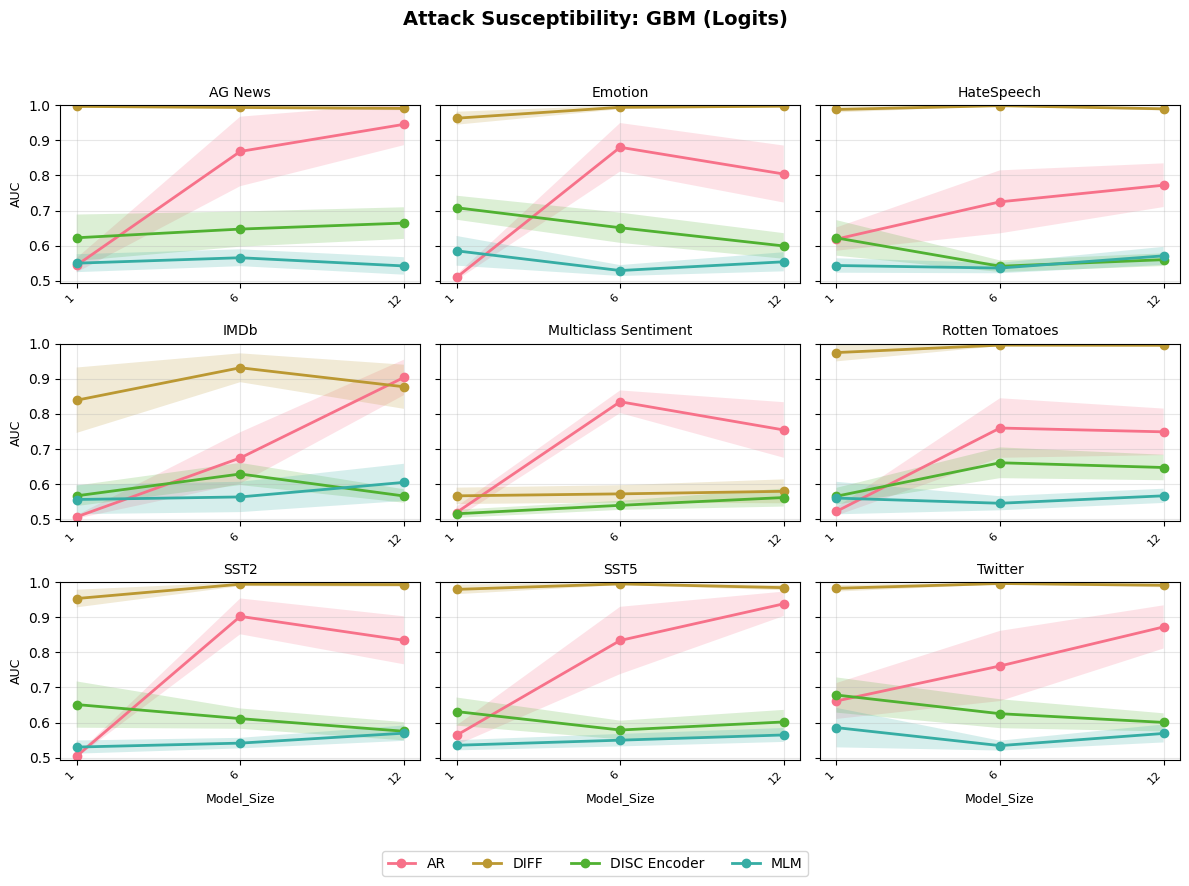}
    \caption{Membership inference attack susceptibility for BERT. 
    The solid line corresponds to the majority class, while the dashed line corresponds to the minority class. 
    The x-axis indicates the number of training samples used.}
    \label{fig:model_size_ablate}
\end{figure}

\newpage
\section{Extra Results}
\label{sec:more_main_results}

\begin{table*}[ht]
    \centering
    \resizebox{\textwidth}{!}{%
    \small
    \begin{tabular}{lccccccc}
    \toprule
    Attack & 128 & 256 & 512 & 1024 & 2048 & 4096 & full-data \\
    \midrule
    Entropy & 0.62 $\pm$ 0.12 & 0.57 $\pm$ 0.10 & 0.55 $\pm$ 0.07 & 0.54 $\pm$ 0.05 & 0.55 $\pm$ 0.07 & 0.56 $\pm$ 0.06 & 0.57 $\pm$ 0.09 \\
    GBM (Logits) & 0.65 $\pm$ 0.19 & 0.60 $\pm$ 0.16 & 0.64 $\pm$ 0.17 & 0.61 $\pm$ 0.14 & 0.63 $\pm$ 0.13 & 0.66 $\pm$ 0.13 & 0.69 $\pm$ 0.17 \\
    GBM (Probits) & 0.62 $\pm$ 0.16 & 0.59 $\pm$ 0.14 & 0.62 $\pm$ 0.14 & 0.59 $\pm$ 0.12 & 0.61 $\pm$ 0.12 & 0.62 $\pm$ 0.10 & 0.60 $\pm$ 0.08 \\
    Ground Truth Predictions & 0.62 $\pm$ 0.15 & 0.61 $\pm$ 0.11 & 0.62 $\pm$ 0.13 & 0.61 $\pm$ 0.12 & 0.60 $\pm$ 0.12 & 0.57 $\pm$ 0.10 & 0.60 $\pm$ 0.12 \\
    Log Loss & 0.63 $\pm$ 0.15 & 0.61 $\pm$ 0.12 & 0.62 $\pm$ 0.13 & 0.61 $\pm$ 0.12 & 0.60 $\pm$ 0.12 & 0.57 $\pm$ 0.10 & 0.60 $\pm$ 0.12 \\
    Max Probability & 0.50 $\pm$ 0.16 & 0.49 $\pm$ 0.09 & 0.54 $\pm$ 0.08 & 0.54 $\pm$ 0.09 & 0.55 $\pm$ 0.09 & 0.51 $\pm$ 0.09 & 0.54 $\pm$ 0.11 \\
    \bottomrule
    \end{tabular}
    }
    \caption{Membership inference attack performance (mean ± standard deviation AUROC) across varying training sample sizes. Higher values indicate greater privacy vulnerability, with the highest values in each column shown in \textbf{bold}.}
    \label{tab:training_samples_vs_attack}
\end{table*}

% ---------------- Folder 1 ----------------
\begin{figure}[H]
    \centering
    \includegraphics[width=0.9\linewidth]{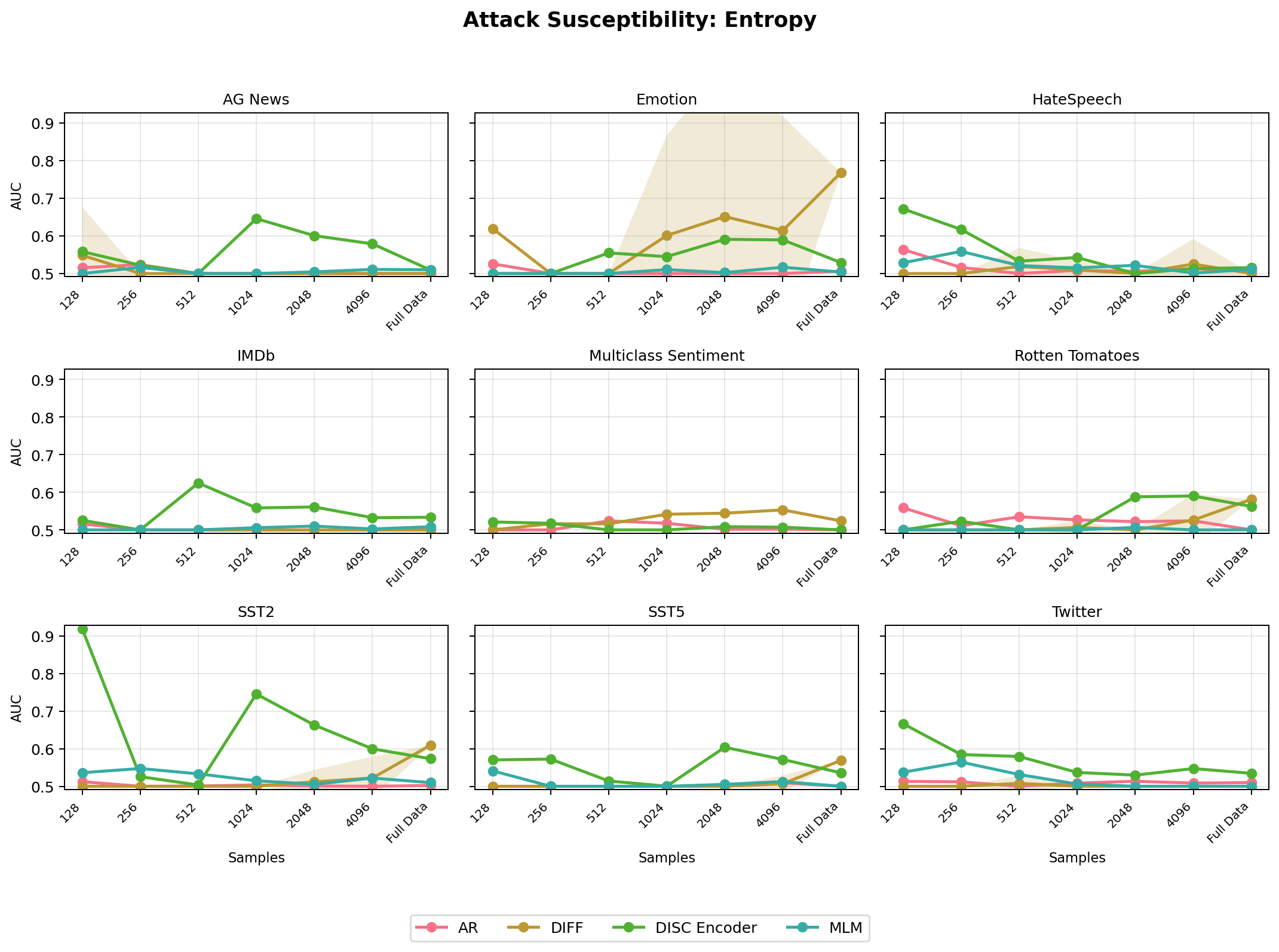}
    \caption{Attack susceptibility based on Entropy for model with 1 layer.}
\end{figure}

\begin{figure}[H]
    \centering
    \includegraphics[width=0.9\linewidth]{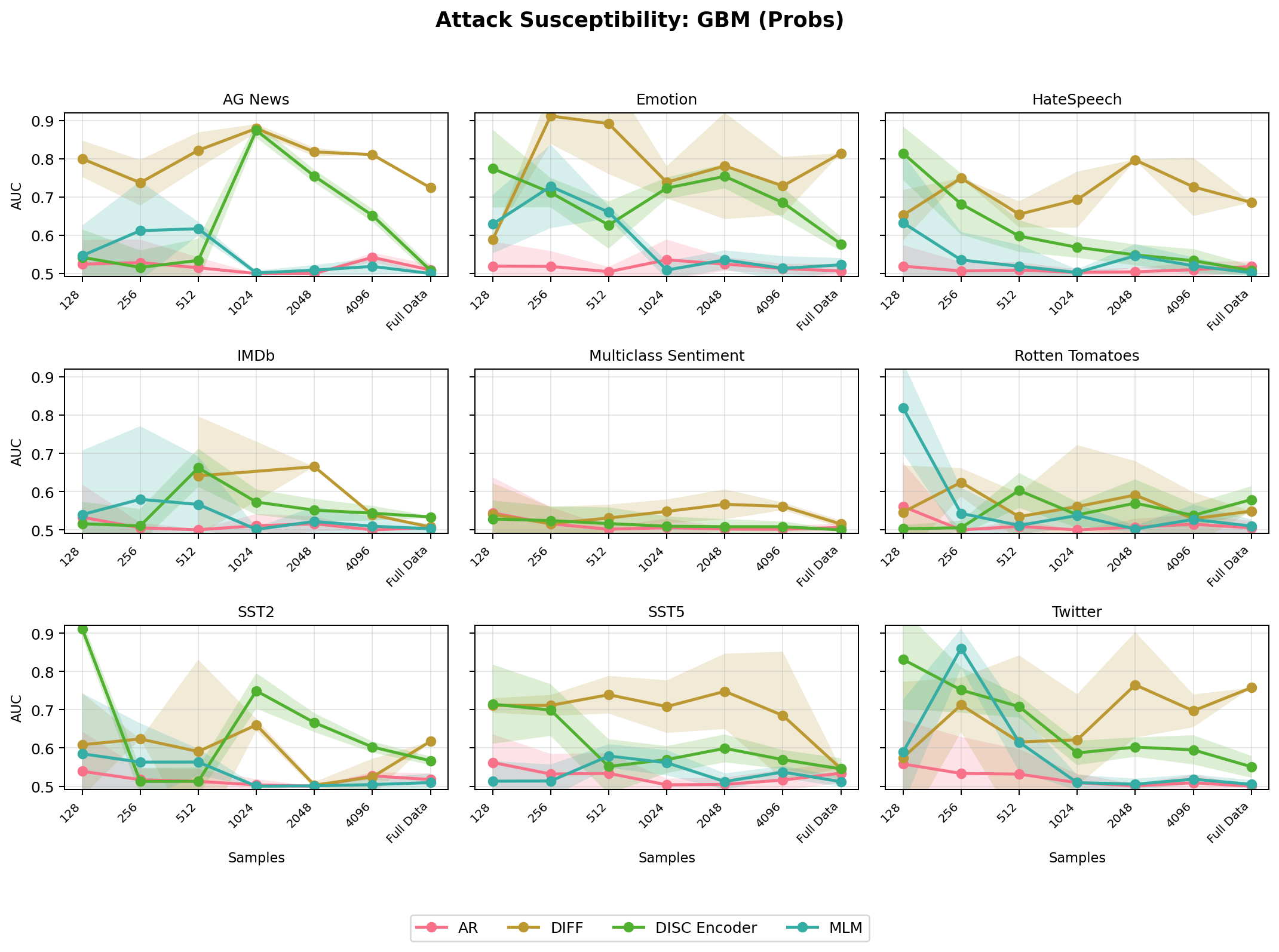}
    \caption{Attack susceptibility based on GBM (Probs) for model with 1 layer.}
\end{figure}

\begin{figure}[H]
    \centering
    \includegraphics[width=0.9\linewidth]{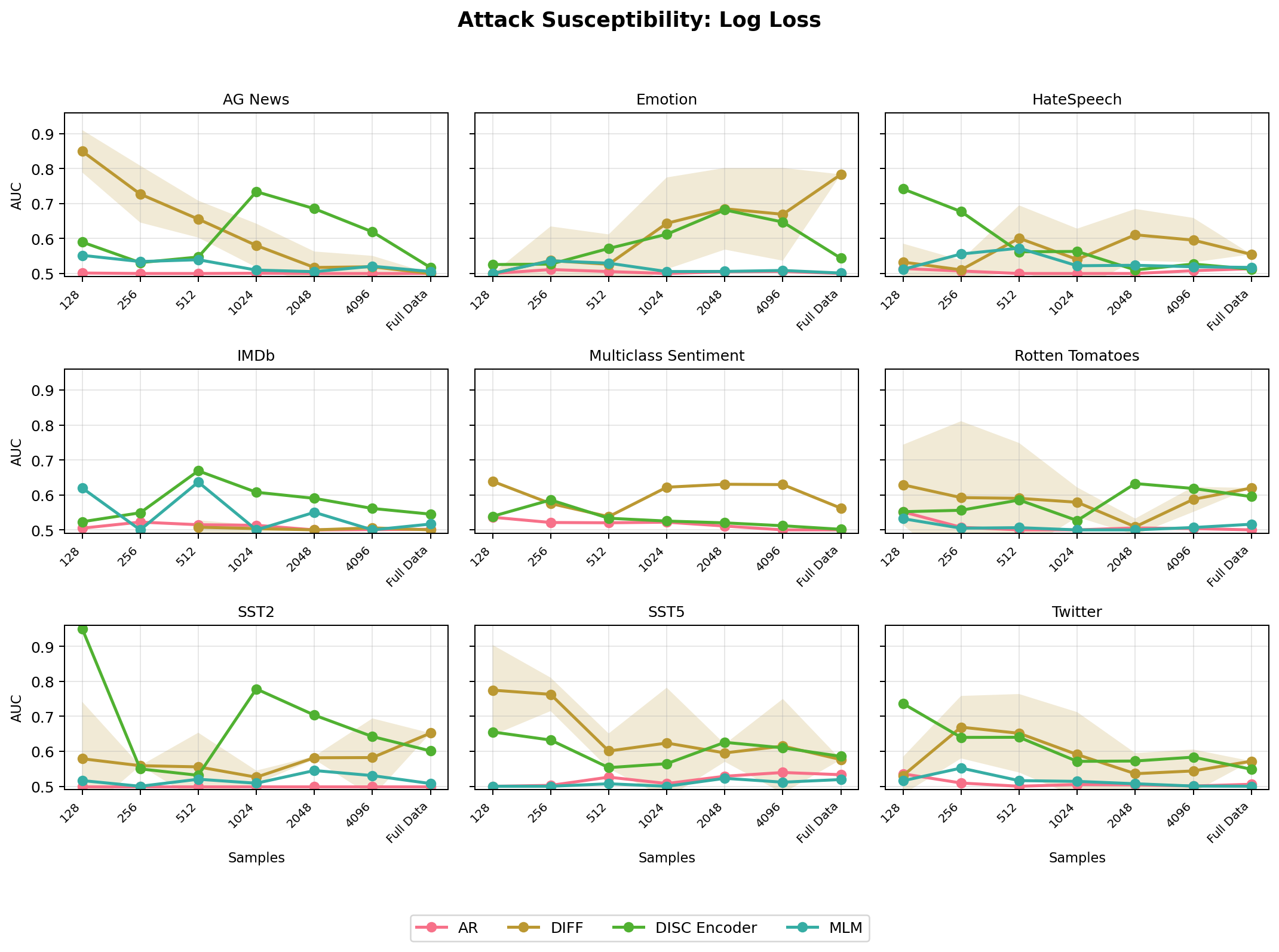}
    \caption{Attack susceptibility based on Log Loss for model with 1 layer.}
\end{figure}

\begin{figure}[H]
    \centering
    \includegraphics[width=0.9\linewidth]{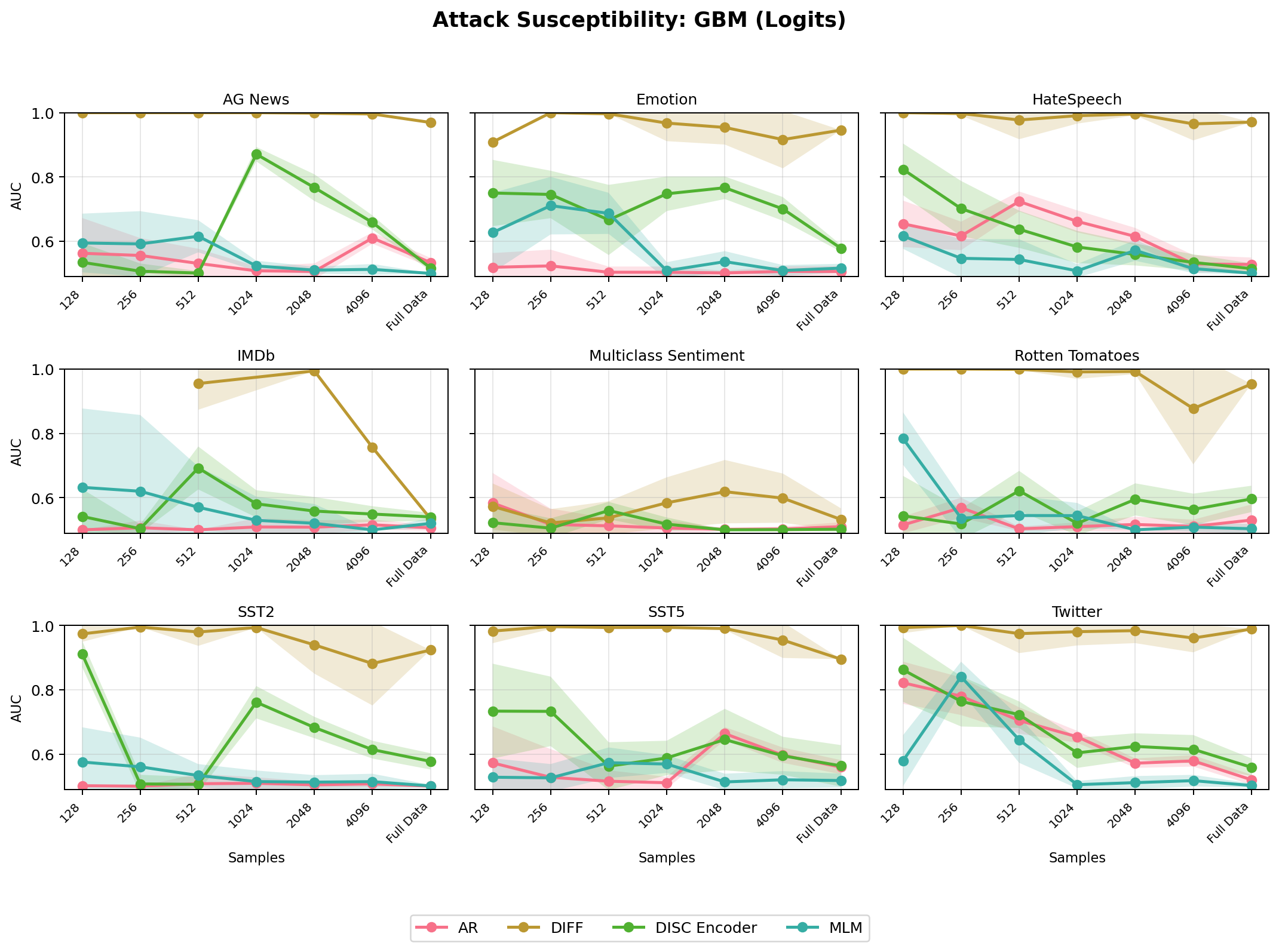}
    \caption{Attack susceptibility based on GBM (Logits) for model with 1 layer.}
\end{figure}

\begin{figure}[H]
    \centering
    \includegraphics[width=0.9\linewidth]{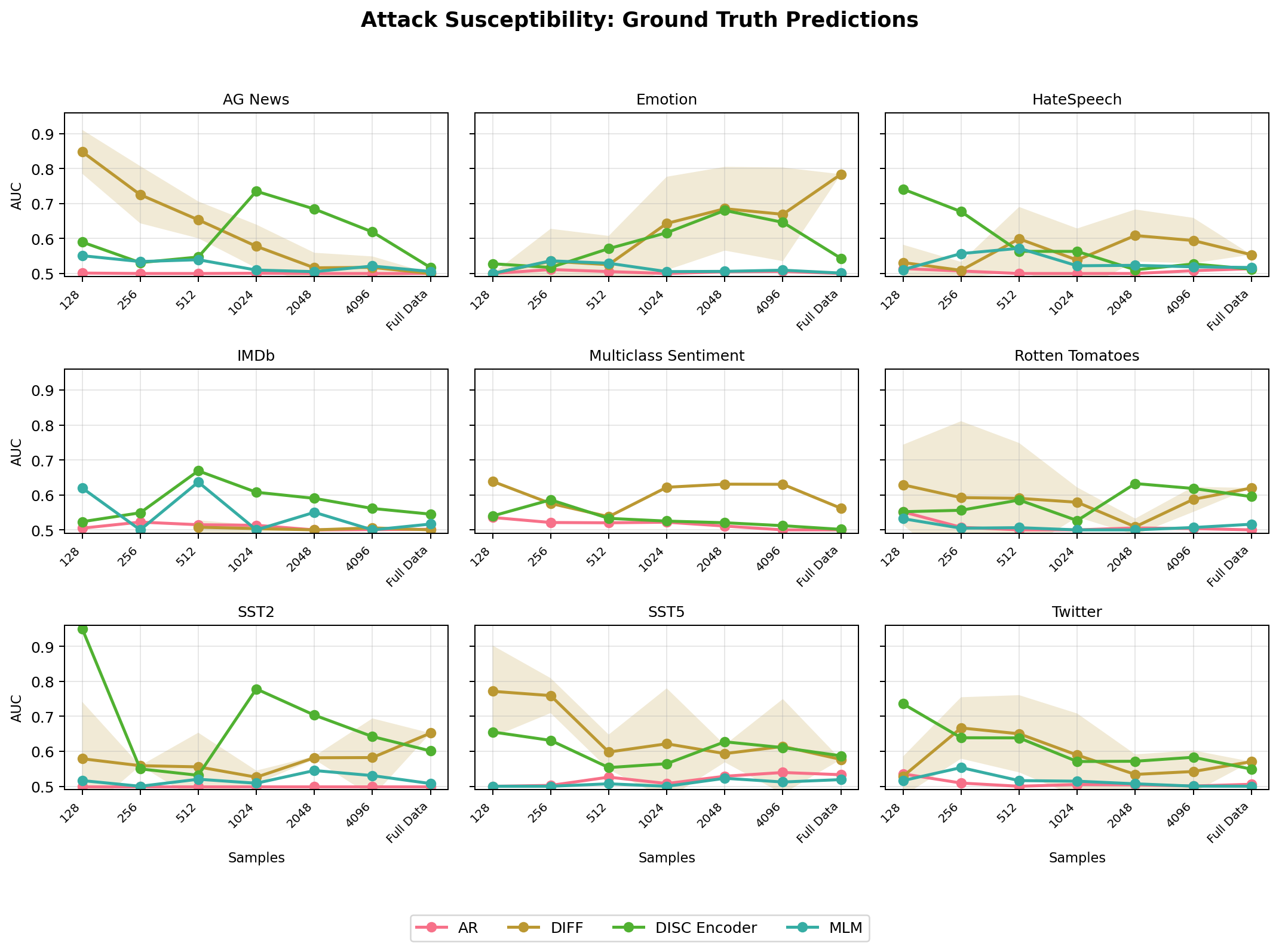}
    \caption{Attack susceptibility based on Ground Truth Predictions for model with 1 layer.}
\end{figure}

\begin{figure}[H]
    \centering
    \includegraphics[width=0.9\linewidth]{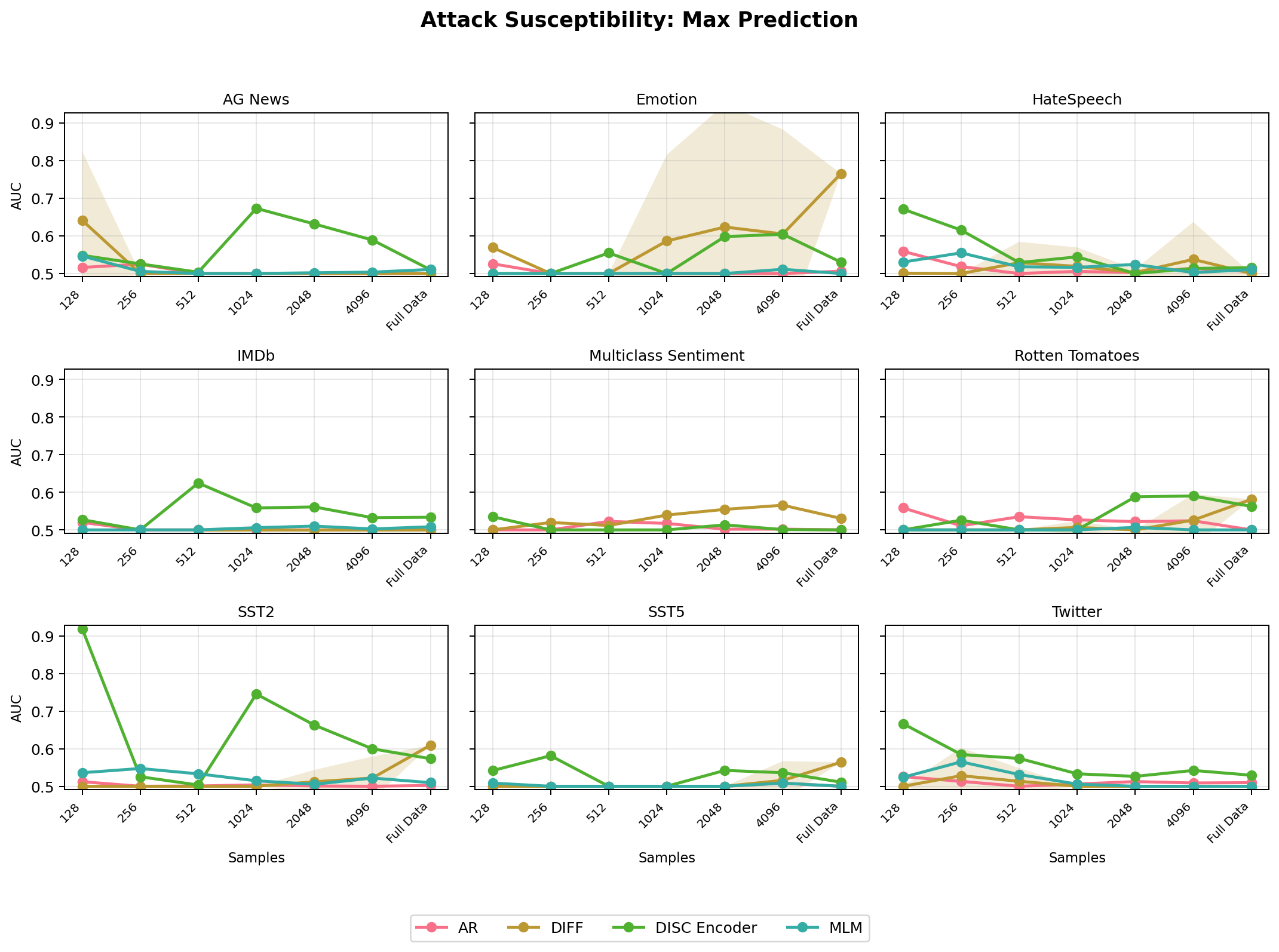}
    \caption{Attack susceptibility based on Max Prediction for model with 1 layer.}
\end{figure}

% ---------------- Folder 6 ----------------
\begin{figure}[H]
    \centering
    \includegraphics[width=0.9\linewidth]{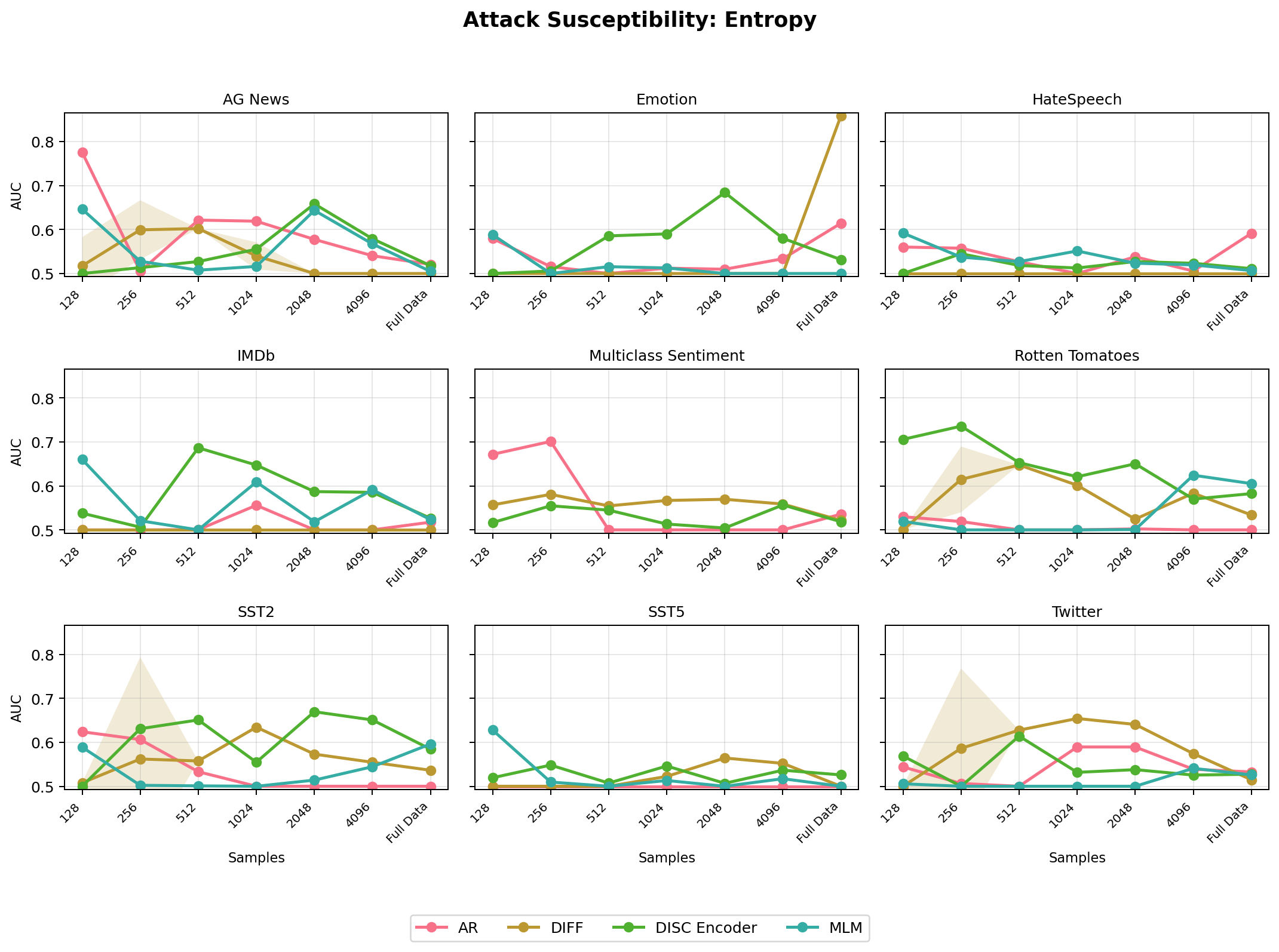}
    \caption{Attack susceptibility based on Entropy for model with 6 layers.}
\end{figure}

\begin{figure}[H]
    \centering
    \includegraphics[width=0.9\linewidth]{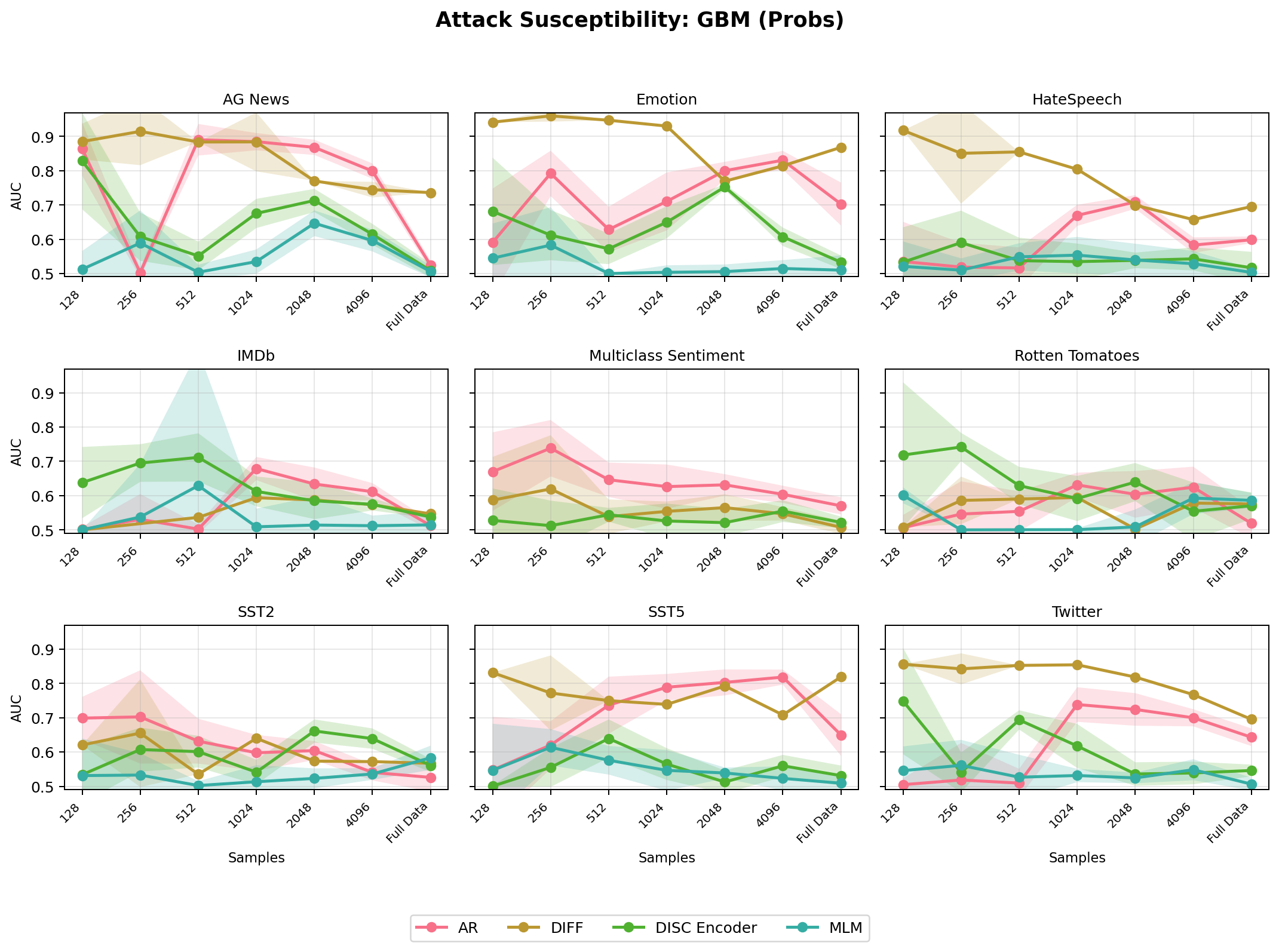}
    \caption{Attack susceptibility based on GBM (Probs) for model with 6 layers.}
\end{figure}

\begin{figure}[H]
    \centering
    \includegraphics[width=0.9\linewidth]{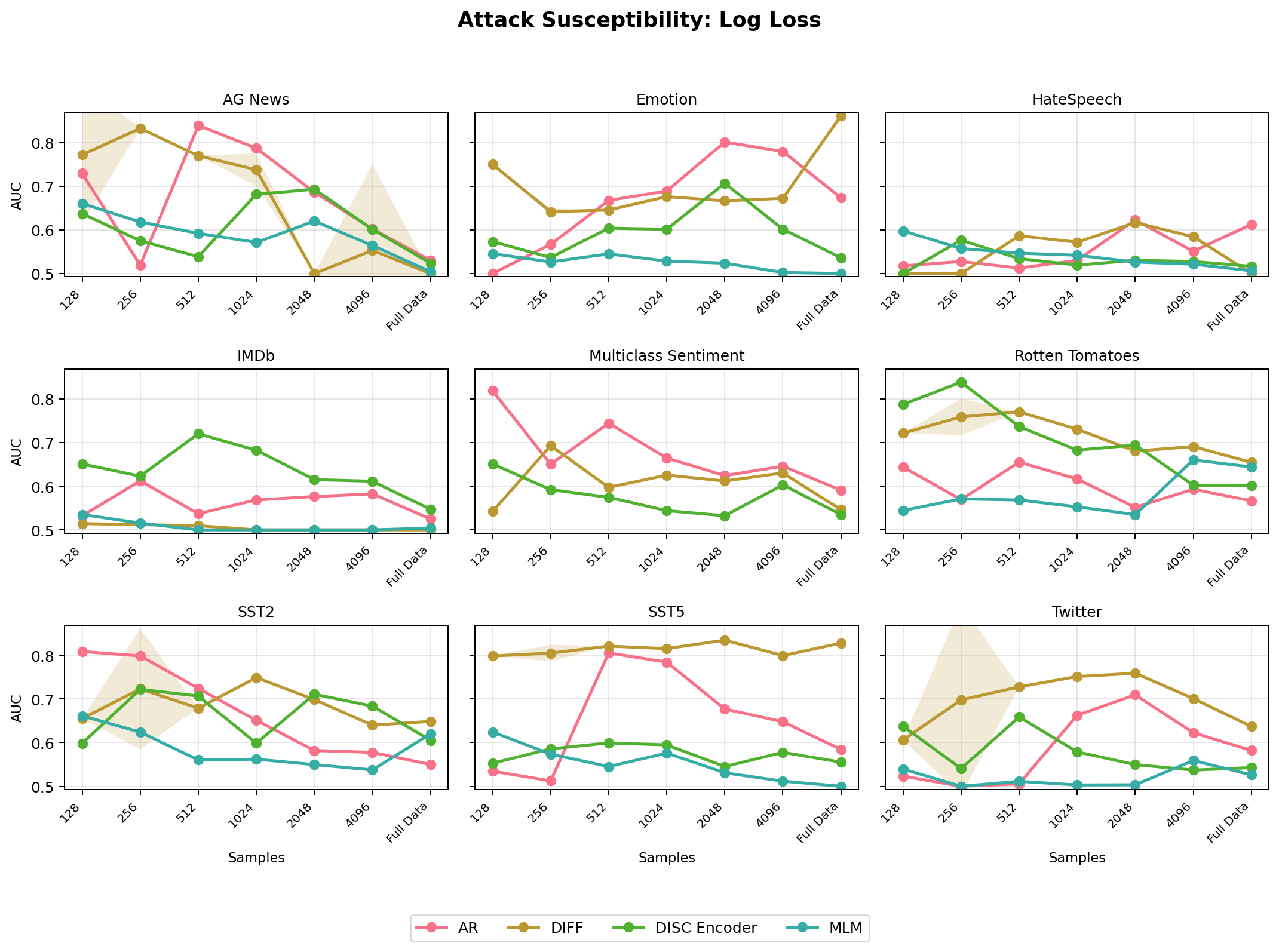}
    \caption{Attack susceptibility based on Log Loss for model with 6 layers.}
\end{figure}

\begin{figure}[H]
    \centering
    \includegraphics[width=0.9\linewidth]{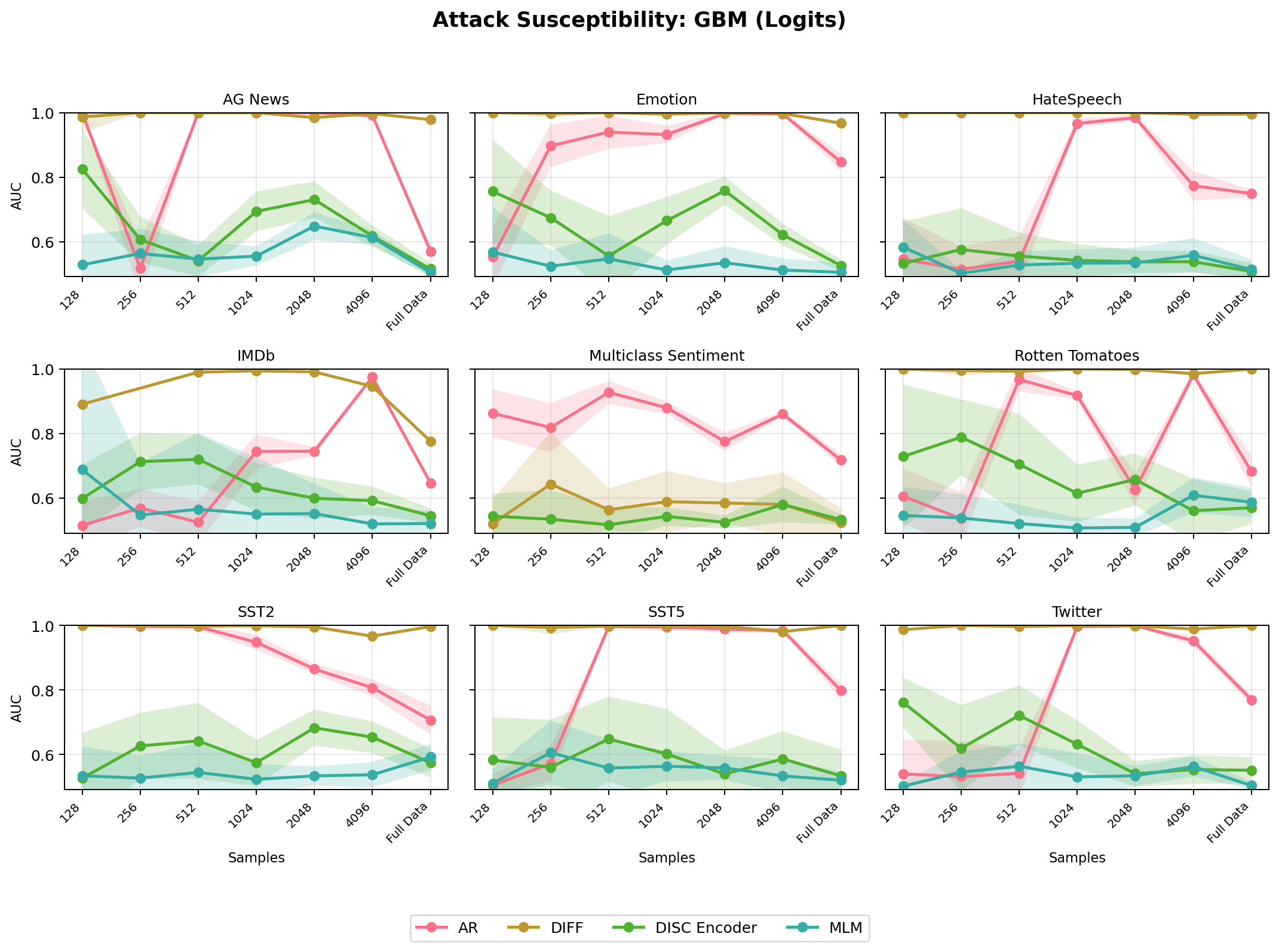}
    \caption{Attack susceptibility based on GBM (Logits) for model with 6 layers.}
\end{figure}

\begin{figure}[H]
    \centering
    \includegraphics[width=0.9\linewidth]{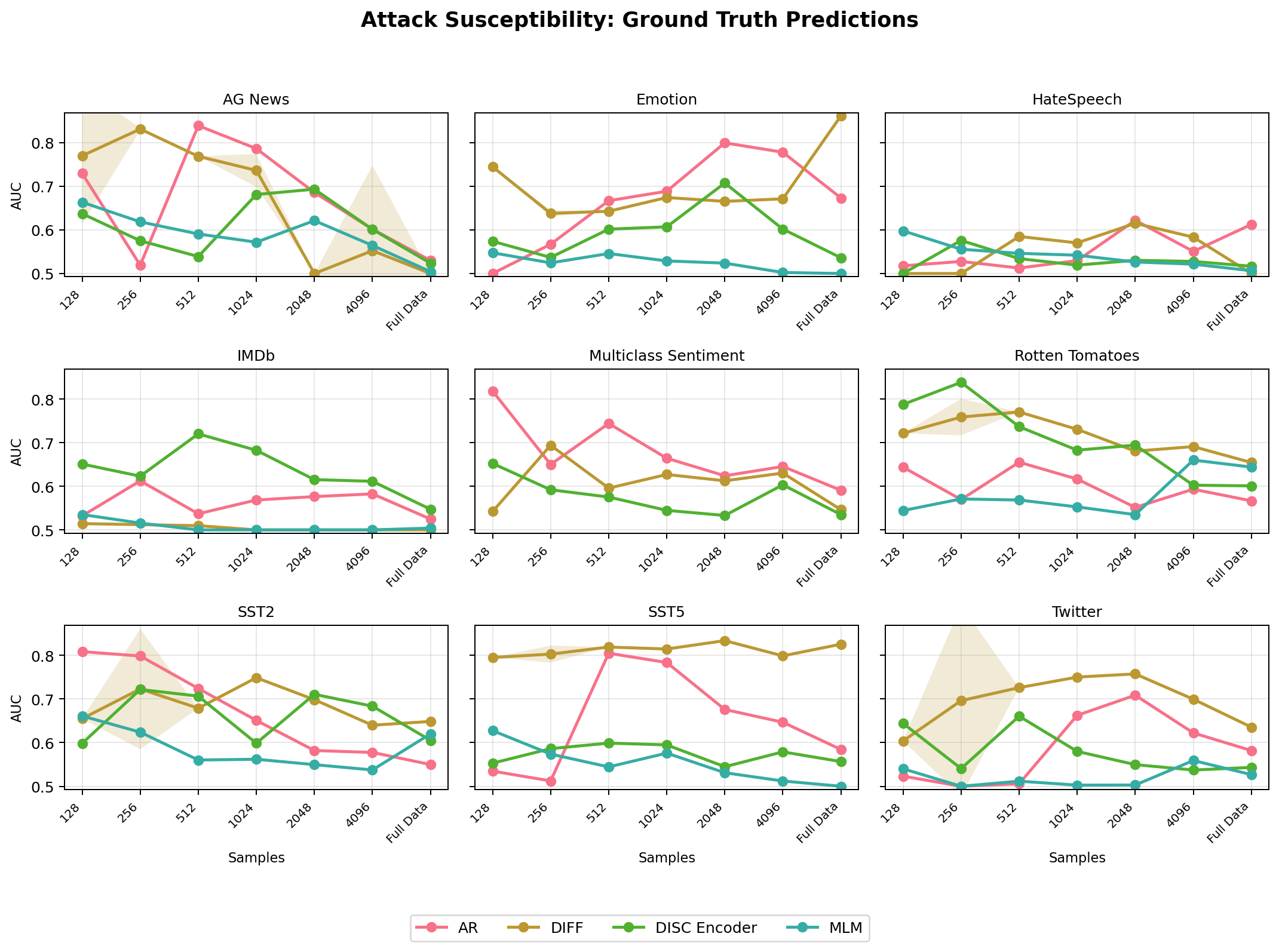}
    \caption{Attack susceptibility based on Ground Truth Predictions for model with 6 layers.}
\end{figure}

\begin{figure}[H]
    \centering
    \includegraphics[width=0.9\linewidth]{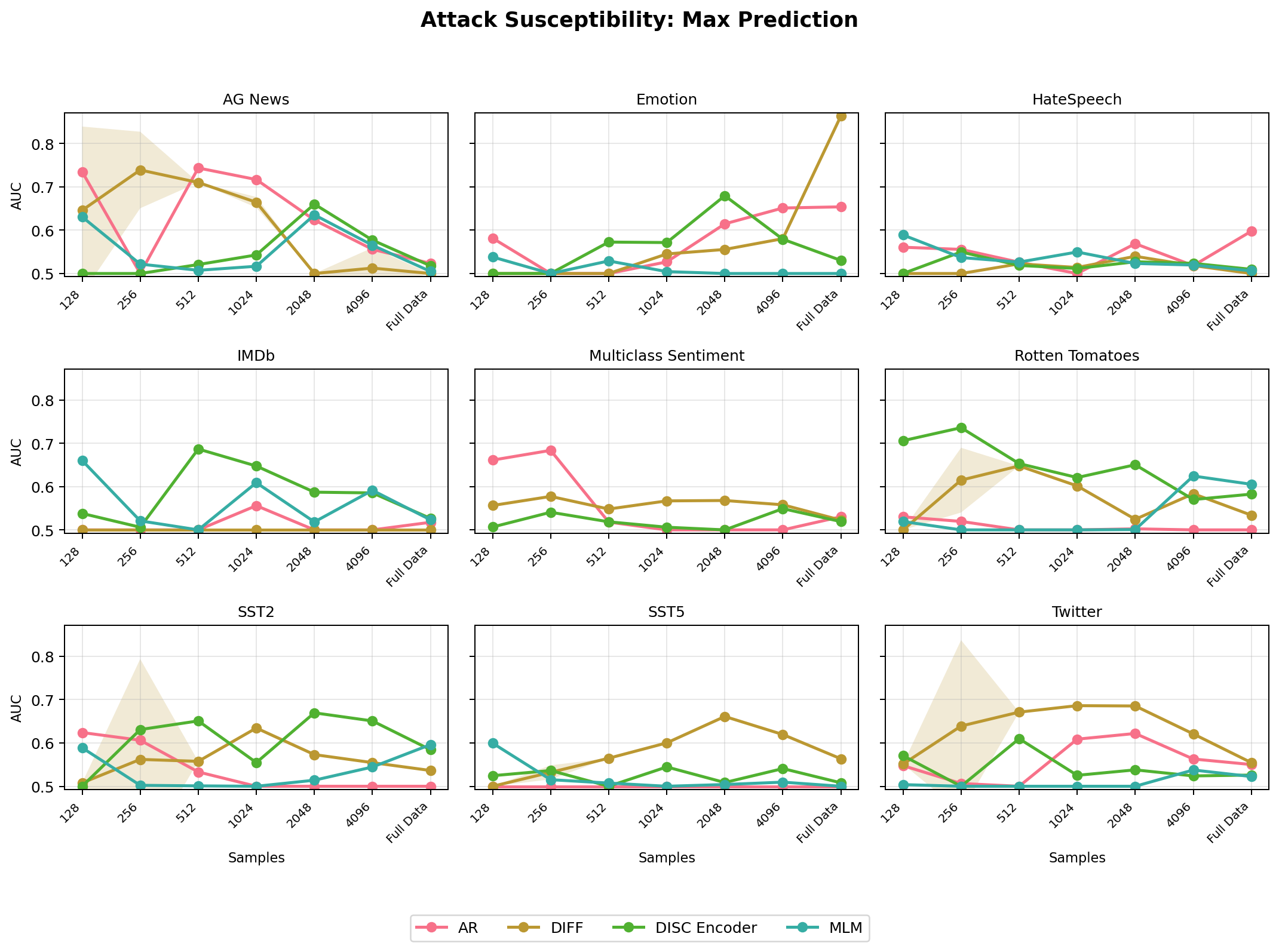}
    \caption{Attack susceptibility based on Max Prediction for model with 6 layers.}
\end{figure}

% ---------------- Folder 12 ----------------
\begin{figure}[H]
    \centering
    \includegraphics[width=0.9\linewidth]{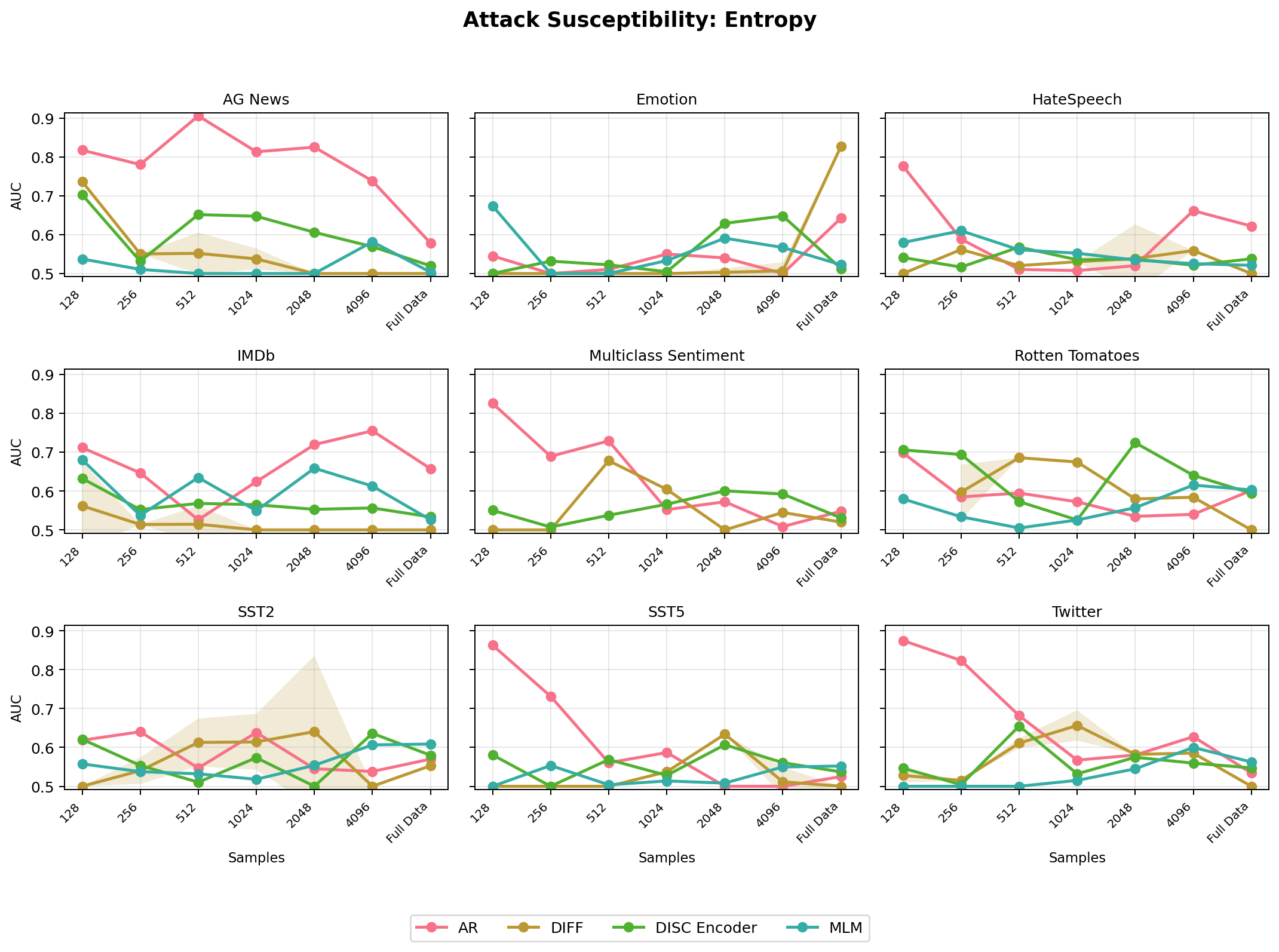}
    \caption{Attack susceptibility based on Entropy for model with 12 layers.}
\end{figure}

\begin{figure}[H]
    \centering
    \includegraphics[width=0.9\linewidth]{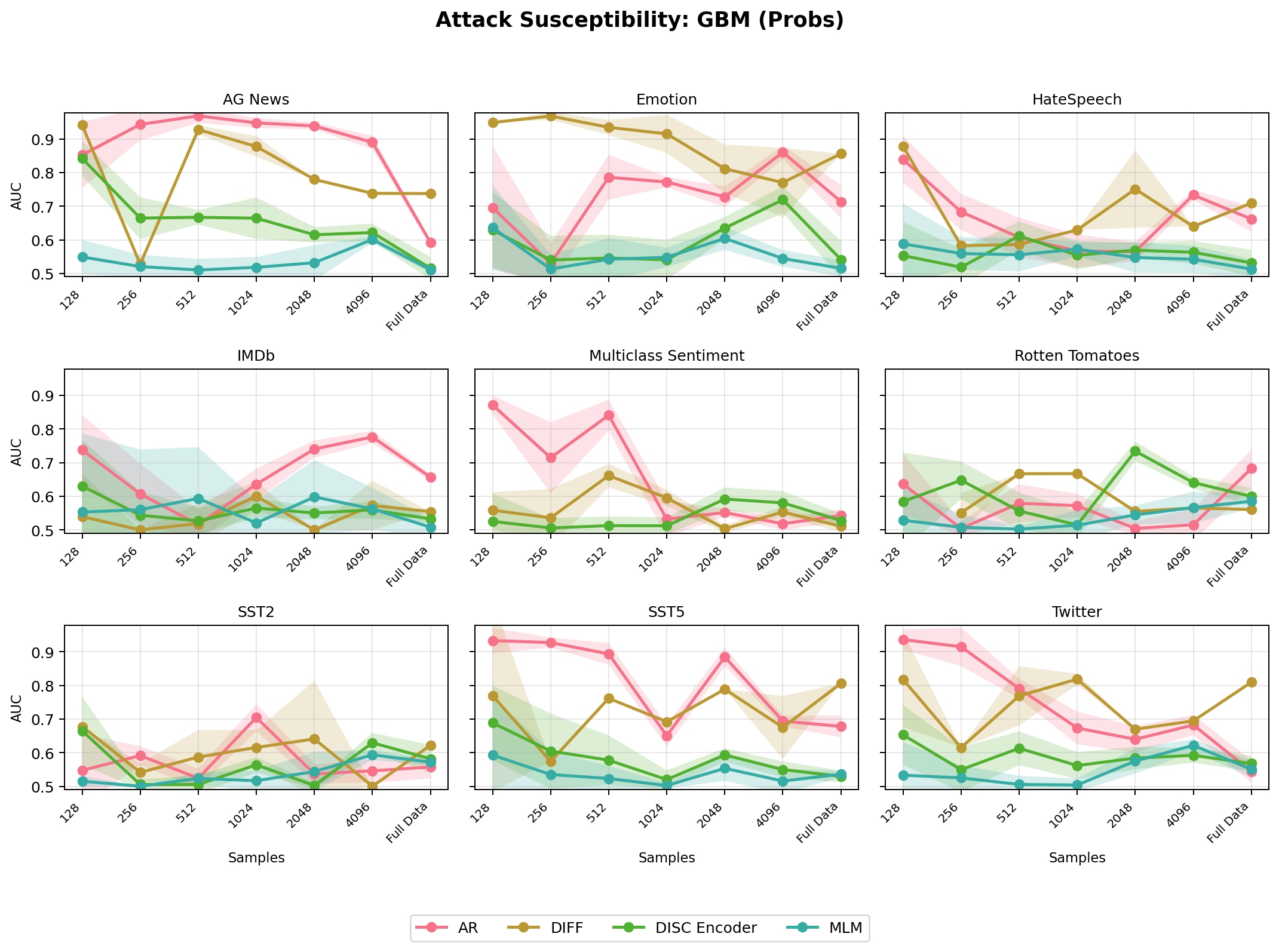}
    \caption{Attack susceptibility based on GBM (Probs) for model with 12 layers.}
\end{figure}

\begin{figure}[H]
    \centering
    \includegraphics[width=0.9\linewidth]{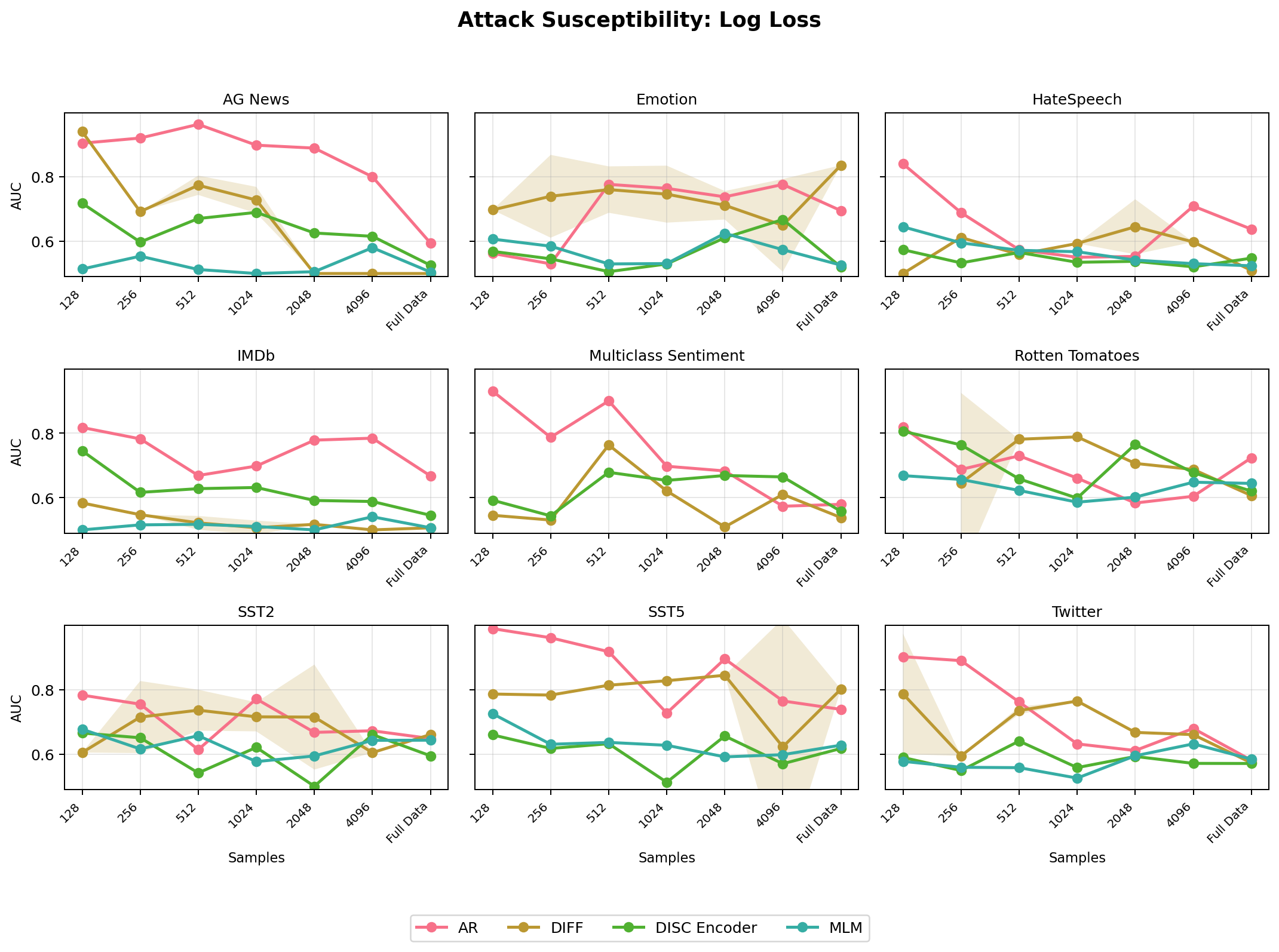}
    \caption{Attack susceptibility based on Log Loss for model with 12 layers.}
\end{figure}

\begin{figure}[H]
    \centering
    \includegraphics[width=0.9\linewidth]{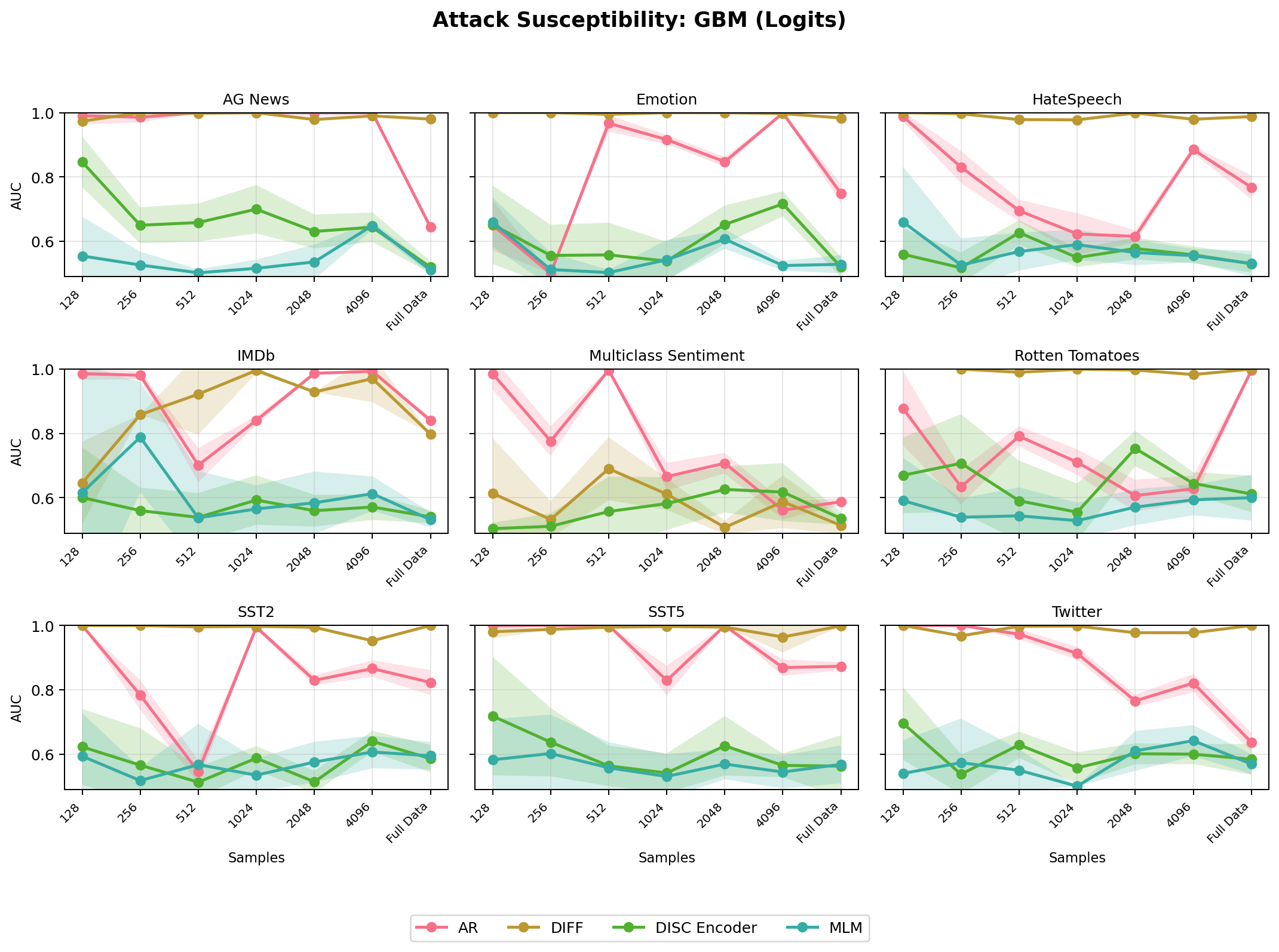}
    \caption{Attack susceptibility based on GBM (Logits) for model with 12 layers.}
\end{figure}

\begin{figure}[H]
    \centering
    \includegraphics[width=0.9\linewidth]{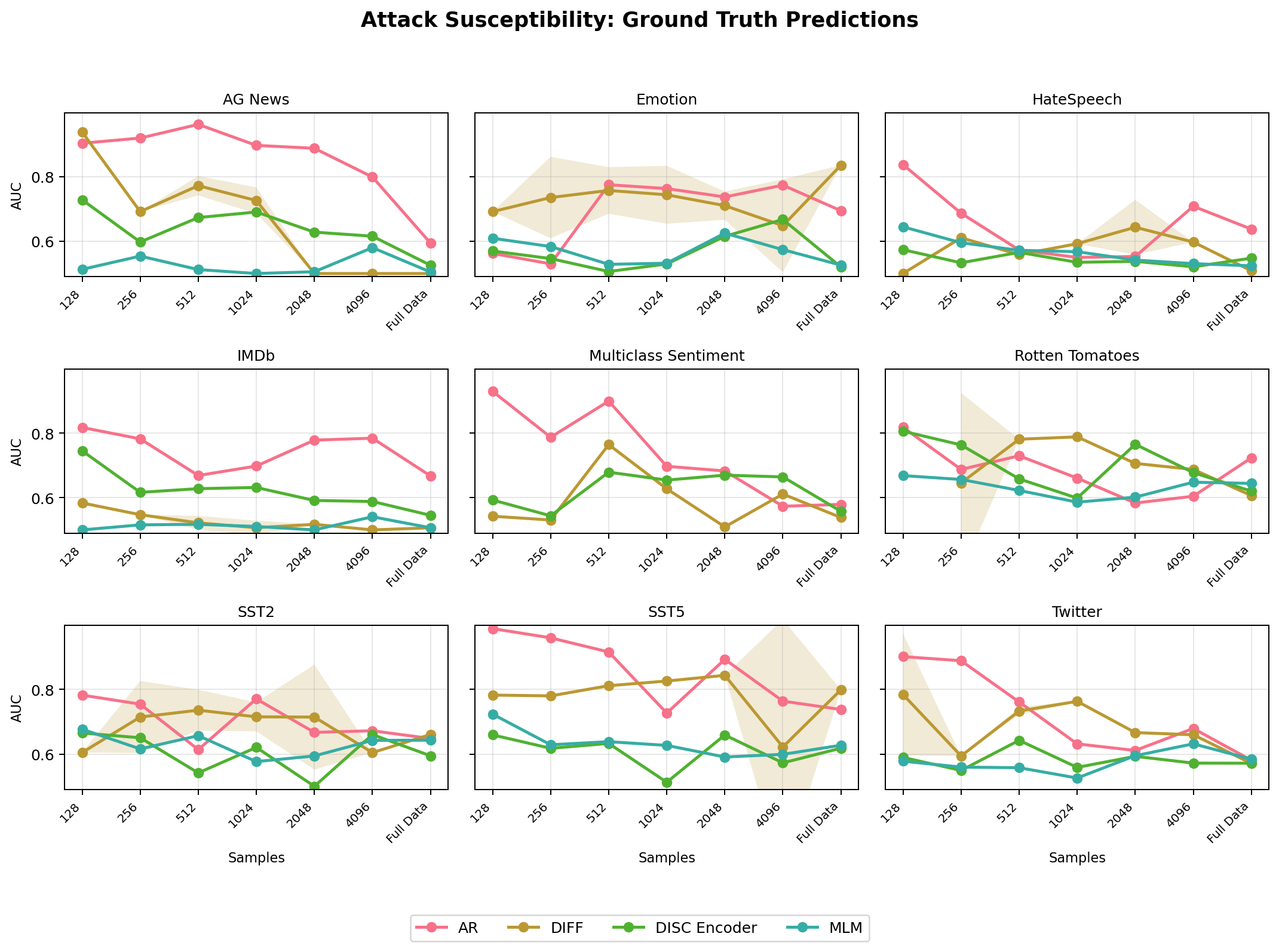}
    \caption{Attack susceptibility based on Ground Truth Predictions for model with 12 layers.}
\end{figure}

\begin{figure}[H]
    \centering
    \includegraphics[width=0.9\linewidth]{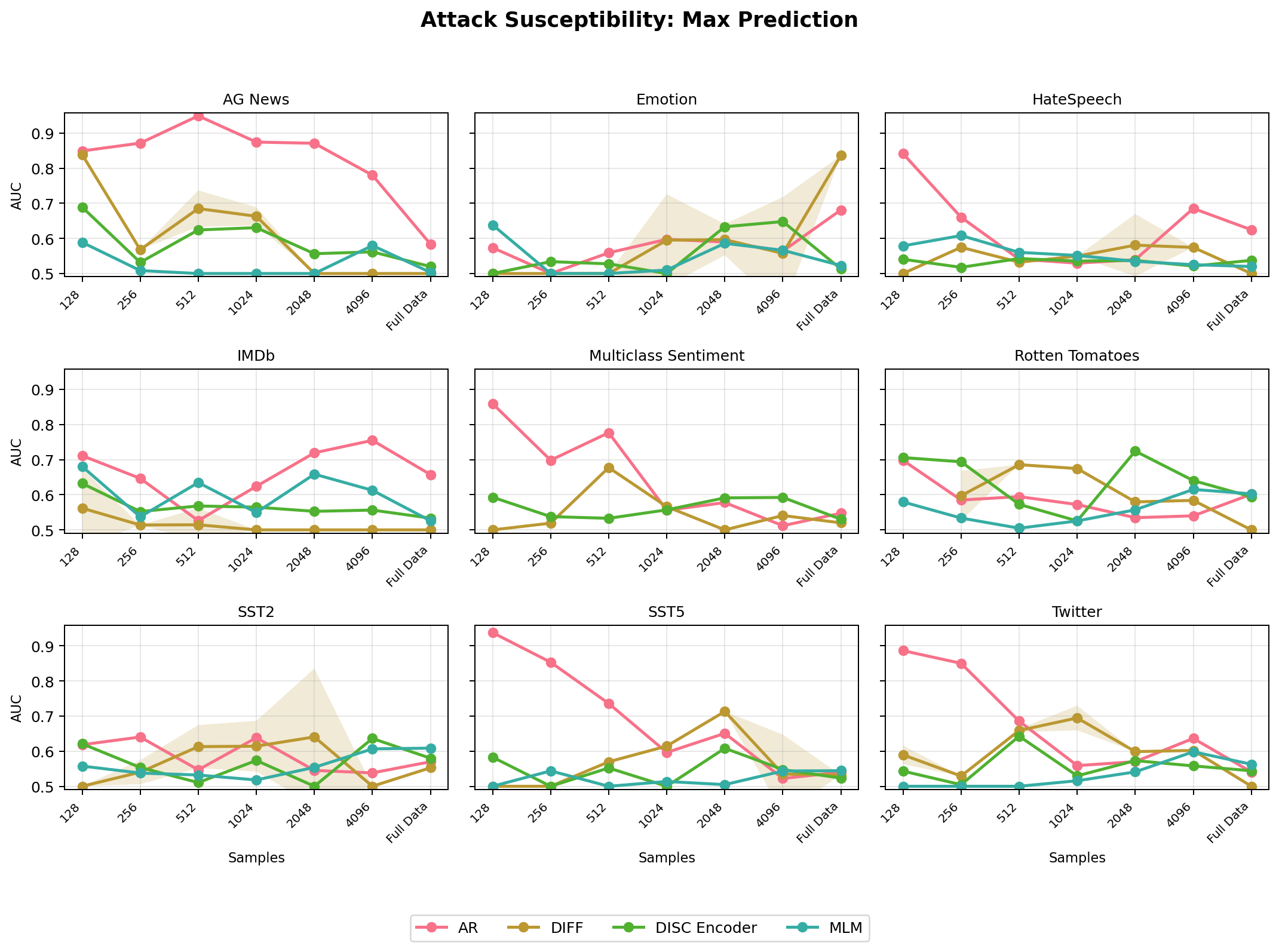}
    \caption{Attack susceptibility based on Max Prediction for model with 12 layers.}
\end{figure}

\section{Toy Illustration}
\label{sec:toy_illustration}

\subsection{Experimental Setup}

We study membership inference in a controlled synthetic setting where each input \(x \in \mathbb{R}^d\) is composed of
\[
x = \big[x_{\text{core}},\; 
% x_{\text{spu}},\; 
x_{\text{noise}}\big].
\]

\paragraph{Labels.}
Binary labels \(y \in \{-1,+1\}\) are drawn from
\[
P(y=+1)=w,\qquad P(y=-1)=1-w.
\]

\paragraph{Core feature (signal).}
The one-dimensional core feature correlates directly with the label:
\[
x_{\text{core}} \sim \mathcal{N}\!\big(y \cdot \mu,\; \sigma^2\big),
\]
where \(\mu\) (\texttt{core\_scale} in code) controls class separation and \(\sigma\) controls within-class spread. \emph{For the membership-inference experiments we match the train and test distributions}, i.e., we use the same \((\mu,\sigma)\) for both sets so that members and non-members are drawn i.i.d.\ from the same distribution.

% \paragraph{Spurious feature (disabled in plots).}
% An optional spurious cue is
% \[
% x_{\text{spu}} = \big(1 - 2\cdot \mathrm{flip}\big)\cdot y \cdot B,\quad \mathrm{flip}\sim\mathrm{Bernoulli}(1-\rho),
% \]
% with magnitude \(B\) and correlation \(\rho\). In our main figures we set \(B{=}0\) to disable this feature.

\paragraph{Noise features.}
The remaining \(d-1\) coordinates are independent Gaussian clutter:
\[
x_{\text{noise}} \sim \mathcal{N}\!\big(0,\, \sigma_{\text{noise}}^2 I_{d-1}\big).
\]

\paragraph{Training/Test sizes and sweeps.}
We generate \(n_{\text{train}}\) training samples and \(n_{\text{test}}{=}4000\) test samples. We sweep
\[
\mu \in \{0.05, 0.10, \ldots, 0.50\},\quad
n_{\text{train}} \in \{50, 200, 2000\},\quad
d \in \{16, 64, 256\},
\]
and evaluate three class-prior settings \(w\in\{0.1,0.3,0.5\}\). Unless stated otherwise, figures fix \(w{=}0.5\), \(\sigma_{\text{noise}}{=}1.0\), and \(\sigma{=}0.15\).

\begin{table}[t]
\centering
\small
\begin{tabular}{ll}
\toprule
\textbf{Parameter} & \textbf{Description} \\
\midrule
\(w\) & Class prior for \(y{=}+1\) (imbalance) \\
\(\mu\) & Core mean shift (class separation) \\
\(\sigma\) & Core feature standard deviation (train $=$ test) \\
\(\sigma_{\text{noise}}\) & Noise level for the $d{-}1$ nuisance dims \\
% \(B, \rho\) & Spurious cue magnitude and correlation (set \(B{=}0\) in main plots) \\
\(d\) & Dimensionality (\(1\) core \(+\) \(d{-}1\) noise) \\
\(n_{\text{train}}, n_{\text{test}}\) & Train/test sample counts \\
\bottomrule
\end{tabular}
\caption{Synthetic data parameters. For MIA we use matched train/test distributions.}
\label{tab:exp-params}
\end{table}

\paragraph{Models and training.}
We compare (i) Logistic Regression (LBFGS, \texttt{max\_iter} \(=10{,}000\)) and (ii) LDA (\texttt{solver=lsqr}, \texttt{shrinkage=auto}). Each configuration is run with \(5\) random seeds; we report means and shaded uncertainty bands.

\subsection{Motivation}

Our toy setup is designed to cleanly \emph{tease apart} the drivers of membership inference without architectural or optimization confounds. By controlling a few interpretable knobs, we can test how membership signals scale with statistical difficulty:

\begin{itemize}
  \item \textbf{Dimensionality ($d$):} Increasing $d$ adds nuisance directions and dilutes per-sample information, stressing generalization and potentially amplifying member--nonmember score gaps.
  \item \textbf{Sample size ($n$):} Larger $n$ reduces estimator variance and overfitting; smaller $n$ increases memorization pressure. The ratio $n/d$ serves as an effective \emph{signal budget} per parameter.
  \item \textbf{Decision boundary separation ($\mu$):} Larger $\mu$ widens class separation, boosting accuracy and confidence. This lets us study whether membership advantage tracks confidence or generalization.
  \item \textbf{Signal strength ($\mu$ and $n/d$):} Together, geometric margin ($\mu$) and sample complexity ($n/d$) summarize how much reliable signal the model can extract relative to noise.
  \item \textbf{Imbalance (class weight $w$):} Varying the class prior via a weight $w\in(0,1)$ shifts the decision threshold and posterior calibration, directly affecting confidence-based and generative scores used by MIAs.
\end{itemize}

We keep train and test \emph{i.i.d.} to isolate membership effects from distribution shift, and average over multiple seeds to separate systematic trends from randomness. This controlled regime exposes how membership advantage scales with $(d,n,\mu,w)$ and provides intuition that transfers to real datasets.

\subsection{Membership Inference Scores}

For each trained model we compute member scores on the training set and non-member scores on an i.i.d.\ test set, and report AUROC.

\paragraph{Max-probability (\texttt{auroc\_prob}).}
Given posterior estimates \(\hat p(y\mid x)\),
\[
s_{\text{prob}}(x) \;=\; \max_{y\in\{-1,+1\}} \hat p(y\mid x).
\]
This is the standard, label-agnostic confidence attack we plot for both Logistic Regression and LDA.

\paragraph{Log-joint (LDA only; \texttt{auroc\_logjoint}).}
For LDA with class priors \(P(y)\), means \(\mu_y\), and shared covariance \(\Sigma\),
\[
s_{\text{logjoint}}(x) \;=\; \max_{y} \big\{ \log P(y) + \log \mathcal{N}\!\big(x \,\big|\, \mu_y, \Sigma\big) \big\}.
\]
This generative score often differs from max-probability and is shown in our AUROC plots.

% \subsection{Membership Inference Scores}

% Given a trained model, we compute member scores on its training set and non-member scores on a disjoint test set drawn from the \emph{same} distribution. We report AUROC for each attack and also use the direction-invariant ``advantage''
% \[
% \mathrm{Adv}(\mathrm{AUROC}) \;=\; \max\{\mathrm{AUROC},\, 1-\mathrm{AUROC}\}
% \]
% when summarizing membership strength.

% \paragraph{Max-probability (\texttt{auroc\_prob}).}
% For posterior estimates \(\hat p(y\mid x)\),
% \[
% s_{\text{prob}}(x) \;=\; \max_{y\in\{-1,+1\}} \hat p(y\mid x).
% \]

% % \paragraph{Margin (\texttt{auroc\_margin}).}
% % For binary classifiers with class logits \(\ell_{+1}(x), \ell_{-1}(x)\),
% % \[
% % s_{\text{margin}}(x) \;=\; \big|\ell_{+1}(x) - \ell_{-1}(x)\big|.
% % \]
% % For Logistic Regression this is \(|w^\top x+b|\); for LDA it is the absolute difference of its linear discriminant scores.

% % \paragraph{Label-aware loss (\texttt{auroc\_ll}).}
% % Given the true label \(y\),
% % \[
% % s_{\text{ll}}(x,y) \;=\; -\big(-\log \hat p(y\mid x)\big) \;=\; \log \hat p(y\mid x).
% % \]

% \paragraph{Log-joint (LDA only; \texttt{auroc\_logjoint}).}
% For LDA with class priors \(P(y)\), class means \(\mu_y\), and shared covariance \(\Sigma\),
% \[
% \log p(x\mid y) \;=\; -\tfrac{1}{2}(x-\mu_y)^\top \Sigma^{-1} (x-\mu_y) - \tfrac{1}{2}\log\!\det(2\pi\Sigma),
% \]
% and the score is
% \[
% s_{\text{logjoint}}(x) \;=\; \max_{y} \big\{ \log P(y) + \log p(x\mid y) \big\}.
% \]

\subsection{Findings}

\begin{figure}[htbp]
\centering
\begin{subfigure}[t]{\linewidth}
  \centering
  \includegraphics[width=0.85\linewidth]{mu_trends_by_d_seeds_p05.pdf}
  \caption{$w=0.5$}
\end{subfigure}

% \medskip

\begin{subfigure}[t]{\linewidth}
  \centering
  \includegraphics[width=0.85\linewidth]{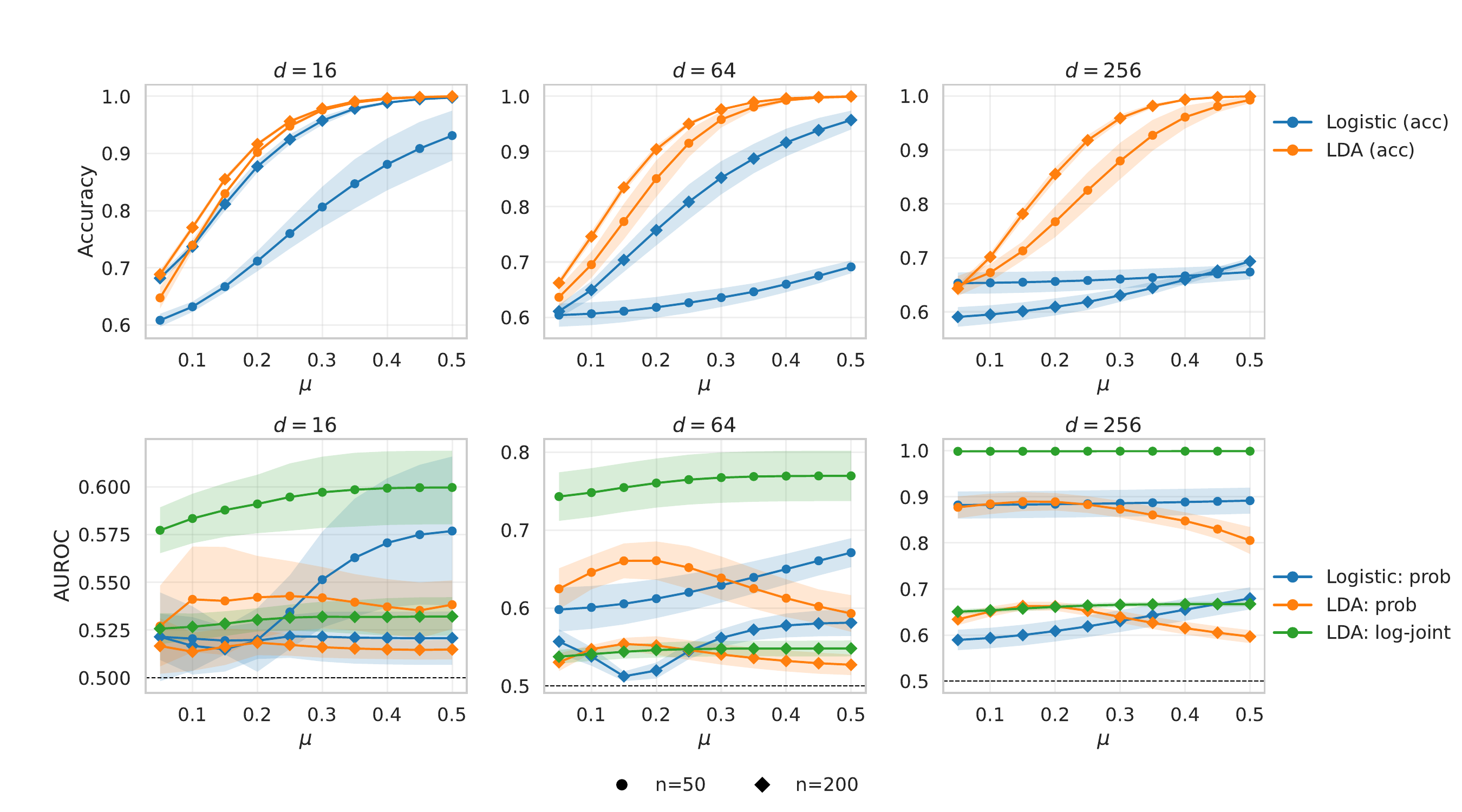}
  \caption{$w=0.3$}
\end{subfigure}

% \medskip

\begin{subfigure}[t]{\linewidth}
  \centering
  \includegraphics[width=0.85\linewidth]{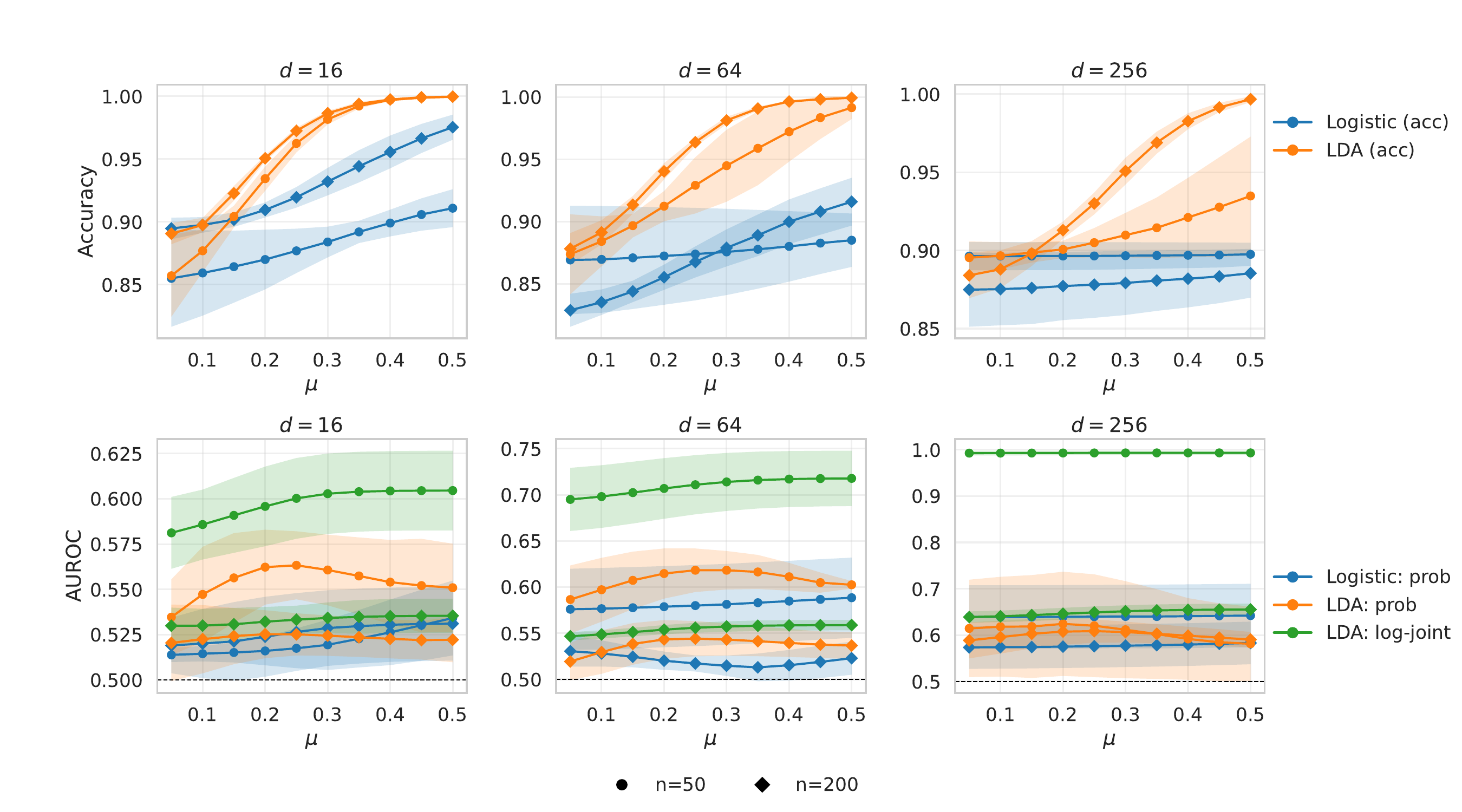}
  \caption{$w=0.1$}
\end{subfigure}

\caption{\textbf{Mean $\pm$ SEM across 5 seeds.} The three subfigures correspond to varying degree of imbalance, with $w = 0.5$ corresponding to the balanced case. Each subfigure shows: top row = test accuracy vs.\ $\mu$, bottom row = MIA $(\mathrm{AUROC})$ vs.\ $\mu$; columns are $d\in\{16,64,256\}$. Markers denote $n_{\text{train}}\in\{50,2000\}$. 
% We fix $B{=}0$, $\sigma{=}0.15$, and $\sigma_{\text{noise}}{=}1.0$.
Within each panel, color denotes series (Accuracy: Logistic/LDA; AUROC: Logistic max-prob, LDA max-prob, LDA log-joint)}
\label{fig:mu-trends-all}
\end{figure}

\textbf{Protocol \& visualization.}
For each configuration \((\mu, n, d, w)\) we train Logistic Regression and LDA on i.i.d.\ train/test data with shared core variance \(\sigma=0.15\), noise level \(\sigma_{\text{noise}}=1.0\), and no spurious cue \((B=0)\).
We run \(5\) seeds and plot means with shaded bands showing \(\pm 1.96\times\mathrm{SEM}\).
Membership is reported via AUROC and, when summarizing trends, the direction-invariant advantage \(\AdvAUROC=\max\{\mathrm{AUROC},\,1-\mathrm{AUROC}\}\).
AUROC panels include a reference line at \(0.5\). The results are given in Figure \ref{fig:mu-trends-all}.

\paragraph{Notation.}
\LRprob{} denotes the \emph{max-probability (confidence)} score \(s_{\mathrm{prob}}(x)=\max_{y}\hat p(y\mid x)\) computed from a Logistic Regression model; \LDAprob{} is the same score computed from an LDA posterior; and \LDAjoint{} denotes the \emph{log-joint} score \(s_{\mathrm{logjoint}}(x)=\max_{y}\{\log P(y)+\log \mathcal{N}(x\mid \mu_y,\Sigma)\}\) from LDA. 
All are label-agnostic membership scores; unless stated, AUROC panels report the direction-invariant advantage \(\AdvAUROC\).

\paragraph{Dimensionality ($d$) and signal per parameter ($n/d$).}
Holding \(n\) fixed, increasing \(d\) reduces test accuracy while \emph{increasing} membership advantage.
This is consistent with weaker signal per parameter (\(n/d\)): estimation error grows and models lean more on idiosyncrasies of the training set, widening member–nonmember score gaps.
Across-seed variability (std) of both accuracy and AUROC \emph{shrinks} as \(d\) rises, indicating more concentrated (though worse) accuracy and a more consistently elevated membership signal in high dimensions.

\paragraph{Geometric separation ($\mu$).}
Larger \(\mu\) (wider class separation) monotonically increases accuracy and also increases membership susceptibility:
as margins grow, both models become more confident; training points attain slightly higher confidence (and, for LDA, higher log-joints) than i.i.d.\ test points, making member/nonmember scores easier to separate.

\paragraph{Imbalance (class weight $w$).}
Moving away from balance (\(w\neq 0.5\)) improves accuracy for both methods by shifting the optimal threshold toward the minority class.
For membership, \LRprob\ exhibits a \emph{dampened} susceptibility under imbalance—posteriors saturate toward the majority, compressing train–test score gaps—whereas \LDAprob\ remains comparatively stable and often higher in \(\AdvAUROC\) across \(\mu\).
Imbalance tends to increase across-seed variability, reflecting reduced effective sample size for the minority class.

\paragraph{Generative vs.\ discriminative sample efficiency.}
Even at \(n=50\), LDA substantially outperforms Logistic Regression in accuracy; this gap persists (and often widens) as \(d\) increases (i.e., smaller \(n/d\)), reflecting the classic sample-efficiency advantage of a correctly specified generative model with shrinkage.

\paragraph{\LDAjoint\ vs.\ \LDAprob.}
Across essentially all \((d,n,\mu,w)\), \LDAjoint\ yields higher \(\AdvAUROC\) than \LDAprob.
The log-joint exposes modeled density scale: training points lie closer to estimated class means and receive larger \(\log p(x\mid y)\), hence larger \(\log P(y)+\log p(x\mid y)\), than i.i.d.\ test points.
Posteriors \(\hat p(y\mid x)\) partially compress this scale information, making \LDAprob\ consistently less susceptible.
The gap typically widens as \(d\) increases or \(n/d\) decreases, underscoring the added risk of releasing joint/likelihood values.

\paragraph{\LDAprob\ vs.\ \LRprob\ across separation.}
At small \(\mu\), \LRprob\ shows both \emph{lower} accuracy and \emph{lower} membership susceptibility than \LDAprob, matching LDA’s sample-efficiency advantage when \(n/d\) is small.
As \(\mu\) grows, \LRprob\ confidence rises steeply with margin and its \(\AdvAUROC\) increases; it can meet or exceed \LDAprob\ at larger separations.
Under stronger imbalance, this rise is \emph{dampened} for \LRprob, while \LDAprob\ remains comparatively high.

\paragraph{\LDAjoint\ vs.\ \LRprob.}
Except in a single benign regime (balanced \(w=0.5\), good separation \(\mu\), and low \(d\)), \LDAjoint\ exceeds \LRprob\ in membership advantage.
Practical takeaway: even when discriminative posteriors appear relatively less susceptible, exposing generative joint/likelihood scores can be markedly more revealing.

\paragraph{Summary.}
Stronger signal (larger \(\mu\), larger \(n/d\)) improves accuracy but also strengthens confidence-based membership cues; higher \(d\) at fixed \(n\) hurts accuracy yet sharpens membership separation.
Explicit prior modeling amplifies accuracy gains under imbalance without a commensurate reduction in susceptibility.
Generative LDA is more sample-efficient than LR, and its log-joint scores are the most vulnerable among the considered outputs.

\begin{figure}[t]
\centering
\includegraphics[width=\linewidth]{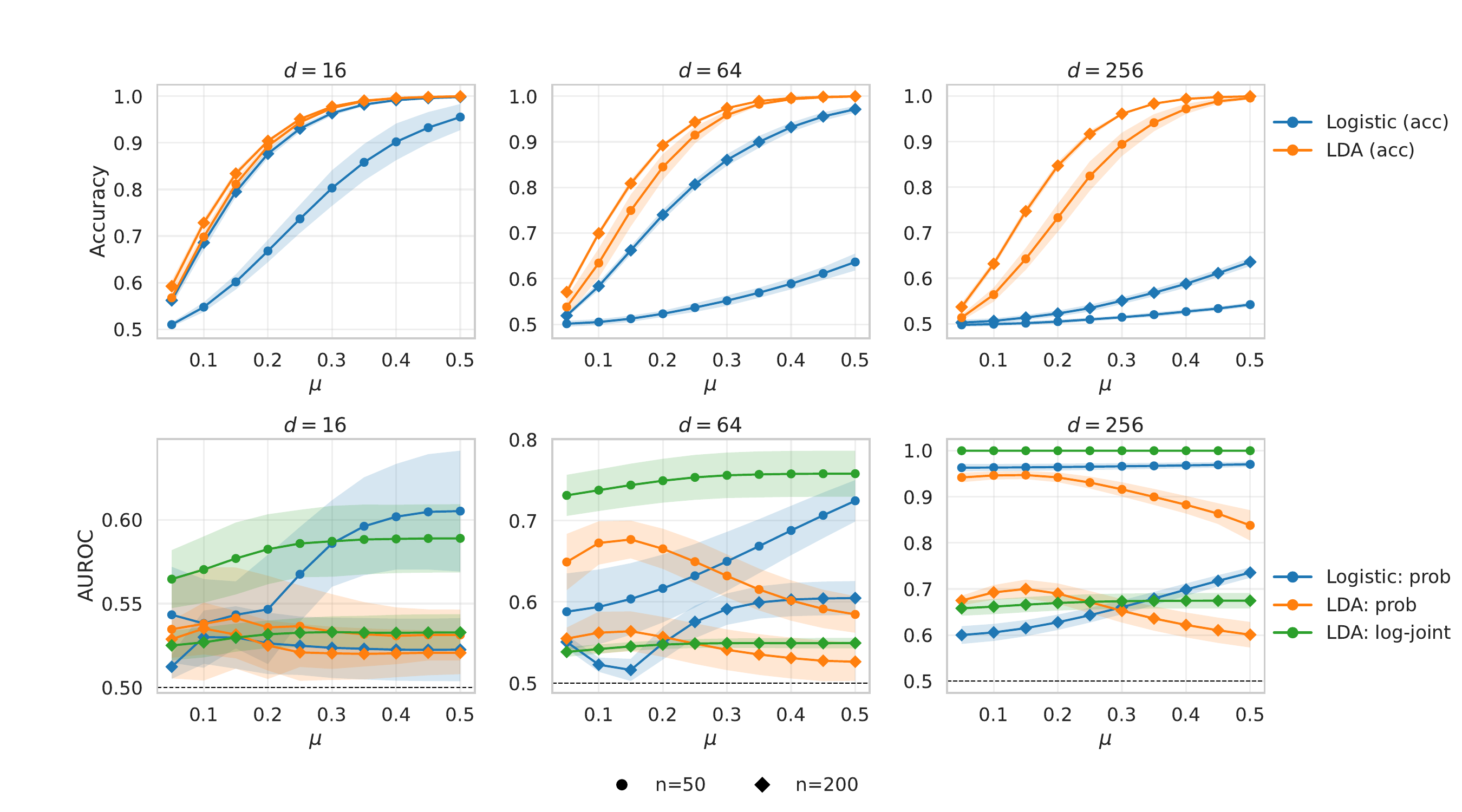}
\caption{\textbf{Mean $\pm$ SEM across 5 seeds.} Top row: test accuracy vs.\ core separation \(\mu\). Bottom row: membership \(\mathrm{Adv}(\mathrm{AUROC})\) vs.\ \(\mu\). Columns correspond to \(d\in\{16,64,256\}\). Within each panel, color denotes series (Accuracy: Logistic/LDA; AUROC: Logistic max-prob, LDA max-prob, LDA log-joint), and marker denotes \(n_{\text{train}}\in\{50,200,2000\}\). We fix \(w{=}0.5\), \(B{=}0\), \(\sigma{=}0.15\), and \(\sigma_{\text{noise}}{=}1.0\).}
\label{fig:mu-trends}
\end{figure}

\subsection{Findings under model misspecification}
\label{sec:toy_example_misspecification}

\textbf{Contamination model.}
We introduce misspecification through Huber-style $\varepsilon$–contamination \citep{huber1992robust,kasa2023avoiding} by \emph{replacing} each example (independently in train and test) with probability $\varepsilon$ by an isotropic high-variance draw that is independent of the label:
\[
X \sim \begin{cases}
\text{clean generator (core/spurious/noise)} & \text{w.p. } 1-\varepsilon,\\
\mathcal{N}(0,\tau^2 I_d) & \text{w.p. } \varepsilon,
\end{cases}
\quad \text{with }\ \tau=\texttt{tau\_mult}\cdot\sigma_{\text{noise}}.
\]
We keep the label $y$ unchanged. In our runs we use $\varepsilon=0.02$ and $\texttt{tau\_mult}=10$, yielding empirical contamination rates $\approx2.2\%$ in train and $\approx2.0\%$ in test on average (diagnostics in the CSV).

\begin{figure}[htbp]
\centering
\begin{subfigure}[t]{\linewidth}
  \centering
  \includegraphics[width=0.85\linewidth]{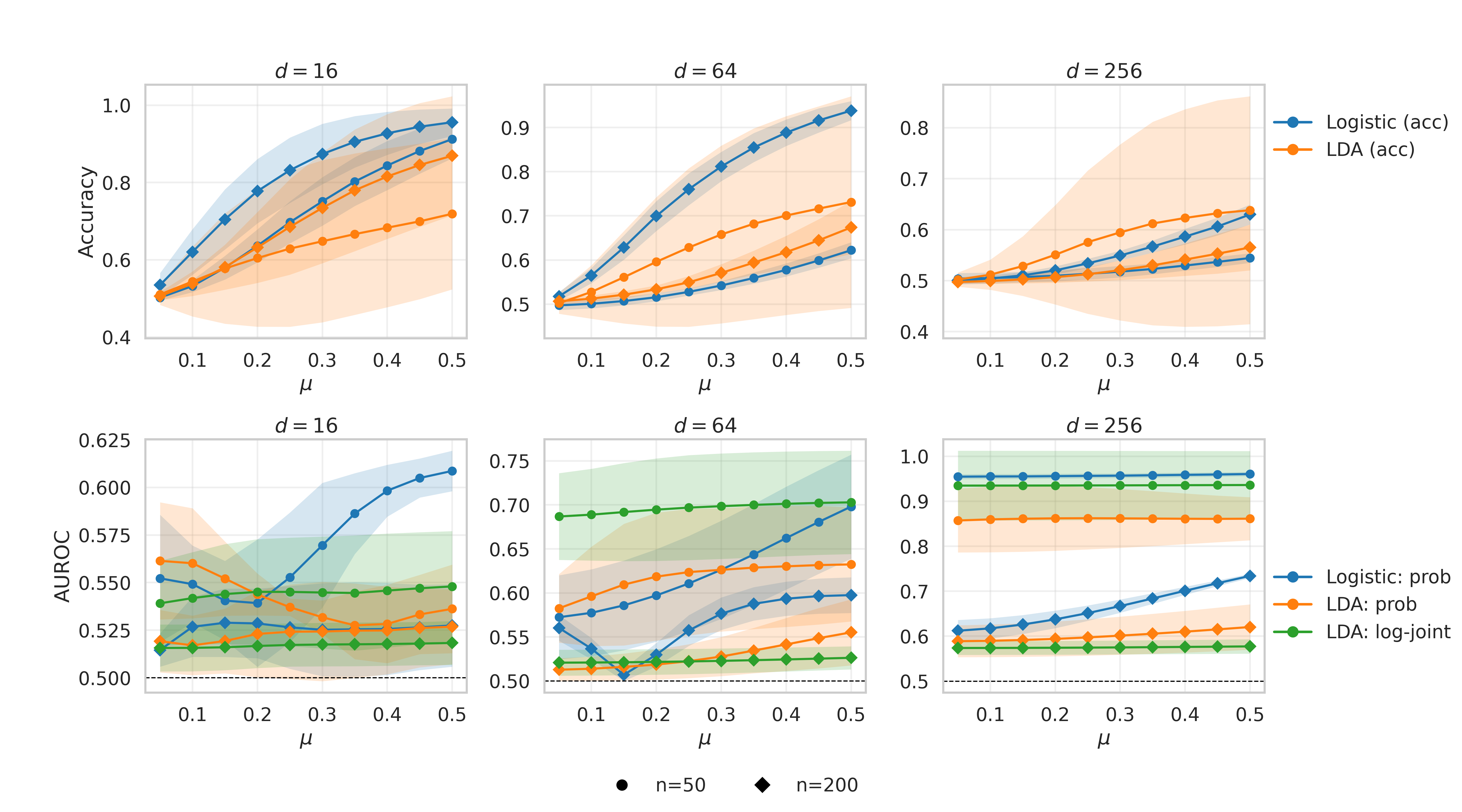}
  \caption{$w=0.5$}
\end{subfigure}

% \medskip

\begin{subfigure}[t]{\linewidth}
  \centering
  \includegraphics[width=0.85\linewidth]{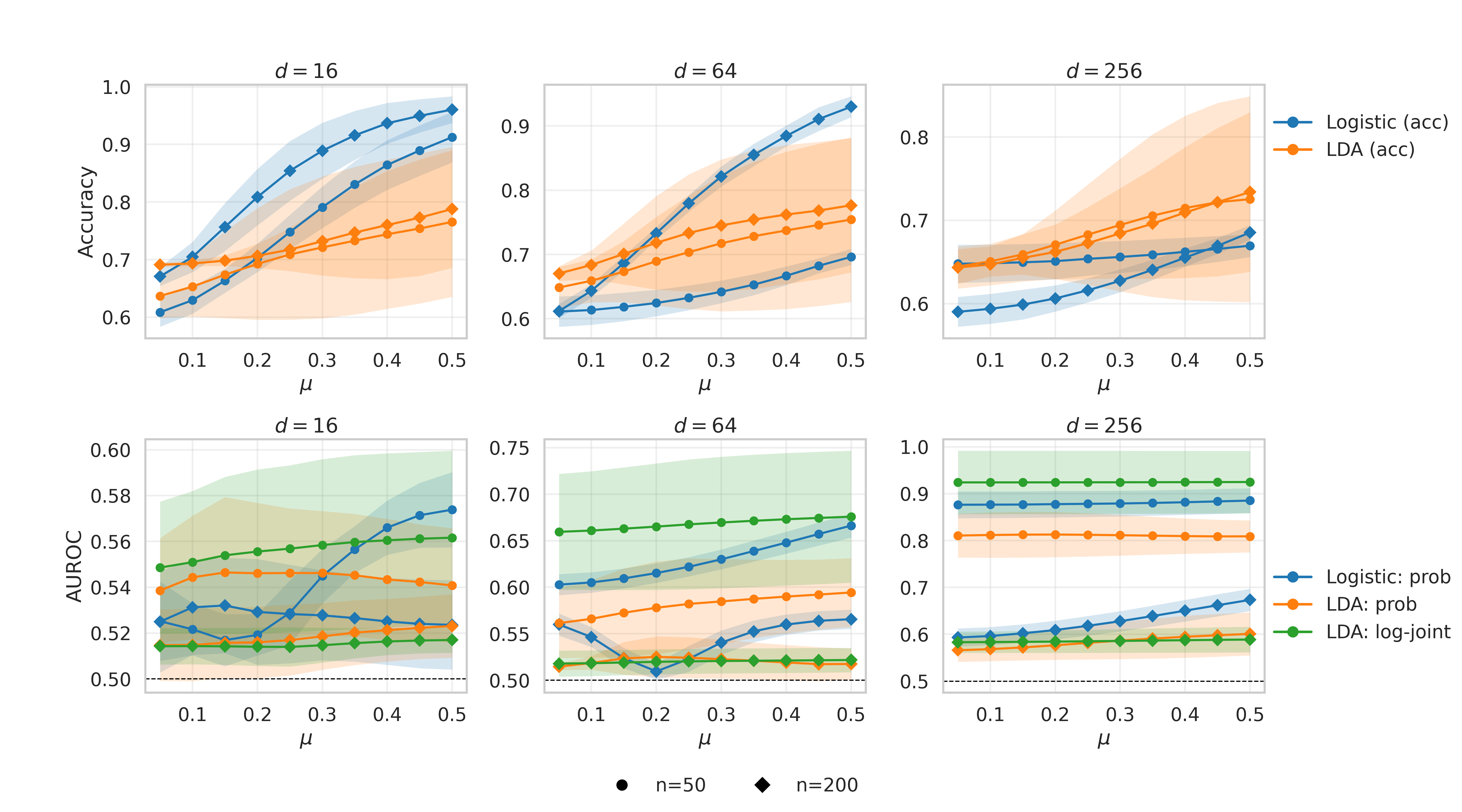}
  \caption{$w=0.3$}
\end{subfigure}

% \medskip

\begin{subfigure}[t]{\linewidth}
  \centering
  \includegraphics[width=0.85\linewidth]{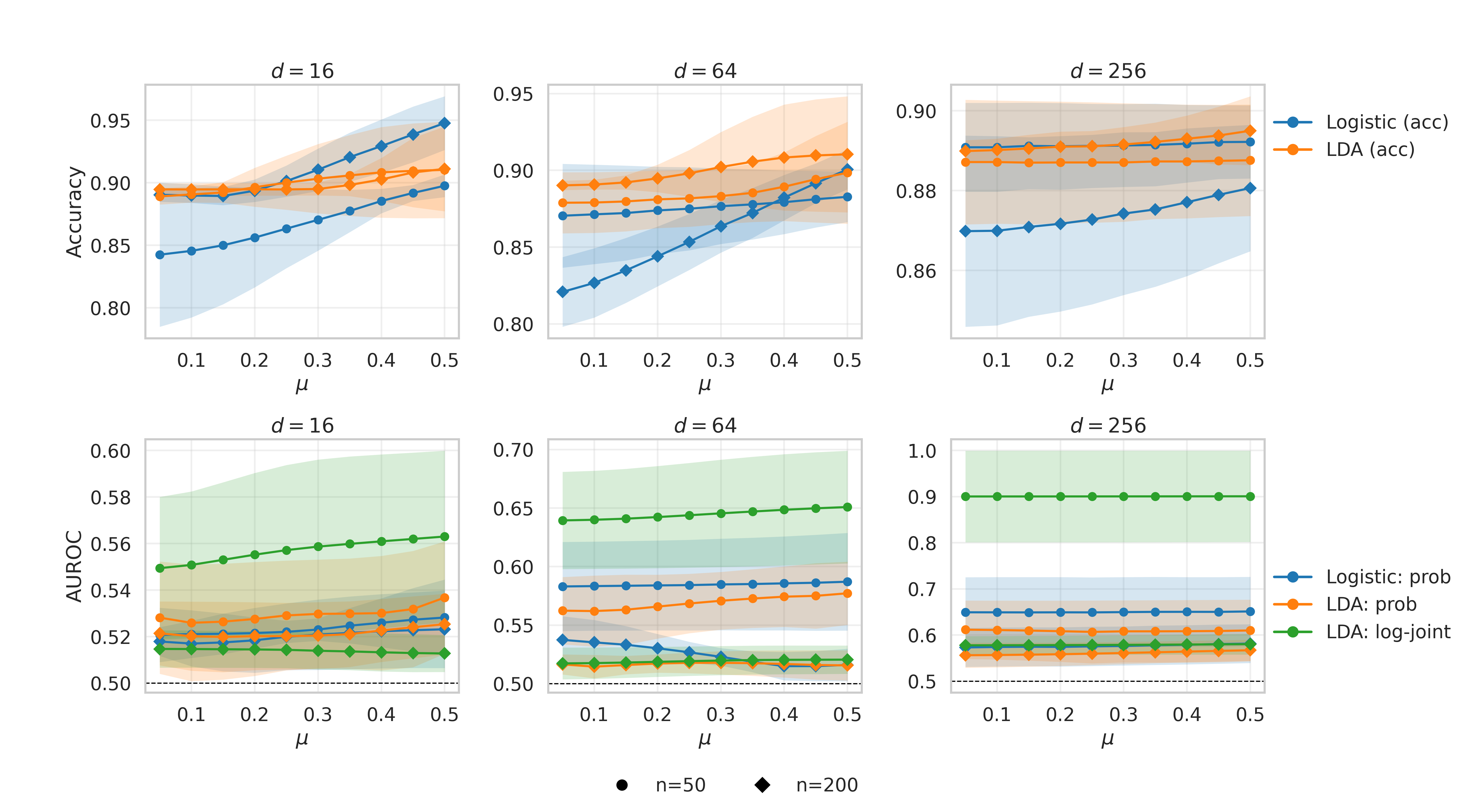}
  \caption{$w=0.1$}
\end{subfigure}

\caption{\textbf{Mean $\pm$ SEM across 5 seeds.} The three subfigures correspond to varying degree of imbalance, with $w = 0.5$ corresponding to the balanced case. Each subfigure shows: top row = test accuracy vs.\ $\mu$, bottom row = MIA $(\mathrm{AUROC})$ vs.\ $\mu$; columns are $d\in\{16,64,256\}$. Markers denote $n_{\text{train}}\in\{50,2000\}$. 
% We fix $B{=}0$, $\sigma{=}0.15$, and $\sigma_{\text{noise}}{=}1.0$.
Within each panel, color denotes series (Accuracy: Logistic/LDA; AUROC: Logistic max-prob, LDA max-prob, LDA log-joint)}
\label{fig:mu-trends-all-contaminated}
\end{figure}

\textbf{Protocol.}
Except for the contamination replacement above, the setup matches the clean case: for each $(\mu,n,d,w)$ we train Logistic Regression and LDA; we fix $\sigma=0.15$, $\sigma_{\text{noise}}=1.0$, and $B=0$; we average over $5$ seeds and summarize membership with AUROC and its direction-invariant advantage $\AdvAUROC=\max\{\mathrm{AUROC},\,1-\mathrm{AUROC}\}$ for the three scores \LRprob, \LDAprob, and \LDAjoint. The results are given in Figure \ref{fig:mu-trends-all-contaminated}.

% Contamination depresses accuracy—especially at larger $d$—while accuracy still improves with geometric separation $\mu$ and sample size $n$.  
% Under this misspecification the discriminative model is more resilient: averaged over our grid, Logistic achieves $\sim\!0.775$ vs.\ LDA’s $\sim\!0.742$ mean accuracy, and accuracy declines from $d{=}16$ to $d{=}256$ for both (e.g., LDA $0.76\!\to\!0.72$, Logistic $0.83\!\to\!0.73$), reflecting leverage from large-norm replacements and inflated covariance estimates.

% Contamination \emph{amplifies} membership signals. Advantages grow with $d$, are strongest at small $n$, and increase with $\mu$. 
% Among the three scores, \LDAjoint{} is consistently most revealing; the gap is pronounced at $n{=}50$ and diminishes toward chance by $n{=}2000$. 
% Imbalance moderates these effects: moving from $w{=}0.1$ to $0.5$ reduces accuracy (less prior help) but \emph{increases} susceptibility for all signals (\LRprob, \LDAprob, \LDAjoint), echoing the clean-data trend that class imbalance dampens confidence-based MIAs.

% \emph{Takeaway:} contamination reverses LDA’s clean-data accuracy edge in favor of Logistic, yet exposing generative density scale (e.g., \LDAjoint) remains particularly risky under outliers, especially at high $d$ and small $n$.

\paragraph{Generative vs.\ discriminative under misspecification.}
Contamination reverses LDA’s clean-data sample-efficiency edge in accuracy—Logistic is typically better—because a few large-norm replacements strongly distort shared-covariance estimation even with shrinkage. 
However, exposing density scale remains risky: \LDAjoint{} is the most susceptible membership score across most regimes we tested, particularly at high $d$ and small $n$.

The introduction of misspecification through contamination depresses accuracy overall and especially at higher $d$; accuracy increases with geometric separation $\mu$ and with sample size $n$. Under contamination the discriminative model is more resilient than LDA: averaged across the grid, Logistic attains $\!\sim\!0.775$ vs.\ LDA $\!\sim\!0.742$ mean accuracy. By dimension, accuracy drops from $(d{=}16)$ to $(d{=}256)$ for both methods (e.g., LDA: $0.76\to0.72$, Logistic: $0.83\to0.73$), consistent with inflated covariance estimates and leverage effects from large-norm points.

Contamination \emph{amplifies} member–nonmember score gaps, with stronger effects at larger $d$, smaller $n$, and larger $\mu$.
Both posterior-based signals rise with $d$ 
% (e.g., at $d{=}256$: \LRprob{} $\AdvAUROC\approx0.653$, \LDAprob{} $\approx0.621$)
, and \LDAjoint{} is consistently the most revealing 
% (e.g., $d{=}256$: $\approx0.671$)
. 
At very small sample sizes ($n{=}50$) the advantage is largest 
% (\LDAjoint{} $\approx0.712$, \LRprob{} $\approx0.664$)
; by $n{=}2000$ these fall back toward chance 
% (\LDAjoint{} $\approx0.505$, \LRprob{} $\approx0.507$)
.

As we move from extreme imbalance ($w{=}0.1$) toward balance ($0.5$), accuracy decreases (less prior help), while membership susceptibility \emph{increases} for all three signals (e.g., \LRprob{} mean $\AdvAUROC$ $\approx0.54\to0.61$, \LDAprob{} $\approx0.54\to0.58$, \LDAjoint{} $\approx0.58\to0.59$), echoing the dampening effect of imbalance on confidence-based MIAs in the clean setting.

\paragraph{Summary.}
Replacing a small fraction of points by high-variance, label-independent outliers simultaneously hurts accuracy and strengthens membership signals, with the sharpest increases at larger $d$, smaller $n$, and larger $\mu$. 
While Logistic is more robust in accuracy, releasing generative \emph{log-joint/likelihood} values (\LDAjoint) is notably more revealing than posteriors, reinforcing the recommendation to avoid exposing such scores under potential contamination.

\newpage
\section{Privacy-Utility Analysis}
\label{sec:privacy-utility-analysis}

We conducted a comprehensive privacy-utility analysis examining how privacy vulnerabilities change with model architecture and training data characteristics. Our analysis focused on four key strategies: ENC (Encoder/DISC), AR (Autoregressive), MLM (Masked Language Model), and DIFF (Diffusion), evaluating their susceptibility to Gradient Boosting Machine (GBM) based membership inference attacks.

\subsection{Methodology}

The analysis examined privacy-utility trade-offs across different model configurations, specifically investigating:
\begin{itemize}
    \item \textbf{Model Size Impact}: Varying the number of transformer layers (1, 6, 12)
    \item \textbf{Training Data Size}: Different sample counts (128, 256, 512, 1024, 2048, 4096, Full Data)
    \item \textbf{Attack Methods}: GBM-based attacks using logits and probability distributions
    \item \textbf{Utility Metric}: F1 scores across multiple text classification datasets
\end{itemize}

\subsection{Results}

\subsubsection{Model Size Analysis}

\begin{figure}[htbp]
    \centering
    \includegraphics[width=\textwidth]{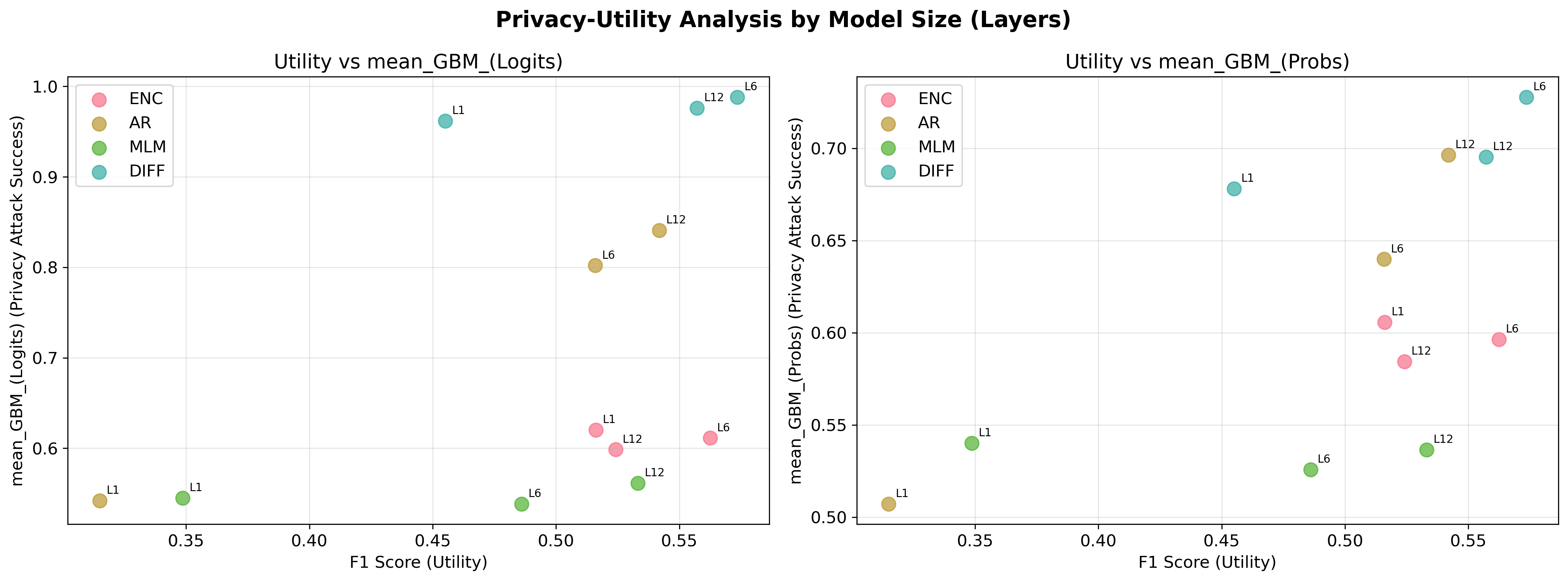}
    \caption{Privacy-utility trade-offs by model size across four strategies. Left panel shows GBM Logits attack success vs. F1 utility scores. Right panel shows GBM Probs attack success vs. F1 utility scores. Lower attack success indicates better privacy protection.}
    \label{fig:privacy-utility-model-size}
\end{figure}

Figure~\ref{fig:privacy-utility-model-size} presents the privacy-utility trade-offs as a function of model size. Each point represents the average performance across datasets, with layer annotations (L1, L6, L12) indicating the model depth.

\subsubsection{Training Sample Size Analysis}

\begin{figure}[htbp]
    \centering
    \includegraphics[width=\textwidth]{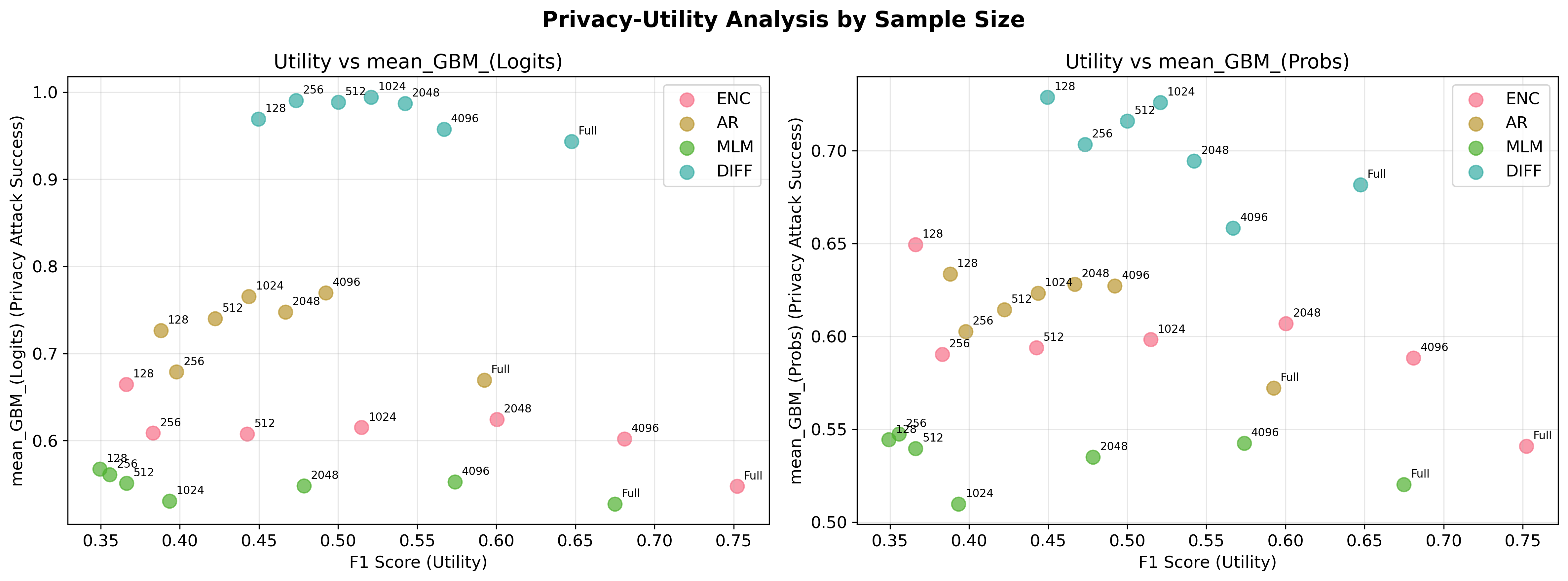}
    \caption{Privacy-utility trade-offs by training sample size across four strategies. Left panel shows GBM Logits attack success vs. F1 utility scores. Right panel shows GBM Probs attack success vs. F1 utility scores. Sample size annotations indicate training data volume.}
    \label{fig:privacy-utility-sample-size}
\end{figure}

Figure~\ref{fig:privacy-utility-sample-size} illustrates how training data size affects the privacy-utility balance. Sample size annotations indicate the number of training examples used.

\subsection{Key Findings}

\subsubsection{Strategy Performance Ranking}

Our analysis reveals significant differences in privacy-utility characteristics across strategies:

\paragraph{Utility Performance (F1 Scores):}
\begin{enumerate}
    \item \textbf{ENC}: 0.534 (±0.233) -- Best overall utility performance
    \item \textbf{DIFF}: 0.529 (±0.171) -- Second best with highest consistency
    \item \textbf{AR}: 0.458 (±0.210) -- Moderate utility with high variance
    \item \textbf{MLM}: 0.456 (±0.238) -- Lowest utility but improving with model size
\end{enumerate}

\paragraph{Privacy Vulnerability (Attack Success Rates):}
For GBM Logits attacks (lower values indicate better privacy protection):
\begin{enumerate}
    \item \textbf{MLM}: 0.548 (±0.056) -- Best privacy protection
    \item \textbf{ENC}: 0.610 (±0.086) -- Good privacy protection
    \item \textbf{AR}: 0.728 (±0.198) -- Moderate vulnerability
    \item \textbf{DIFF}: 0.976 (±0.058) -- Highest vulnerability
\end{enumerate}

\subsubsection{Model Architecture Impact}

The relationship between model size and privacy-utility trade-offs varies significantly across strategies:

\begin{itemize}
    \item \textbf{ENC Strategy}: Demonstrates optimal balance with utility peaking at 6 layers (F1=0.562) while privacy protection improves with model depth. Attack success rates decrease from 0.620 to 0.599 (GBM Logits) as layers increase from 1 to 12.
    
    \item \textbf{MLM Strategy}: Shows the most favorable privacy characteristics with consistent protection across all model sizes. Utility improves substantially with depth (0.349 → 0.533) while maintaining the lowest attack success rates.
    
    \item \textbf{AR Strategy}: Exhibits concerning behavior where utility gains (0.315 → 0.542) come at severe privacy cost, with attack success rates increasing dramatically (0.542 → 0.841) for larger models.
    
    \item \textbf{DIFF Strategy}: Despite achieving good utility, consistently shows the highest vulnerability to privacy attacks ($>95\%$ success rate) across all configurations, making it unsuitable for privacy-sensitive applications.
\end{itemize}

\subsection{Recommendations}

Based on our comprehensive analysis, we provide the following recommendations:

\begin{itemize}
    \item \textbf{General Applications}: Use ENC strategy with 6-12 layers for optimal privacy-utility balance
    \item \textbf{Privacy-Critical Systems}: Deploy MLM strategy with 12 layers for maximum privacy protection
    \item \textbf{High-Risk Scenarios}: Avoid DIFF strategy due to severe privacy vulnerabilities
    \item \textbf{AR Strategy Caution}: Monitor privacy implications carefully when scaling AR models
\end{itemize}

The analysis demonstrates that privacy and utility considerations must be carefully balanced when selecting model architectures and training strategies, with ENC and MLM strategies offering the most favorable trade-offs for privacy-preserving applications.

\section{Limitations}

Despite providing the first systematic analysis of MIAs across \textit{Discriminative}, \textit{Generative}, and \textit{Pseudo-Generative} text classifiers, our study has several limitations. First, our experiments are conducted under standard i.i.d.\ assumptions, and the results may not generalize to real-world scenarios involving distribution shifts, such as covariate or concept drift \citep{bickel2009discriminative, roychowdhury2024tackling}, where both attack success and classifier behavior could differ substantially. Second, we limit our study to only black-box attacks; it would be interesting to study if the same findings translate to white-box attacks on generative classifiers, which we leave for future work. Third, we focus on transformer-based architectures with conventional fine-tuning, omitting emerging paradigms such as few-shot or prompt-based in-context learning \citep{sun2023text, gupta2023robust}, as well as parameter-efficient adaptation techniques like LoRA \citep{hu2022lora}, which may exhibit different privacy-utility trade-offs. Fourth, our analysis is restricted to text classification; multi-modal data—including tabular, visual, or audio modalities \citep{pattisapu2025leveraging, lu2019vilbert, kushwaha2023multimodal}—may yield distinct membership leakage patterns due to richer or correlated feature structures. Fourth, we primarily evaluate standard MIA strategies and do not explore fully adaptive adversaries that could exploit model-specific quirks, ensemble behaviors, or auxiliary side information. Finally, while we study training data volume as a factor influencing vulnerability, other aspects such as pretraining data composition, model calibration, or data augmentation strategies may also impact privacy risks but remain unexplored. These limitations suggest that while our findings provide foundational insights, extending analyses to diverse settings and adaptive attacks is necessary to fully understand and mitigate privacy risks in generative classification systems.

\section{Contributions \& Acknowledgment}
\textbf{Owais} led the experimental efforts, assessing model susceptibility, generating the plots for the paper, contributing to the writing, and participating in theoretical discussions. \textbf{Siva} proposed the research direction, brainstormed initial ideas, developed the toy experiments and theoretical results, and contributed significantly to the paper's writing. \textbf{Sumegh} derived the theoretical upper and lower bounds for the optimal MIA advantage, proposed divergence-based metrics to analyze leakage, generated key results and plots, and wrote the Results \& Discussions and Experimental Setup sections. \textbf{Karan} came up with the intuition as to why memorizing P(x) could be a factor in the vulnerability of generative classifiers, played a crucial coordinating role across various stakeholders, owning the writing for later sections of the paper, ensuring experiment completion, and leading the workshop submission process. \textbf{Nikhil} contributed to the initial draft's abstract, introduction, and approach sections, implemented the pseudo-generative and true generative models used in the study, and proposed linking class imbalance to memorization. \textbf{Santhosh} conducted the diffusion model experiments, notably extracting the negative log-likelihood bound for threshold-based attacks, and ensured the codebase was reproducible. \textbf{Sumit} fostered research curiosity, supported key discussions, and provided valuable feedback on the manuscript.

\end{document}